\documentclass[11pt,a4paper]{article}
\usepackage{amsmath}
\usepackage{amsfonts}
\usepackage{amssymb}
\usepackage[scr=boondoxupr]{mathalpha}
\usepackage{lmodern}
\usepackage{graphicx}
\usepackage[left=2.5cm,right=2.5cm,top=2.5cm,bottom=2.5cm]{geometry}
\usepackage{tensor}
\usepackage{braket}
\usepackage[nottoc]{tocbibind} % Puts references in toc
\usepackage{bbm} % provides \mathbbm{1}
\usepackage{slashed} % provides \slashed{#}
\usepackage{xcolor} % loads all of color + more (i think) -chris
\usepackage{cite} 
\usepackage{multirow}
\usepackage{color} % not to have squares on links
\usepackage{hyperref} % internal links from toc to document
\hypersetup{
	bookmarksnumbered=true,
    colorlinks=true, % make the links colored
    linkcolor= black, % color TOC links in blue
    urlcolor= black, % color URLs in red
    citecolor = black,
    linktoc=all % 'all' will create links for everything in the TOC
}

\allowdisplaybreaks  % allow to break the content of align between pages

\numberwithin{equation}{section} % equation numbering (X.Y)

\DeclareMathOperator{\tr}{tr}

\newcommand{\DD}[1]{\! \mathscr{D}#1\ }
\newcommand{\dd}[1]{d#1\ }
\newcommand{\WW}{ \mathfrak{F} }

\newcommand{\Lagr}{ { \mathcal{L} } }

\newcommand{\1}{ { \mathbbm{1} } }

\newcommand{\cd}{ {\mathcal D} }

\newcommand{\intl }{\int\limits  }
\newcommand{\Riem}{ { \text{Riem} } }

\def\a{\alpha}
\def\b{\beta}
\def\g{\gamma}
\def\d{\delta}
\def\ui{\underline{i}}
\def\uj{\underline{j}}
\def\uk{\underline{k}}
\def\ul{\underline{l}}
\newcommand{\bsubeq}{\begin{subequations}}
\newcommand{\esubeq}{\end{subequations}}

\definecolor{ck}{rgb}{0.858, 0, 0.478} %christian comment color
\definecolor{lc}{rgb}{0, 0.5, 0.3} %lorenzo comment color
\definecolor{gtm}{rgb}{0, 0.478, 0.858} %gabriele comment color

\usepackage[normalem]{ulem} %allows a strikeout of text
\usepackage{float} %here briefly to make table look nicer

\begin{document}

\begin{center}
\vspace*{1cm}

{\Large\bf Conformal anomalies for (maximal) 6d conformal supergravity
}

\vspace{1cm}

{Lorenzo Casarin,\textsuperscript{1,2}
 Christian Kennedy\textsuperscript{3}
and
Gabriele Tartaglino-Mazzucchelli\textsuperscript{3} 
}

\vspace{0.5cm}

{\em
\vspace{0.5cm}
\textsuperscript{\rm 1}Institut f\"ur Theoretische Physik  \\ 
Leibniz Universit\"at Hannover \\
Appelstra\ss{}e 2, 30167 Hannover, Germany \\
\vspace{0.5cm}
\textsuperscript{\rm 2}Max-Planck-Institut f\"u{}r Gravitationsphysik (Albert-Einstein-Institut)  \\
Am M\"u{}hlenberg 1, DE-14476 Potsdam, Germany  \\
\vspace{0.5cm}
\textsuperscript{\rm 3}School of Mathematics and Physics, University of Queensland,
\\
 St Lucia, Brisbane, Queensland 4072, Australia
}
\\[1cm]
 \texttt{lorenzo.casarin@\{itp.uni-hannover.de, aei.mpg.de\}}\\
 \texttt{christian.kennedy@uq.edu.au}\\
 \texttt{g.tartaglino-mazzucchelli@uq.edu.au} 

\end{center}
\vspace{0.9cm}
%
%%%%%%%%%%%%%%%%%%%%%%%%%%%%%%%%%%%%%%%
\begin{abstract}
\noindent 
We compute the conformal anomalies for \(6d\) \((2,0)\) conformal supergravity by direct calculation in component fields.
The main novel results consist of the type-B anomaly coefficients for the gravitino and the 3-form, as well as their explicit quadratic action on some specific backgrounds. We also comment on the graviton contribution, whose Lagrangian is essentially given by the \(\mathscr Q\)-curvature.
We confirm the expectation that, when coupling \((2,0)\) conformal supergravity to \(26\) copies of the \((2,0)\) tensor multiplet, the resulting theory is free of conformal anomalies. We also consider the conformal anomalies for its \((1,0)\) truncation and confirm their relation with the chiral anomaly polynomial recently derived. For calculating the anomalies, we work with an Einstein on-shell background and make a factorised Ansatz for the operators governing the quadratic fluctuations. This reduces the calculation to evaluating heat-kernel coefficients of standard 2-derivative operators. We fix and check our Ansatz against the explicit evaluation of the component-field supergravity action in some cases.

\end{abstract}
%%%%%%%%%%%%%%%%%%%%%%%%%%%%%%%%%%%%%%%
 
\newpage
\tableofcontents

\setcounter{footnote}{0}

\newpage

%%%%%%%%%%%%%%%%%%%%%%%%%%%%%%%%
\section{Introduction and summary}
%%%%%%%%%%%%%%%%%%%%%

Conformal anomalies are fundamental defining properties of conformal field theories. Indeed, they have been the subject of a thorough investigations in 2 and 4d \cite{Osborn:1993cr,Duff:1980qv,Duff:1993wm,Bonora:1985cq}. Similar analyses have been carried out also in 6d  \cite{Bonora:1985cq,Cordova:2015fha,Cordova:2015vwa,Herzog:2013ed,Bastianelli:2000hi}.
The conformal anomaly \(\mathscr A\) in 6 dimensions takes the form\footnote{
For convenience we factor out the universal normalisation \((4\pi)^3\) in our definition of \(\mathscr A\) and therefore of \((a,c_i)\).  We also drop scheme-dependent  total-derivative contributions.} \cite{Bonora:1985cq,Deser:1993yx,Bastianelli:2000hi}
\begin{equation}\label{aaa}
\mathscr{ A}  
:=  g^{mn}\braket{T_{mn}}   \cdot (4\pi)^3
= - a \, \mathbb{E}_6 + c_1 \,  I_1 + c_2 \,  I_2 + c_3 \, I_3
\,,
\end{equation}
where \(\mathbb E_6\) is the  Euler density in \(6\) dimensions and the invariants \(I_i\) are built from the Weyl tensor (\(I_{1},I_2\sim\mathrm{Weyl}^3\), \(I_3\sim \mathrm{Weyl}\,\cd^2\,\mathrm{Weyl} \) -- see Appendix~\ref{app:notQFT} for the explicit expressions). The anomaly \(\mathscr A\)  in \eqref{aaa} also appears in the logarithmically UV divergent part of the effective action, and the anomaly coefficients \(a,c_i\) enter in the stress tensor two-, three- and four-point functions.

Six dimensions are somewhat special for conformal field theories, as no interacting unitary supersymmetric conformal field theory can exist in more than six dimensions \cite{Nahm:1977tg}. 
It is however difficult to study unitary theories in six dimensions due to the lack of perturbative renormalizability for standard 2-derivative actions. Relaxing unitarity, perturbatively renormalizable higher-derivative theories can be constructed and provide formal UV completions of standard 2-derivative theories and can therefore help to shed light on the properties of conformal field theories and of the space of QFTs in higher dimensions, see e.g.\ \cite{Ivanov:2005qf,Buchbinder:2016url,Buchbinder:2018lbd,Buchbinder:2020ovf,Osborn:2016bev,Gracey:2015xmw,Gracey:2016ncd,Gracey:2020kbb}.

Higher-derivative theories, despite being non-unitary, provide concrete examples that allow one to explore the CFT space and produce explicit data potentially useful to learn general lessons \cite{Beccaria:2015uta,Beccaria:2015ypa,Beccaria:2017dmw,Tseytlin:2013fca,Mukherjee:2021alj,Casarin:2023ifl,Safari:2021ocb,Stergiou:2022qqj,Osborn:2016bev}.
A relevant example of information that can be learnt is the linear (affine) relation \cite{Beccaria:2015ypa} determining the conformal anomaly \eqref{aaa} in terms of the numerical (theory-dependent) coefficients in the chiral anomaly polynomial for supersymmetric 6d theories. By looking at the  examples available, some of which consisting of higher derivative models, \cite{Beccaria:2015ypa} obtained explicit expressions in terms of a universal (theory-independent) parameter \(\xi\). At that time, there were not enough additional examples available to compute this parameter directly, which had to be fixed via indirect considerations. In fact, the original value obtained in \cite{Beccaria:2015ypa} turned out to be incompatible with the later results of \cite{Yankielowicz:2017xkf,Beccaria:2017dmw}, which indicated a value \(\xi=-\frac89\).

For generic, non-supersymmetric, field theories the four coefficients in \eqref{aaa} are independent. When  supersymmetry is present there are additional constraints  on the non-topological (type-B, following \cite{Deser:1993yx}) sector  of the anomaly:
 only the following two combinations of \(I_i\) admit a \((1,0)\) supersymmetric extension \cite{Butter:2016qkx} 
\begin{equation}\label{www}
W^{(1,0)}_1 = 4 I_1 - I_2 - I_3 \,, \qquad\qquad
W^{(1,0)}_2 = 2 I_1 + I_2  \,.
\end{equation}
This forces the conformal anomalies to be a linear combination of \(W^{(1,0)}_{1,2}\).
Correspondingly, only two of the \(c_i\)'s coefficients are independent, or equivalently they have to obey a linear relation, that is
\begin{equation}\label{ci1}
(1,0) :
\qquad\qquad
c_1 - 2c_2 +6 c_3 =0
\,.
\end{equation}
Even more restrictive is the \((2,0)\) case. Indeed, only a particular linear combination of \(W^{(1,0)}_{1,2}\) admits a \((2,0)\) superextension \cite{Butter:2017jqu}
\begin{equation}\label{ww2}
W^{(2,0)}=  W^{(1,0)}_1 + 4 W^{(1,0)}_2 = 12 I_1 + 3 I_2 - I_3\,,
\end{equation}
and as a consequence there is only a single \(c\) coefficient,
\begin{equation}\label{aab}
(2,0) : \qquad\qquad 
\mathscr{ A}  
= - a \, \mathbb{E}_6 + c\, W^{(2,0)}
\,.
\end{equation}
For example, in  the case of the \((2,0)\) tensor multiplet \(\text T\) one has, from \cite{Bastianelli:2000hi},
\begin{equation}\label{fef}
\text{\((2,0)\) tensor multiplet T}:
\qquad\qquad
a_{\text T} = \frac{7}{1152}  \,, \qquad
c_{\text T}  =- \frac{1}{36} \,.
\end{equation}
The terms  \(W^{(1,0)}_{1,2}\) and \(W^{(2,0)}\) are therefore the gravitational Lagrangians that admit a supersymmetric completion, namely conformal supergravities in 6d. 
Some quantum aspects of these theories have indeed been investigated in the literature before.
As spin-2 CFTs, conformal supegravities constitute the natural examples to provide additional data not yet available and cross-check indirect results. We also note that similar calculations about the quantum nature of four-dimensional conformal supergravity have been done long ago, see e.g.~\cite{Fradkin:1985am} for a comprehensive review. 
Furthermore, some aspects of the quantum gravitational theory defined by \(W^{(2,0)}\) have been investigated in \cite{Lovrekovic:2015thw,Knorr:2021lll,Martini:2021lcx}.%
\footnote{Quantum properties of standard Einstein theory in six dimensions have also been subject of study for a long time, see e.g.~\cite{VanNieuwenhuizen:1977ca,Dunbar:2002gu,Gibbons:1999qz,Critchley:1978kb,Bastianelli:2023oca}.
} 

More in general, in recent years, higher-derivative supergravity has been instrumental in opening a new era of precision tests of holography. Next to leading order, AdS/CFT relates on the CFT side to non-planar contributions to various observables that have recently been computed by using, e.g., localisation techniques, while on the gravity side, higher-derivative supergravity is used to describe quantum corrections in string theory. A systematic study of higher-derivative supergravity and their quantum properties, including the important example of conformal-supergravity actions (see, e.g., the role played by the Weyl-squared invariant in 4d), is crucial for future developments of precision tests of holography. Hence, our work fits in the challenging research program aimed at classifying higher-derivative supergravities towards novel applications in AdS/CFT.
In particular, understanding the properties of conformal supergravity in six dimensions is an important step towards extending the works of \cite{Bobev:2020egg,Bobev:2021oku} to the higher-dimensional case. \(\text{AdS}_6\) solutions of Poincaré supergravities relevant for \(\text{AdS}_6/\text{CFT}_5\) holography have been constructed in \cite{DHoker:2016ujz,DHoker:2016ysh,DHoker:2017mds,DHoker:2017zwj,Jafferis:2012iv,Brandhuber:1999np}, and display a significantly more complicated structure than those used in lower-dimensional examples.
It is however important to notice that these \(\text{AdS}_6\) solutions preserve only 16 supercharges, so that the maximal theory discussed here is to be supplemented by less supersymmetric terms.
A simpler set-up relevant for \(\text{AdS}_3/\text{CFT}_2\) concerns maximally supersymmetric \(\text{AdS}_3\times \text{S}^3\) solutions corrected by higher-derivative terms \cite{Baggio:2014hua,Butter:2018wss,Butter:2017jqu}. It is then of interest to understand if the application of conformal supergravity in capturing quantum effects in an Effective Field Theory setting, as in e.g.~\cite{Bobev:2020egg,Bobev:2021oku} for $d=4$, can be applied at the quantum level extending the heat kernel analysis of \cite{Bobev:2023dwx} to six dimensions. However, an important point that needs to be dealt with is the presence of ghost states due to the higher-derivative nature of the theory. We will comment again on this. For more results and discussions related to new avenues in precision tests of holography by using higher-derivative supergravity see also \cite{Bobev:2020zov,Bobev:2021qxx,Liu:2022sew,Hristov:2021qsw,Hristov:2022lcw,Bobev:2022bjm,Cassani:2022lrk,Cassani:2023vsa,Gold:2023ymc,Cassani:2024tvk,Ma:2024ynp}, the reviews \cite{Ozkan:2024euj,Hristov:2024cgj}, and references therein.

The purpose of this paper is to compute the conformal anomaly of \((2,0)\) conformal supergravity in six dimensions by direct calculation of the contributions of its component fields.
On the basis of   \(a\)-anomaly and Casimir-energy cancellation,  Beccaria and Tseytlin conjectured in \cite{Beccaria:2015uta} that \((2,0)\) conformal supergravity coupled to \(26\) copies of the \((2,0)\) tensor multiplet \(\mathrm T\) should be anomaly-free. This was then argued in \cite{Beccaria:2015ypa} comparing the chiral anomaly polynomials. However, a direct calculation was still missing. In particular, the component-field contributions to the \(c_i\)'s of the gravitino and the 3-form had not been obtained yet. Furthermore,  \cite{Pang:2012rd} computed the graviton anomaly on an Einstein background with the additional assumption of parallel curvature, under which only two of the invariants  \(I_i\)'s are independent, thus determining the \(c_i\)'s up to a numerical parameter.  Then,  \cite{Beccaria:2017dmw} appears to fix its value by considering an additional background, which, however, turns out to be a subcase of those used in \cite{Pang:2012rd},  thus the validity of the result was unclear.\footnote{More in detail, \cite[Appendix D]{Beccaria:2017dmw}
reproduces the calculation of \cite{Pang:2012rd}  in the particular case of a Ricci-flat background (which is a subcase of the Einstein backgrounds considered in \cite{Pang:2012rd}) with the same assumption of parallel curvature (thus only two independent \(I_i\)'s). However, this appears to produce an additional piece of information that fixes the values of all \(c_i\)'s, in seeming contradiction with the fact that the calculation should be a subcase of  \cite{Pang:2012rd}. The solution to this puzzle is the fact that the parallel curvature condition is not needed to obtain a factorized operator on Einstein backgrounds in the first place.
We thank Arkady Tseytlin for clarifications on the calculation of~\cite[Appendix D]{Beccaria:2017dmw}.
} However, \cite{Aros:2019tjw} revisited the calculation on a generic Einstein background, showing that the parallel curvature condition is not necessary and confirmed the anomaly coefficients.\footnote{We became aware of \cite{Aros:2019tjw} only while preparing this manuscript, when we had already independently reproduced all their results. This is a reassuring independent consistency check of both the results of \cite{Aros:2019tjw} and our analysis.}

Computing   conformal anomaly is, in principle, a standard calculation usually performed via the evaluation of the heat-kernel coefficient of the quadratic operator on a geometric background, see e.g.\ the reviews \cite{Duff:1993wm,Vassilevich:2003xt,Birrell:1982ix} for historical remarks and references.
In fact, one can express the UV logarithmically divergent part of the effective action, and equivalently the conformal anomaly, in terms of the determinant of the quadratic operator via the heat-kernel method. By providing a representation of the determinant of a differential operator preserving background covariance, the heat kernel is particularly suited to study 1-loop effects.
In the present case, the  relevant terms are captured by
\begin{equation}\label{iab} 
\Gamma_\infty 
=  - \frac{   \log   \Lambda  }{(4\pi)^{3}}
 \int \!  \sqrt{g }  \ b_6 
 \,,
 \qquad     
 \mathscr{ A} =   b_6\,,
 \qquad       
 b_6 = b_6(\Delta_{\text{b}})-b_6(\Delta_{\text{f}}) \pm b_6(\Delta_{\text{gh}})\,,
\end{equation}
where \(b_6\) is a combination of the heat-kernel coefficients  \( b_6(\Delta) \) of the operators \( \Delta \)  governing the quadratic fluctuations, while \(\Lambda\) is a UV cutoff and the renormalisation scale \(\mu\) is suppressed. In writing \eqref{iab} we assumed real bosons (b) and Weyl or Majorana fermions (f) in gamma-matrix representation.  The last term schematically represents ghost (gh) contributions.

Due to the significant increase in computational complexity for the six-dimensional case compared to the four-dimensional one, the supersymmetric completions of \(W^{(1,0)}_{1,2}\) have been only relatively recently constructed in conformal superspace \cite{Butter:2016qkx} and then evaluated in components in \cite{Butter:2017jqu}. This was partially uplifted to the \((2,0)\) case (to a level that is sufficient for the evaluation of the anomaly), and the intermediate expressions show remarkable complexity. 
The quadratic operators associated with the one-loop fluctuations for the higher-rank fields are of order higher than five or involve non-minimal principal symbols, for which the heat-kernel coefficients have not been worked out yet.  However, it has been observed in several examples \cite{Fradkin:1983tg,Fradkin:1985am,Casarin:2019aqw,Beccaria:2017dmw,Pang:2012rd}, that higher-derivative operators factorize for all the fields in the supermultiplet on specific types of on-shell backgrounds, 
therefore reducing the effective action (and thus the conformal anomaly) to combinations of determinants of simpler lower-order operators. The typical gravitational example of such a background is Einstein manifolds, namely \(R_{mn} = R g_{mn}/d\) with constant \(R\), and with all other fields set to zero.
We shall therefore try to exploit this phenomenon to constrain the form of the operators and reduce the calculation to known heat-kernel coefficients; combining this with a certain Ansatz for the factorisation, for the first time, we are able to completely fix the kinetic operators on geometric Einstein background and thereby compute the anomaly coefficients.
This Ansatz for the explicit form of the \(W^{(2,0)}\) Lagrangian expanded to quadratic order on the fluctuations over an Einstein background is also a nontrivial by-product of our analysis, given the complexity of the Lagrangian. We motivate and test our Ansatz at length and hope to come back to a more complete derivation in the future.

Table~\ref{tab} summarises\footnote{ \label{foot:ferm} 
Note that the constructions of \cite{Butter:2016qkx,Butter:2017jqu,Butter:2018wss} are in the Lorentzian setting and the 6d spinors are subject to a Symplectic-Majorana-Weyl (SMW) condition. In part of this paper, including this introduction, we adopt an Euclidean notation to have formally convergent functional integrals. For convenience, we translate Lorentzian SMW fermions into Euclidean Majorana fermions rather than using Euclidean Weyl fermions.  Conformal anomalies, which are parity-even and sit in the real part of the effective action, are insensitive to this choice. This also justifies our use of both self-dual and anti-self-dual parts of the Euclidean three-form in intermediate stages of calculations.  
} 
the contributions to the anomaly coefficients of the different fields,\footnote{We thank D.\ Díaz for spotting a typo in the original version of the table. 
} including their multiplicities \(m^{(\mathcal{N},0)}\) for \((\mathcal{N},0) \) conformal supergravity (CSG), \(\mathcal{N}=1,2\), as well as the number of dynamical (on-shell) degrees of freedom.\footnote{%
In some references, notably  \cite{Beccaria:2015uta,Beccaria:2015ypa,Beccaria:2017dmw}, formal Majorana-Weyl fermions are used. In comparing with them, one has therefore to double the multiplicities and halve the degrees of freedom, anomalies, etc.}
For the graviton, gravitino and vector fields, where gauge symmetry is present, we give the values for the physical fields, i.e.\ transverse and (gamma-)traceless. All the \(a\)-coefficients   are known from \cite{Beccaria:2015uta}. The scalar, vector and spinor cases have been investigated in the literature, see e.g.\  \cite{Bastianelli:2000hi,Beccaria:2017lcz,Casarin:2023ifl,Mukherjee:2021alj}. The graviton was studied in \cite{Pang:2012rd,Beccaria:2017dmw,Aros:2019tjw} (but cf.\ discussion above); our work confirms and justifies their results. 
The \(c_i\) coefficients that we obtain in our paper for gravitino and 3-form appear here for the first time.

Additionally, we  perform an extensive analysis   on the 3-form. We  construct the most general four-derivative Weyl-invariant action for a generic (non self-dual)  3-form, which happens to uniquely fix the action on the sphere \(S^6\). Furthermore, we show that, for a self-dual 3-form, a factorisation of the kinetic operator is necessarily of the form assumed in our Ansatz.

The strategy we follow in our calculation is, as said, based on  factorisation of the kinetic operators on on-shell Einstein backgrounds. 
Indeed, the theory defined by \(W^{(2,0)}\), namely the (six-derivative) conformal gravity compatible with \((2,0)\) supersymmetry, admits Einstein solutions, and we show that in this case the operator of the fluctuations factorizes in 2-derivative operators. 
We show this without making any additional assumption, thus generalising the result of \cite{Pang:2012rd} and providing an explanation for its extension in \cite{Beccaria:2017dmw};  
this parallels and confirms the analysis of \cite{Aros:2019tjw}. 
The effective action is therefore finally given in terms of determinants of second-order operators, for which general expressions for the heat-kernel coefficients  are well-established.

\begin{table}
\centering
\begin{tabular}{lcccccccc}
	                                 & dof's & \(m^{(1,0)}\) & \(m^{(2,0)}\) &         \(7!a\)          &       \(7!c_1\)       &     \(7!c_2\)     & \(7!c_3\)  &  \\[1em]
	\(\partial^6\) graviton          &  36   &       1       &       1       &   \(- \frac{3005} 2 \)   &      \(168784\)       &     \(25400\)     & \(-19668\) &  \\[0.25em]
	\(\partial^5\) gravitino (Majorana)  &  72   &       1       &       2       &   \(\frac{4643}{8} \)    &      \(-72128\)       &     \(-8800\)     &  \(9096\)  &  \\[0.25em]
	\(\partial^4\) 3-form (anti self-dual) &  20   &       1       &       5       & \( - \frac{166 }{9}   \) & \( -\frac{6160 }{3}\) &     \(-1320\)     &  \(-100\)  &  \\[0.25em]
	\(\partial^4\) vector  (real)          &   9   &       3       &      10      &   \(- \frac{275}{8}\)    &       \(2716\)        &      \(911\)      &  \(-150\)  &  \\[0.25em]
	\(\partial^3\) fermion (Majorana)    &  12   &       1       &       8       &    \(-\frac{39}{16}\)    &   \(\frac{896} 3 \)   & \(\frac{220} 3 \) &  \(-24 \)  &  \\[0.25em]
	\(\partial^2\) scalar  (real)          &   1   &       1       &      14       &     \(\frac 5 {72}\)     &   \(-\frac{28}3 \)    &  \(\frac{5}3 \)  &   \(2\)    &  \\[0.75em]
\end{tabular}
\label{tab}
\caption{In this table we list the component fields of the conformal supergravity multiplet and their conformal anomalies. We also give the order of the kinetic operator, the number of dynamical degrees of freedom, the multiplicities
\(m\) in the \((2,0)\) case and  its \((1,0)\) truncation. All \(a\)-anomalies, as well as the \(c_i\) anomalies for scalar, fermion and vector were known (see text for references). The \(c_i\)'s for the graviton were also given in the literature (but see text for references and discussion). The values of the \(c_i\)'s anomalies for the gravitino and the 3-form are original results.  }
\end{table} 

An aspect that plays a key role in our proof of the factorisation of the graviton operator is that we use a particular rewriting of the action that  differs from \(W^{(2,0)}\) given in \eqref{ww2} by a term proportional  the Euler density \(\mathbb E_6\), which does not influence the classical equations of motion nor the one-loop effects.  This alternative form of the action is at most linear in the generic Riemann tensor and thus vanishes on Ricci-flat backgrounds.\footnote{This form of the action seems to appear for the first time in \cite{Metsaev:2010kp}. It is the form used in \cite{Beccaria:2017lcz,Aros:2019tjw}, but not in \cite{Pang:2012rd}, which uses \(W^{(2,0)}\) directly.} This form of the action is remarkably easier to manipulate, e.g.\ because no \(\Riem^3\)-term appears in the quadratic operator and manifestly admits Einstein solutions. Intriguingly, this particular combination of \(W^{(2,0)}\) and \(\mathbb E_6\) is  exactly the six-dimensional \(\mathscr Q\)-curvature (see e.g.\ \cite{Branson1985,Juhl2009,graham2007holographic})
\begin{equation}\label{Q6}
\mathscr Q_6 = \frac 1 {48} \mathbb E_6 -2 I_1- \frac12 I_2 + \frac16 I_3 
= \frac 1 {48} \mathbb E_6  - \frac16 W^{(2,0)}\,,
\end{equation} an observation that has a four-dimensional analogue\footnote{\label{foot:Q4}%
In four dimensions \(\mathscr Q_4  =\frac14 \mathbb E_4 -\frac14 \text{Weyl}^2 - \frac16 \Box R = -\frac12 R_{mn}R^{mn}+ \frac16 R^2 - \frac16 \Box R \), which has no \(\Riem^2\)-term, vanishes on Ricci-flat, manifestly admits Einstein solutions and provides a factorised fluctuation operator on Einstein background. It has been extensively used in conformal supergravity calculations \cite{Fradkin:1985am}.
} and that was anticipated in \cite{Aros:2019tjw}.\footnote{\label{foot:Q}%
When writing this paper we learned that the factorisation properties of the \(\mathscr Q\)-curvature  fluctuations  in terms of 2-derivative Lichnerowicz-like operators have been investigated in the mathematical literature in \cite{Matsumoto_2013}. We thank Arkady Tseytlin for pointing out this paper to us.} We hope to explore further this connection between conformal supergravity and  \(\mathscr Q\)-curvature in the future.\footnote{See also \cite{Chernicoff:2018apt} for other examples of usage of the \(\mathscr Q\)-curvature in gravitational physics and in \cite{Bugini:2016nvn,Nakayama:2017eze,Bugini:2018def} for applications to Weyl anomalies.}

For the \((2,0)\) case we can combine the contributions of the different component fields according to the multiplicities to obtain
\begin{equation}\label{20a}
(a,c_i)_{(2,0) \text{CSG} } 
= \left( -\frac{91}{576}   , \frac{26}3    ,  \frac{13}6     ,  -\frac{13}{18}     \right)
 =  -26 \, (a,c_i)_{\text T} \,,
\end{equation}
where the value for the \((2,0)\) tensor multiplet \(\mathrm T\) is given in \eqref{fef}.
As a result, the combined system of \((2,0)   \) conformal supergravity coupled with \(26\) copies of the \((2,0)\) tensor multiplet has vanishing conformal anomaly, which confirms the expectation of \cite{Beccaria:2015uta,Beccaria:2017dmw}. Moreover, the component-field results of this paper provide a new explicit proof of anomaly cancellation. It is also clear from \eqref{20a} that the structure \eqref{aab} is respected.
As observed in \cite{Beccaria:2015uta}, this cancellation is analogous to the  \(4d\) cancellation of \(\mathcal{N}=4\) conformal supergravity coupled to four \(\mathcal{N}=4\) vector multiplets \cite{Fradkin:1985am,Fradkin:1983tg}. Furthermore, when six dimensional \((2,0)\) conformal supergravity is coupled to \(m\) tensor multiplets, \(5\) taken to be ghost-like compensators used to break conformal symmetry, it reduces in an IR limit to \((2,0)\) Poincaré supergravity coupled to \(m-5\) tensor multiplets \cite{Bergshoeff:1999db}. With \(m-5=21\), i.e.\ \(m=26\) as to cancel \eqref{20a}, the resulting theory is known to be free of gravitational anomalies   and to arise from the compactification of type IIB supergravity on K3 \cite{Townsend:1983xt,Witten:1995em}.

The analogous calculation in the \((1,0)\) truncation gives
\begin{equation}\label{10a}  
(a,c_i)_{(1,0) \text{CSG} } 
 =   		\left(    -	\frac{797}{3840} ,
 	 		\frac{184}{9} 	,
 	 			\frac{323}{90}	,
 	 			-\frac{199}{90}  \right)
				\,,
\end{equation}
and this matches\footnote{\label{foot:xi}%
We are referring here to the values discussed in v4 of \cite[footnote 8]{Beccaria:2015ypa} (cf.\ discussion before \eqref{www}). The coefficients of the chiral anomaly polynomial for the truncated \((1,0)\) conformal supergravity can be found  in      \cite[(B.20)]{Beccaria:2015ypa}. Our result \eqref{10a} indeed matches    the prediction of \cite{Beccaria:2015ypa} computed with  \(\xi = -\frac89\) therefore providing a direct verification of these relations and a so-far missing direct piece of information about this parameter.
} the expectation of \cite{Beccaria:2015ypa}. These values of the \(c_i\) coefficients satisfy \eqref{ci1}, or equivalently, the \(I_i\) contributions to the  anomaly \(\mathscr A\) \eqref{aaa}  form a combination of \(W^{(1,0)}_{1,2}\), as required by consistency. Notice that \eqref{10a} are the anomalies for the \((1,0)\) conformal supergravity that arises from truncation of the \((2,0)\) case (or equivalently it is the particular combination of \((1,0)\) actions that admits a \((2,0)\) extension). More generic cases, where one has an arbitrary linear combination of the supersymmetric extensions of \(W^{(1,0)}_{1,2}\), will be characterised by \(c_i\) coefficients that conceivably depend on the relative weight between these two contributions and cannot be captured by the present analysis.

The anomaly cancellation in the \((2,0)\) case, and its apparent relation to anomaly-free Poincar\'e supergravity arising from string theory, are quite remarkable. The resulting extended Poincar\'e supergravity multiplet in 6d can precisely be engineered by the system described by \((2,0)\) conformal supergravity plus 26 tensor multiplets, as mentioned in the previous paragraphs. However, though expected thanks to the arguments given in \cite{Beccaria:2015uta,Beccaria:2015ypa}, there is no immediate reason why the anomalies of the up to 6-derivative \((2,0)\) conformal supergravity action should cancel the anomalies of the 26 tensor multiplet actions, that are quadratic in derivatives.
This hints at possible similarities between the partition functions of the Poincaré and conformal supergravity theories. A closer scrutiny of the explicit form of the one-loop corrections might therefore reveal hidden structures and suggest a physical intepretation of the conformal construction of supergravity theories.
Indeed, in the 6d higher-derivative gauge theory case with conformal invariant action \((DF)^2+F^3 + \text{susy}\), a nontrivial factorisation of the one-loop partition function takes place in terms of ordinary derivative supermultiplet contributions \cite{Casarin:2019aqw}.
The proof of such factorisation in the supergravity case is then an important step and possible consistency test that might be applicable to other supergravity-matter systems, which include the \((2,0)\) conformal supergravities as building blocks.
Furthermore, understanding the precise form of the quadratic term for the fluctuations on a background might help in clarifying the structure of the degrees of freedom useful in an EFT analysis, for example, hinting at possible ways of cancelling the the non-unitary modes arising from the higher derivative terms.\footnote{We thank the referee of our paper for bringing this point to our attention.}

With our work, we have 
made progress towards understanding the structure of the Lagrangian for \((2,0)\) conformal supergravity. More remains to be done to understand the theory at higher-than-quadratic order and to extend the study to the less supersymmetric \((1,0)\) case. These aspects are likely to require a direct evaluation of the superspace expressions \cite{Butter:2016qkx} along the lines of \cite{Butter:2017jqu} since it seems unlikely for a labour-saving factorisation to be available due to the fact that generic \((1,0)\) Lagrangians do not admit Einstein solutions. Furthermore, 
the  expressions for the standard heat-kernel technique relevant for  studying the quantum properties of these six-derivative theories  have not been developed yet, but they can be conceivably obtained with the strategy used for fourth-order differential operators discussed in \cite{Fradkin:1981iu,Casarin:2019aqw,Casarin:2021fgd,Casarin:2023ifl}.
This has to be then supplemented with a more complete evaluation of the component-field actions following  \cite{Butter:2017jqu} or, possibly, through direct calculations in superspace.  
Finally, it would be interesting to understand better the mechanism underlying the factorisation of the kinetic operator and, in particular, its relation with the on-shellness of the background as well as supersymmetry.

\vspace{1em}
The rest of the paper is organised as follows.
In Section~\ref{sect:confsugra} we discuss the \((2,0)\) conformal supergravity action and its \((1,0)\) truncation, focusing on  the contributions that are relevant for the anomaly calculation. We describe the multiplets and present the evaluations of the component-field actions that we used to fix and validate our factorisation Ansatz.
 Additional details and expressions for the \((1,0)\) conformal supergravity Lagrangians are given in Appendix~\ref{app:l10}.
In Section \ref{sect:quadrlagra} we present in detail the quadratic actions on an Einstein background and discuss the factorisation Ansatz. 
With this information, in  Section \ref{sec:partf} we construct the partition functions on Einstein backgrounds and evaluate the conformal anomaly coefficients for each component field, whose outcome we have presented in Table~\ref{tab}.
Appendix~\ref{app:not} presents notation and conventions. Notice that we use two sets of conventions, for the supergravity and QFT calculations respectively.
Appendix~\ref{app:HK} summarises facts about the heat kernel that are relevant for the paper and gives some explicit formulae that are used in the main text. 
Appendix~\ref{app:jac} discusses the Jacobian factors that enter the path integral when restricting to physical fields and the calculation of determinants of differential operators on them.
Appendix~\ref{app:T3} presents the additional analysis that we have performed on the consequences of Weyl invariance for the 3-form.

\section{\texorpdfstring{Structure of \((2,0)\) conformal supergravity}{Structure of (2,0) conformal supergravity}}
\label{sect:confsugra}

In this section we introduce the bosonic \((2,0)\) conformal supergravity action, the \((2,0)\) Weyl multiplet and its truncation to \((1,0)\) conformal supersymmetry following \cite{Butter:2017jqu}. Though we present some results about fermions, in this section we focus on providing the quadratic bosonic sector of the \((1,0)\) Lagrangians and relegate to Appendix~\ref{app:l10} the full bosonic expressions together with the complete quadratic order in fermions.  Indeed, for our purpose of calculating the Weyl anomaly to 1-loop, we need only quadratic order Lagrangians in each field in a pure gravitational background. Under such conditions, R-symmetry connections and curvatures are ignored so that a given field's R-symmetry indices must contract with itself.
As a consequence, all the identical copies of \(\mathrm{Spin}(1,5)\) fields contained in a given \(\mathrm{Spin}(1,5) \times \mathrm{USp}(4)\) representation must contribute equally, thus the only relevant information for the calculation of the anomaly is the number of independent \(\mathrm{Spin}(1,5)\) fields, which is given by the dimension of the USp(4) representation up to enforcing reality conditions.
Thus, since no information is lost, we analyse this in the known \((1,0)\), SU(2), R-symmetry truncation, where one now treats \(\mathrm{Spin}(1,5) \times \mathrm{SU}(2)\) in the same way.
To conclude the section, we describe our analysis of the \((1,0)\) truncation and expansion of superspace results into components of each field's quadratic Lagrangian.

\subsection{\texorpdfstring{Truncation to \((1,0)\)}{Truncation to (1,0)}}
For   consistency with the original conformal supergravity  papers, we adopt the notation of \cite{Butter:2016qkx,Butter:2017jqu} (see also \cite{Butter:2018wss}) solely for this section and Appendix \ref{app:l10}; the only exceptions being that we underline R-symmetry \(\mathrm{USp}(4)\) indices and drop the superscript minus from \(T^-_{abc}\to T_{abc}\). We briefly document relevant parts of this notation in Appendix \ref{app:notSugra} and explain its relation to the Euclidean QFT notation of Appendix \ref{app:notQFT} which is used in the other sections of our paper.

The \((2,0)\) Weyl multiplet \cite{Bergshoeff:1999db,Butter:2017jqu} consists of the vielbein \(e_m{}^a\), gravitino \(\psi_{a \ui}{}^\a\), the R-symmetry connection \(V_a{}^{\underline{ij}}=V_a{}^{(\underline{ij})}\), the dilatation connection \(b_a\), and the covariant matter fields given by \(T_{abc}{}^{\underline{ij}}=T_{[abc]}{}^{[\underline{ij}]}\), 
\(\chi^{\a \ui, \underline{jk}}=\chi^{\a \ui, [\underline{jk}]}\), 
and \(D^{\underline{ij}\underline{kl}}=D^{[\underline{ij}][\underline{kl}]}\). The matter field \(T_{abc}{}^{\underline{ij}}\) is an anti-self-dual 3-form. Moreover, all fields live in irreducible representations of the R-symmetry group \(\mathrm{USp}(4)\) with  \(\Omega_{\underline{ij}} = \Omega_{[\underline{ij}]}\) being the symplectic invariant: the 3-form is traceless (\(T_{abc}{}^{\underline{ij}}\Omega_{\underline{i}\underline{j}}=0\)), 
\(\chi^{\a [\ui, \underline{jk}]} = 0\) and is traceless 
(\(\chi^{\a \ui, \underline{jk}} \Omega_{\underline{ij}}= 0\), 
\(\chi^{\a \ui, \underline{jk}}\Omega_{\underline{jk}} = 0\)), while 
\(D^{[\underline{ijk}]\ul} = 0\) and is also traceless 
(\(D^{\underline{ijkl}} \Omega_{\underline{ij}}= 0\),
\(D^{\underline{ijkl}}\Omega_{\underline{ik}} = 0\),
\(D^{\underline{ijkl}} \Omega_{\underline{kl}}= 0\)). 
The matrices of \(\mathrm{USp}(4) \subset \mathrm{U}(4)\) are complex and generically act on \(\mathbb{C}^4\), thus one requires the charged bosonic fields satisfy natural \(\mathrm{USp}(4)\) reality conditions
\begin{equation}\label{maj1-0}
\overline{(V_a{}^{\underline{ij}})}=V_a{}_{\underline{ij}}
~,~~~
\overline{(T_{abc}{}^{\underline{ij}})}=T_{abc}{}_{\underline{ij}}
~,~~~
\overline{(D^{\underline{ijkl}})}=D_{\underline{ijkl}}
~,
\end{equation}
which ensure that the number of real degrees of freedom matches the complex dimension of the \(\mathrm{USp}(4)\) representation. The fermions obey  Symplectic-Majorana-Weyl (SMW) conditions given by 
\begin{equation}\label{maj1}
\overline{(\psi_{a \ui}{}^{\a})} = \psi_{a}{}^{\ui \a}
\,, \qquad 
\overline{(\chi^{\a \ui,\underline{jk}})} = \chi_{\a}{}_{\ui, \underline{jk}}
\,.
\end{equation}
The $\chi$ spinor's SMW condition, which involves three copies of the symplectic form \(\Omega_{\underline{ij}}\), ensures it contains eight complex Weyl spinors and their charge conjugates; its \(\mathrm{USp}(4)\) representation is 16-dimensional.

The Lagrangian for \((2,0)\) conformal supergravity is the supersymmetric completion of the Weyl invariant gravitational term \(W^{(2,0)}\) defined in \eqref{ww2}. We denote this Lagrangian as \(\Lagr_{(2,0)}\).
Although a first principle complete derivation of \(\Lagr_{(2,0)}\), either by using $(2,0)$ superspace techniques or directly by using component space-time fields, is missing,  most of its bosonic sector was determined in \cite{Butter:2017jqu} by using consistency conditions on an uplift of $(1,0)$ results. 
The \((1,0)\) Lagrangians were uniquely constructed by reducing the superspace results of \cite{Butter:2016qkx} to component fields.
Up to an undetermined constant parameter \(\alpha\) and quartic terms in the anti-self-dual field \(T_{abc}{}^{\underline{ij}}\), the bosonic sector of \(\Lagr_{(2,0)}\)  is given by\footnote{There is a typo in the \((2,0)\) Lagrangian in equation (5.5) in arXiv v3 of \cite{Butter:2017jqu} and prior. 
The coefficient of the \(R(V)^3\)-term should be \(-4\) instead of \(-2\), and we have corrected this here. This ensures that the (2,0) result correctly truncates to match the \((1,0)\) Lagrangians, as seen later in this section. To explicitly distinguish between the conformally and the superconformally covariant derivatives, we have also replaced in this result the \(\hat{\nabla}_a\) used in \cite{Butter:2017jqu} with its purely bosonic counterpart \(\check{\nabla}_a\).}
\begin{align}\label{bl20}
\Lagr_{(2,0)}^{(\text{b})} &=
\frac{1}{3}\, C_{a b c d} \check{\nabla}^2 C^{abcd}
+ C_{a b}{}^{c d} C^{a b}{}^{e f} C_{c d e f} 
- 4\, C_{a b c d} C^{a e c f} C^b{}_e{}^d{}_f 
\nonumber\\
&\quad\,
- R(V)_{ab}\,^{\underline{i j}} \check{\nabla}^2 R(V)^{ab}\,_{\underline{i j}}
- 4 \, R(V)_{a}{}^{b}\,_{\ui}{}^{\uj} R(V)^a{}_{c}\,_{\uj}{}^{\uk} R(V)_{b}{}^{c}\,_{\uk}{}^{\ui}
+ C^{a b c d} \,R(V)_{a b}\,^{\underline{i j}} R(V)_{cd}\,_{\underline{i j}} 
\nonumber\\
&\quad\,
+ f_{a}{}^{b} \Big(
	\frac{32}{3}\, C^{acde} C_{b c d e} 
    - 8\,R(V)_{bc}\,^{\underline{i j}} R(V)^{ac}\,_{\underline{i j}}
	\Big)
- 4\, f_{a}{}^{a} (
    C_{bcde} C^{bcde} 
    - R(V)_{bc}\,^{\underline{i j}} R(V)^{bc}\,_{\underline{i j}}
    ) 
\nonumber\\
&\quad\,
+ \frac{1}{225} D^{\underline{ij}}{}_{\underline{kl}} \check{\nabla}^2 D^{\underline{kl}}{}_{\underline{ij}}
- \frac{2}{3375} D^{\underline{ij}}{}_{\underline{kl}} D^{\underline{kl}}{}_{\underline{i_1 i_2}} D^{\underline{i_1 i_2}}{}_{\underline{ij}}
- \frac{2}{15} D^{\underline{i j}}{}_{\underline{kl}} R(V)_{a b}{}^{\uk}{}_{\ui} R(V)^{a b}{}^{\ul}{}_{\uj} 
\nonumber\\
&\quad\,
+ 4 T_{abc}{}^{\underline{ij}} \check{\nabla}_{d} R(V)^{ab}{}_{\underline{jk}} R(V)^{cd}{}^{\uk}{}_{\ui}
+ 8 T_{abc}{}^{\underline{ij}} \check{\nabla}_{d} R(V)^{ad}{}_{\underline{jk}} R(V)^{bc}{}^{\uk}{}_{\ui}
\nonumber\\
&\quad\,
- \, T_{a b c}{}^{\underline{ij}} \Delta^4 T^{a b c}{}_{\underline{ij}}
+ \frac{8}{3}\, C_{a b c d} T^{a b e}{}_{\underline{ij}} \check{\nabla}_{e} \check{\nabla}_{f} T^{c d f\, \underline{ij}}  
+ \frac{4}{3}\, C_{a b c d} T^{a b e}{}_{\underline{ij}} \check{\nabla}_{f} \check{\nabla}_{e}T^{c d f\, \underline{ij}}
\nonumber\\
&\quad\,
- \frac{8}{3}\, C^{a b c d} T_{a e f\, \underline{ij}} \check{\nabla}_{b} \check{\nabla}_{c} T_{d}{}^{ef\,\underline{ij}}
+ 2\, C_{a b}{}^{c d} \check{\nabla}^{a} T^{b e f}{}_{\underline{ij}}\,  \check{\nabla}_{c} T_{d e f}{}^{\underline{ij}}\, 
+ 3\, C_{a b}{}^{c d} \check{\nabla}_{e} T^{a b f}{}_{\underline{ij}}\,  \check{\nabla}_{f} T^{c d e}{}^{\underline{ij}}\,   
\nonumber\\
&\quad\,
- \frac{4}{3}\, C_{a b e f} C^{c d e f} T_{a b g}{}^{\underline{ij}} T^{c d g}{}_{\underline{ij}}
+ 4 \alpha \, T_{abc}{}^{\underline{ij}} T^{ade\, \underline{kl}} R(V)^{bc}{}_{\underline{ik}} R(V)_{de}{}_{\underline{jl}} 
\nonumber\\
&\quad\,
+ 2 (1 - \alpha) T_{abc}{}^{\underline{ij}} T^{ade}{}_{\underline{ij}} R(V)^{bc}{}_{\underline{kl}} R(V)_{de}{}^{\underline{kl}}
\nonumber\\
&\quad\,
+ \frac{2}{15} D^{\underline{i j}}{}_{\underline{k l}} \Big(
	T_{a b c}{}^{\underline{kl}} \check{\nabla}^a \check{\nabla}_d T^{b c d}{}_{\underline{i j}}
	- \frac{1}{2} \check{\nabla}^a T_{a b c}{}^{\underline{kl}} \check{\nabla}_d T^{b c d}{}_{\underline{i j}}
	\Big)
\nonumber\\
&\quad\,
- \frac{1}{60} D^{\underline{i j}}{}_{\underline{kl}} T_{a b c}{}^{\underline{kl}} T^{a b d}{}_{\underline{ij}}
	T^{c e f\,}{}_{\underline{i_1 i_2}} T_{d e f}{}^{\underline{i_1 i_2}}
+ \mathcal{O}(T^4) 
\,,
\end{align}
where   \(\check{\nabla}^2 = \check{\nabla}^a \check{\nabla}_a\) and we define the \(K\)-invariant combination
\begin{align}\label{3fop}
T^{abc}{}_{\underline{ij}} \Delta^{4} T_{a b c}{}^{\underline{ij}} &:= T^{abc}{}_{\underline{ij}} \Big(\check{\nabla}_{a} \check{\nabla}^{d} \check{\nabla}^2 T_{b c d}{}^{\underline{ij}}
+ \check{\nabla}^2 \check{\nabla}_{a} \check{\nabla}^{d} T_{b c d}{}^{\underline{ij}}
\nonumber\\
&\quad\,
+ \tfrac{1}{3}\, \check{\nabla}_{a} \check{\nabla}^2 \check{\nabla}^{d} T_{b c d}{}^{\underline{ij}}
- \tfrac{4}{3}\, \check{\nabla}_{e} \check{\nabla}_{a} \check{\nabla}^{d} \check{\nabla}^{e} T_{b c d}{}^{\underline{ij}}\Big)
\,.
\end{align}
Here we have also introduced the Weyl tensor \(C_{abcd}\), the \(\mathrm{USp}(4)\) curvature \(R(V)_{ab}{}^{\underline{ij}}\) and the special conformal connection \(f_a{}^b\). The explicit definitions for the curvatures are given in \eqref{Lcurv} and \eqref{Rcurv}. We define the conformal covariant derivative with USp(4) R-symmetry as
\bsubeq 
\begin{align}\label{20rccd}
\check{\nabla}_a &= e_a - \frac{1}{2} \omega(e,b)_a{}^{bc} M_{bc} - V_a{}^{\underline{ij}} J_{\underline{ij}} - b_a \mathbb{D} - f_a{}^b K_b
\,,
\\
[\check{\nabla}_a ,\check{\nabla}_b]
&=
-\frac{1}{2} C_{ab}{}^{de}M_{de}
-R(V)_{ab}{}^{\underline{ij}}J_{\underline{ij}}
-\frac{1}{6}\check{\nabla}^d C_{abcd}K^d
~,
\end{align}
\esubeq
where \(e_a = e_a{}^m \partial_m\) is the vielbein vector field and we have generators \(M_{ab}\) (Lorentz), \(J_{\underline{ij}}\) (\(\mathrm{USp}(4)\)), \(\mathbb{D}\) (dilatation), \(K_a\) (special conformal). The Lorentz connection 
\begin{equation}
\omega(e,b)_{a}{}^{bc}=\omega(e)_a{}^{bc}-2\d_{a}^{[b}b^{c]}~,
\label{oeb}
\end{equation}
reduces to the Levi-Civita connection \(\omega(e)\) when dilatation is gauge-fixed to \(b_a=0\) via a special conformal gauge transformation. In this case, \(\mathcal{D}_a = e_a - \frac{1}{2}\omega(e)_a{}^{bc}M_{bc}\) is the usual Levi-Civita covariant derivative corresponding to the Riemann curvature. There is also a dependence of \(f_a{}^b\) on \(\omega(e,b)\), and when we set \(b_a=0\), it holds that \(f_{ab} = - \frac{1}{2} S_{ab}\), where \(S_{ab} = \frac{1}{4}(\mathcal{R}_{ab} - \frac{1}{10}\eta_{ab} \mathcal{R})\) is the Schouten tensor and \(\mathcal{R}_{ab}\) and \(\mathcal{R}\) are the Ricci and scalar curvatures, respectively.

Conformal covariant derivatives are reviewed in \cite{Butter:2016qkx} 
--- 
see also \cite{Freedman:2012zz,Lauria:2020rhc,Kuzenko:2022skv,Kuzenko:2022ajd} for pedagogical reviews of superconformal approaches to supergravity
---
and are useful tools for writing Weyl-invariant actions in a manifest conformally covariant way. In particular, after gauge-fixing \(b_a=0\) via a \(K\)-gauge transformation, Weyl transformations correspond to combined \(K,\mathbb{D}\)-gauge transformations which preserve \(b_a=0\). One can then extract the \(K\)-connection from covariant derivatives
by making use of soft algebra \([K_a,\check{\nabla}_b]\) relations. Such a process is typically called degauging the connection from a covariant derivative. In the case of the special conformal connection, it holds
\begin{subequations} \label{kdeg}
\begin{align} 
\check{\nabla}_a &= (\mathcal{D}_a - V_a) - f_a{}^b K_b 
\,,\\
\check{\nabla}_a \check{\nabla}_b &= 
(\mathcal{D}_a - V_a) \check{\nabla}_b - 2f_{ab}\mathbb{D} -2f_a{}^e M_{eb} - f_a{}^e \check{\nabla}_b K_e
\,,
\end{align}
\end{subequations}
where \(V_a = V_a{}^{\underline{ij}} J_{\underline{ij}}\) and similar relations hold for higher-derivatives. Similarly, one can degauge the \(\mathrm{USp}(4)\)-connection and then arrive at a Lagrangian in terms of Levi-Civita covariant derivatives only. 
Some extra information regarding Lorentz connections and curvatures is given in \eqref{Lconn} and \eqref{Lcurv}. In general, more details about this formalism for the six-dimensional case are provided in \cite{Butter:2016qkx,Butter:2017jqu,Butter:2018wss}.

In relating the minimal and extended conformal supergravity actions, it is important to elaborate on how the \((2,0)\) Weyl multiplet decomposes into the following \((1,0)\) multiplets: a Weyl multiplet, two gravitini multiplets, and a \(\mathrm{SU}(2)\) vector multiplet --- see \cite{Butter:2017jqu}. We ignore the gravitini and vector multiplets as all the information relevant for the calculation of the \((2,0)\) conformal anomaly is encoded in the \((1,0)\) Weyl multiplet by the R-symmetry arguments above. To do this, one splits the \(\mathrm{USp}(4)\) indices \(\ui =1,\dots,4\) into \(i=1,2,i'=1,2\) and switches off the third and fourth gravitini \(\psi_{a i'}{}^\a =0\). The only \(\mathrm{USp}(4)\) matrices preserving this condition are block-diagonal and, choosing the symplectic form as follows
\begin{equation}\label{symp}
\Omega^{\underline{ij}} = 
\begin{pmatrix}
\varepsilon^{ij} & 0 \\
0 & \varepsilon^{i'j'}
\end{pmatrix}
\,, \qquad 
\Omega_{\underline{ij}} = 
\begin{pmatrix}
\varepsilon_{ij} & 0 \\
0 & \varepsilon_{i'j'}
\end{pmatrix}
\,,
\end{equation}
fixes the subgroup to be \(\mathrm{SU}(2)_{\mathrm{R}} \times \mathrm{SU}(2)\)  
where the subscript \(\mathrm{R}\) indicates where the \((1,0)\) R-symmetry transformations arise from. The other ``primed'' \(\mathrm{SU}(2)\) forms the gauge transformations of the additional $(1,0)$ vector multiplet.
The gravitini supersymmetry transformations then require that \(V_a{}^{ij'}=0\) and discarding the $(1,0)$ vector multiplet requires \(V_a{}^{i'j'}=0\), leaving only \(V_a{}^{ij}=\mathcal{V}_a{}^{ij}\). The non-trivial part of the subgroup is now only \(\mathrm{SU}(2)_{\mathrm{R}}\) acting on the \(i,j,k,l,\ldots\) indices, and the various covariant fields of the \((2,0)\) Weyl multiplet decompose as follows
\begin{subequations}
\begin{align}\label{trunc1}
T_{abc}{}^{i j} &= \varepsilon^{i j} T_{abc}\,, \qquad T_{abc}{}^{i'j'} = -\varepsilon^{i'j'} T_{abc} \,, \\ \label{trunc2}
\chi_i{}^{jk} &= \varepsilon^{jk} \chi_i \,, \quad \chi_i{}^{j'k'} = - \varepsilon^{j'k'} \chi_i \,, \quad
\chi_{i'}{}^{j'k} =  \frac{1}{2}\d_{i'}^{j'} \chi^k \,, \\ \label{trunc3}
D^{ij}{}_{kl} &= - \varepsilon^{ij} \varepsilon_{kl} D \,,~~
D^{ij}{}_{k'l'} = \varepsilon^{ij} \varepsilon_{k'l'} D \,, ~~
D^{i'j'}{}_{k'l'} = - \varepsilon^{i'j'} \varepsilon_{k'l'} D \,,~~
D^{ij'}{}_{kl'} = - \frac{1}{2} \d^i_k \d^{j'}_{l'} D \,.
\end{align}
\end{subequations}
The \((1,0)\) Weyl multiplet consists of the vielbein \(e_m{}^a\), gravitino \(\psi_{a i}{}^\a\), the R-symmetry connection \(\mathcal{V}_a{}^{ij}\), the dilatation connection \(b_a\), and the covariant matter fields \(T_{abc}\), \(\chi^{\a i}\), \(D\). The anti-self-dual 3-form \(T_{abc}\) and the scalar field $D$ are real. The SMW conditions and bosonic reality conditions in the truncated case are
\begin{equation}\label{maj2}
\overline{(\psi_{a i}{}^{\a})} =\psi_{a}{}^{i \a}
\,, \qquad 
\overline{(\chi^{\a i})} = \chi^{\a}{}_i \,
\,, \qquad
\overline{(\mathcal{V}_a{}^{ij})} = \mathcal{V}_{a ij}
\,.
\end{equation}

We now present the quadratic order bosonic sectors of the two \((1,0)\) Lagrangians. We abuse notation and use the conformal covariant derivatives with \((1,0)\) \(\mathrm{SU}(2)\) R-symmetry given by
\bsubeq\begin{align}\label{10rccd}
\check{\nabla}_a &= e_a - \frac{1}{2} \omega(e,b)_a{}^{bc} M_{bc} - \mathcal{V}_a{}^{ij} J_{ij} - b_a \mathbb{D} - f_a{}^b K_b
\,,
\\
[\check{\nabla}_a ,\check{\nabla}_b]
&=
-\frac{1}{2} C_{ab}{}^{de}M_{de}
-\mathcal{R}_{ab}{}^{ij}J_{ij}
-\frac{1}{6}\check{\nabla}^d C_{abcd}K^d
\,,
\end{align}
\esubeq 
which are the same as \eqref{20rccd}, but with \(\mathrm{SU}(2)\) R-symmetry replacing the \(\mathrm{USp}(4)\) R-symmetry.

The so-called \(C\Box C\)   Lagrangian, \(\Lagr_{C\Box C}\), is the \((1,0)\) supersymmetric completion of the Weyl invariant  \(W^{(1,0)}_1\) defined in \eqref{www}. Focusing on the bosonic sector, the relevant contributions for the quadratic expansion on a gravitational background are\footnote{We stress that, like in the \((2,0)\) case, we have replaced \(\hat{\nabla}\) of \cite{Butter:2017jqu} by its purely bosonic counterpart \(\check{\nabla}\). Moreover, we have rewritten some terms with a single derivative acting on a conformal primary by using that \(K_a(\text{primary}) = 0\).
}
\begin{align}
\Lagr_{C\Box C}^{(2,\text{b})}
	&=
\frac{1}{3}\, C_{a b c d} \check{\nabla}^2 C^{abcd}
- \frac{1}{3} C_{a b}{}^{c d} C_{c d}{}^{e f} C_{e f}{}^{a b}
- \frac{4}{3} C_{a b c d} C^{a e c f} C^b{}_e{}^d{}_f 
\nonumber\\
&\quad\,
- \mathcal{R}_{ab}\,^{i j} \check{\nabla}^2 \mathcal{R}^{ab}\,_{i j}
+ 2 \, C^{a b c d} \,\mathcal{R}_{a b}\,^{i j} \mathcal{R}_{c d}\,_{ij}
\nonumber\\
&\quad\,
+ \hat{\mathfrak f}_{a}{}^b (
	\frac{32}{3}\, C^{acde} C_{b c d e} 
	- 8\, \mathcal{R}_{bc}\,^{i j} \mathcal{R}^{ac}\,_{ij}
	) 
- 4\, \hat {\mathfrak f}_{a}{}^a (C_{bcde} C^{bcde} - \mathcal{R}_{bc}\,^{i j} \mathcal{R}^{bc}\,_{ij})
\nonumber\\
&\quad\,
+ \frac{4}{45}\, D \check{\nabla}^2 D
+ \frac{2}{15} D\, C_{abcd} C^{abcd}
+ \frac{20}{3} T^{a b e} C_{a b}{}^{c d} \check{\nabla}^f C_{f e c d}
+ 4  \,T^{abe} \check{\nabla}^f C_{ab}{}^{cd} C_{fecd}
\nonumber\\
&\quad\,
- 4 \, T_{a b c} \Delta^4 T^{a b c}
- \frac{16}{3}  C_{a b c d} T^{a b e} \check{\nabla}_{e}{\check{\nabla}_{f}{T^{c d f}}\, }\,  
- \frac{8}{3}  C_{a b c d} T^{a b e} \check{\nabla}_{f}{\check{\nabla}_{e}{T^{c d f}}\, }\,  
\nonumber\\
&\quad\,
+ \frac{16}{3}  C_{a b}{}^{c d} T^{a e f} \check{\nabla}^{b}{\check{\nabla}_{c}{T_{d e f}}\, }\,
- 4  \, C_{a b}{}^{c d} \check{\nabla}^{a}{T^{b e f}}\,  \check{\nabla}_{c}{T_{d e f}}\, 
- 6 \, C_{a b c d} \check{\nabla}_{e}{T^{a b f}}\,  \check{\nabla}_{f}{T^{c d e}}\,  
\nonumber\\
&\quad\,
- \frac{4}{3} \, C_{a b e f} C^{c d e f} T^{a b g} T_{c d g}
\,,
\end{align}
where \(T^{abc} \Delta^4 T_{abc}\) retains the same form as \eqref{3fop}.

The  \(C^3\) Lagrangian, denoted \(\Lagr_{C^3}\), is the \((1,0)\) supersymmetric completion of \(W^{(1,0)}_2\) given in \eqref{www}.  Focusing on the bosonic sector,
the terms that contribute to the quadratic expansion on a gravitational background are
\begin{align}
\Lagr_{C^3}^{(2,\text{b})}
&=
\frac{8}{3}\,{C}_{a bc d} {C}^{a b e f} {C}^{c d}{}_{e f} 
  - \frac{16}{3}\, {C}_{a b c d} {C}^{aecf} {C}^{b}{}_{ e}{}^{d}{}_{f} 
-2\, {C}_{a b c d} {\mathcal{R}}^{a b}\,^{i j} {\mathcal{R}}^{c d}\,_{ij}  
\nonumber\\
&\quad\,
  - \frac{4}{15} \, D {C}_{a b c d} { C}^{a b c d} 
- \frac{32}{3}\, {T}_{a b c} { C}^{a b d e} {\check{\nabla}}^{f}{{ C}^{c}{}_{d e f}}
   + \frac{16}{3}\, { C}_{a b c d} { C}^{a b e f} {\check{\nabla}}^{c}{{T}^{d}{}_{e f}}
\nonumber\\
&\quad\,
 + 16 \, {T}_{a b c} { C}^{a b d e} {\check{\nabla}}^{c}{{\check{\nabla}}^{f}{{T}_{d e f}} }
 - 16\, {T}_{a b c} {C}^{a b d e} {\check{\nabla}}_{d}{{\check{\nabla}}^{f}{{T}^{c}{}_{e f}}}
 - 4\, {C}_{a b c d} {\check{\nabla}}_{e}{{T}^{a b e}} {\check{\nabla}}_{f}{{T}^{c d f}}
\nonumber\\
&\quad\,
 +8\, { C}_{a b c d} {\check{\nabla}}_{e}{{T}^{a b f}}\,  {\check{\nabla}}_{f}{{T}^{c d e}}
- \frac{64}{3}\, {T}^{f b}{}_{ d} \,{\check{\nabla}}^{e}{{C}_{e a b c}}\,  {\check{\nabla}}_{f}{{T}^{a c d}}
+32\, {T}^{a b}{}_{ d} \,  {\check{\nabla}}^{e}{{ C}_{e a b c}}\,  {\check{\nabla}}_{f}{{T}^{f c d}}
\,.
\end{align}

We give the full bosonic expressions, \(\Lagr_{C^3}^{(\text{b})}\) and \(\Lagr_{C\Box C}^{(\text{b})}\), in Appendix~\ref{app:bosonic10lag}. We also provide
in Appendix~\ref{app:c3fermquadlag} and \ref{app:cbcfermquadlag}, respectively, the expressions \(\Lagr_{C^3}^{(2)}\) and \(\Lagr_{C\Box C}^{(2)}\) which also include fermions up to quadratic order.

Consider now the full Lagrangians, $\Lagr_{(2,0)}$, i.e.\ with all bosons and fermions to all orders. Under the truncation described above with extra \((1,0)\) vector and gravitini multiplets switched off, it has to hold
\begin{equation}\label{l20}
\Lagr_{(2,0)} 
\qquad\longrightarrow\qquad
\Lagr_{(2,0)\to(1,0)} = \Lagr_{C\Box C} + \frac{1}{2} \Lagr_{C^3}
\,,
\end{equation}
where \(\Lagr_{(2,0)\to(1,0)}\) stands for the unique gravitational \((1,0)\) Lagrangian which permits an uplift to \((2,0)\) conformal supersymmetry.
If one were to keep the extra \((1,0)\) vector multiplet, a third Yang--Mills contribution  (denoted  \(\mathcal L_{F\Box F}\)) is also present --- see \cite{Butter:2017jqu} for detail.

\subsection{Multiplet structure}

Before describing some more details of the structures of the conformal supergravity multiplet, a remark is in order. As mentioned in the introduction, and in Table~\ref{tab} in particular, we adopt a Euclidean notation where Lorentzian SMW fermions are translated into Euclidean Majorana fermions which ensures the degrees of freedom are the same. Classically, at the linear level of the multiplet restricted to a truncation where only arbitrary dependence upon the metric is kept, this is simply a conventional choice.  
At the quantum level, this  does not alter results for parity even quantities such as the conformal anomalies~\eqref{aaa}, cf.\ footnote~\ref{foot:ferm}. However, we stress that the construction of the fully non-linear Euclidean conformal supergravity multiplet, the conformal supergravity invariants, and their relation (`Wick rotation') with the Lorentzian counterpart is not immediate and deserves further study beyond the scope of our paper --- see for example the discussion of Euclidean four-dimensional supergravity of \cite{deWit:2017cle}. 
In the 6d case, replacing our choice of Euclidean Majorana fermions with Euclidean Weyl fermions and proceeding with the Euclidean approach of \cite{Nicolai:1978vc,Frohlich:1974zs,Osterwalder:1973zr} would fix the chirality of the fermions. The formulation of Euclidean 6d conformal supergravities might be an interesting direction to look at in the future.

\begin{table}[h]
\centering
\begin{tabular}{cccccccc}
&	&&  \multicolumn{2}{c}{\((2,0)\) multiplet } & & \multicolumn{2}{c}{ \((1,0)\)  multiplet } \\
field & notes & &symbol & \(\mathrm{USp}(4)\) dim &  & symbol & \(\mathrm{SU}(2)\)  dim \\[0.5em]
graviton & & & \(e_m{}^a\) & 1  & & \(e_m{}^a\)  & 1  \\[0.25em]
gravitino & SMW && \(\psi_{a \ui}{}^{\a}\) &  4&   & \(\psi_{a i}{}^{\a}\) &  2  \\[0.25em]
3-form & anti-self-dual && \(T_{abc}{}^{[\underline{ij}]}\)&  5 & & \(T_{abc}\)&  1  \\[0.25em]
vector  &  real  &   & \(V_a{}^{(\underline{ij})}\) & 10  & & \(\mathcal{V}_a{}^{({ij})}\) & 3  \\
fermion & SMW   & & \(\chi^{\a \, \ui, [\underline{jk}]}\)&     16 &   &\(\chi^{\a i}\)  &  2   \\[0.25em]
scalar &  real     &    &  \(D^{[\underline{ij}][\underline{kl}]}\) &    14 &  & \(D\)  &1  \\[0.75em]
\end{tabular}
\caption{R-symmetry representation dimensions of \((1,0)\) and \((2,0)\) Weyl multiplets in the Lorentzian notations of \cite{Butter:2016qkx,Butter:2017jqu}. The fermionic fields obey SM conditions, which is a reality condition  imposed on top of the  R-symmetry representation. Therefore the  number of truly independent fields is reduced by a factor 2, producing the multiplicites quoted in Table~\ref{tab}.}
\label{tab2}
\end{table}

With this in mind, in Table~\ref{tab2} we give the complex dimensions of the R-symmetry representations, which makes it easy to compare to the multiplicities in Table~\ref{tab}. The SM condition is not automatically imposed by the R-symmetry representation. Therefore, the number of independent real fermionic components (i.e.\ the multiplicities) is half of it. For bosonic fields, their reality conditions ensure that the number of independent real components equals the complex dimension.

\subsection{Component reduction}
We now turn to extracting the terms of the truncated \((2,0)\) Lagrangian that contribute to quadratic order in a pure gravitational background. In particular, using the truncation, we expand both \((1,0)\) conformal supergravity Lagrangians into components up to quadratic order in all fields except the vielbein, which we allow to all orders, and take the linear combination \eqref{l20}. All covariant derivatives are also rewritten in terms of Levi-Civita covariant derivatives, so that the resulting quadratic Lagrangians are living on some pure gravitational background given by a pseudo-Riemannian manifold. We give an overview of how this is practically done and then specialise to the case of each field in the \((1,0)\) Weyl multiplet.

We aim to keep the pure gravitational background general where possible, but in the more complex cases, namely the 3-form and gravitino, we give some results only in the special cases of: (i) the sphere \(S^6\) with Riemann curvature \(\mathcal{R}_{abcd} = \frac{1}{30}(\eta_{ac}\eta_{bd} - \eta_{ad}\eta_{bc}) \mathcal{R}\); (ii) a generic Ricci-flat background with parallel curvature. The extension of these results to more general backgrounds will be given elsewhere.

\subsubsection{Component reduction procedure}
We summarise the general procedure followed to obtain the component results, more detail can be found in \cite{Butter:2017jqu}. First, we introduce some relevant notation. The component form of the \((1,0)\) superconformal covariant derivative is given by\footnote{We work in the traceless frame of \cite{Butter:2017jqu}.}
\begin{align}\label{sccov}
\hat{\nabla}_{a} 
&= e_{a} - \frac{1}{2} \psi_{a i}{}^{\alpha} Q_{\alpha}^{i} - \frac{1}{2} \hat{\omega}_{a}{}^{bc} M_{bc} - \mathcal{V}_{a}{}^{ij} J_{ij} - b_{a} \mathbb{D} - \hat{\mathfrak{f}}_{a}{}^{b} K_b - \frac{1}{2} \hat{\phi}_{a \alpha}{}^{i} S^{\alpha}_{i}
\nonumber\\
&= \mathcal{D}_{a}  - \frac{1}{2} \psi_{a i}{}^{\alpha} Q_{\alpha}^{i} - \frac{1}{2} (\hat{\omega}_{a}{}^{bc} - \omega(e)_{a}{}^{bc} ) M_{bc} - \mathcal{V}_{a}{}^{ij} J_{ij} - b_{a} \mathbb{D} - \hat{\mathfrak{f}}_{a}{}^{b} K_b - \frac{1}{2} \hat{\phi}_{a \alpha}{}^{i} S^{\alpha}_{i} \,,
\end{align}
where \(e_a = e_a{}^m \partial_m\) is the vielbein vector field, \(\psi_{a i}{}^{\a}\) is the gravitino, $\mathcal{D} = d - \frac{1}{2} \omega(e)^{bc} M_{bc}$ is the Levi-Civita covariant derivative, and we have connection forms for \(Q\)-supersymmetry \((Q_\a^i)\), Lorentz \((M_{ab})\), \(\mathrm{SU}(2)\) R-symmetry \((J_{ij})\), dilatation \((\mathbb{D})\), special conformal \((K_{a})\) and \(S\)-supersymmetry \((S_i^\a)\). In particular, due to a set of conventional torsion-curvature constraints on the $(1,0)$ superconformal geometry, the connections \(\hat{\omega}, \hat{\mathfrak{f}}, \hat{\phi}\) are composite in the sense that they have implicit spinor and gravitino contributions and their expansions are given in \eqref{cscconn}.

Following \cite{Butter:2017jqu}, we find that, to quadratic order in pure gravitational backgrounds, the \((1,0)\) Lagrangians take the form\footnote{The \(C\Box C\) Lagrangian is not primary and transforms into a total derivative under \(K,S\)-gauge transformations, as suggested by its explicit dependence upon the \(K,S\)-connections \(\hat{\mathfrak{f}},\hat{\phi}\).}
\bsubeq\begin{align}
\Lagr_{C^3}^{(2)} &= F - \frac{\mathrm{i}}{4} \psi_{a i}{}^{\a} \Omega'{}_\a{}^{a i} 
\,, \\
\Lagr_{C\Box C}^{(2)} &= \hat{F} 
+ \frac{\mathrm{i}}{2}\psi_{a i}{}^\a \hat{\Omega}_{\a}{}^{a i} 
- 16\hat{\mathfrak{f}}^{ab} C_{ab} 
- 8\mathrm{i}\psi_{ai}{}^{\a} (\g^{ab})_{\a}{}^{\b} \Lambda_{\b c}{}^i \hat{\mathfrak{f}}_b{}^c
+ 2\hat{\phi}_{a i \a} \rho^{a \a i}
\,,
\end{align}
\esubeq
where \(F\) and \(\Omega^\prime{}_\a{}^{ai}\) are defined in Appendix~\ref{app:c3fermquadlag}, while  \(\hat{F}, \hat{\Omega}_\a{}^{ai}, C_{ab}, \Lambda_{\b c}{}^i\) and \(  \rho^{a \a i}\) are given in Appendix~\ref{app:cbcfermquadlag}. In particular, each of these fields are built in terms of the components of the \((1,0)\) standard Weyl multiplet \cite{Bergshoeff:1985mz,Butter:2017jqu}. The  \((1,0)\) standard Weyl multiplet can be efficiently described in terms of a single super-Weyl tensor \(W^{\a\b}=W^{(\a\b)}\) in the $(1,0)$ conformal superspace of \cite{Butter:2016qkx}. The independent descendant superfields that generate the standard Weyl multiplet
are the following
\begin{subequations}
\begin{align}
X^{\a i} 
	&:= 
	-\frac{\mathrm{i}}{10}\hat{\nabla}_\b^i W^{\a\b}
\,,\quad
X_\g^k{}^{\a\b} 
	:= 
	-\frac{\mathrm{i}}{4}\hat{\nabla}_\g^k W^{\a\b} 
	- \d_\g^{(\a} X^{\b)k}
\,, \\
Y
	&:=
	\frac{1}{4}\hat{\nabla}_\g^k X_k^\g 
\,,\quad
Y_\a{}^{\b ij}
	:=
	-\frac{5}{2}\left( \hat{\nabla}_\a^{(i} X^{\b j)}
	-\frac{1}{4}\d_\a^\b \hat{\nabla}_\g^{(i}X^{\g j)} \right)
\,, \\
Y_{\a\b}{}^{\g\d}
	&:=
	\hat{\nabla}_{(\a}^kX_{\b) k}{}^{\g\d}
	-\frac{1}{3}\hat{\nabla}_\rho^k X_{(\a k}{}^{\rho(\g}\d_{\b)}^{\d)}
\,,
\end{align}
\end{subequations}
where \(\hat{\nabla}_\a^i\) is the conformal superspace spinor covariant  derivative, which projects onto the base manifold as \(Q_\a^i\)-supersymmetry generators acting on covariant fields.
The component fields obtained from projecting these onto the base manifold are defined as
\begin{subequations}
\begin{align}
T_{abc} 
	&:= -2 W_{abc} \vert 
\,,\quad
\chi^{\a i} 
	:= 
	\frac{15}{2}X^{\a i} \vert
\,,\quad
\mathcal{X}_{\a}^i{}^{\b\g}
	:=
	X_{\a}^i{}^{\b\g} \vert 
\,, \\
D
	&:=
	\frac{15}{2}Y \vert
\,,\quad
\mathcal{Y}_\a{}^\b{}^{kl}
	:=
	Y_\a{}^\b{}^{kl} \vert 
\,,\quad
\mathcal{Y}_{\a\b}{}^{\g\d}
	:=
	Y_{\a\b}{}^{\g\d} \vert	 
\,,
\end{align}
\end{subequations}
where the vertical bar denotes setting all fermionic coordinates \(\theta_\a^i\) to zero and we used that \(W_{abc} = \frac{1}{8} (\g_{abc})_{\a\b} W^{\a\b}\). The component fields \(\mathcal{X}_\a^i{}^{\b\g}, \mathcal{Y}_\a{}^{\b kl}, \mathcal{Y}_{\a\b}{}^{\g\d}\) are composite and their expansions in terms of the \((1,0)\) Weyl multiplet \((e_m{}^a, \psi_{a i}{}^{\a}, \mathcal{V}_a{}^{ij}, b_a, T_{abc}, \chi^{\a i}, D)\) are given in Appendix~\ref{app:l10}. These three fields directly correspond to the \(Q\)-supersymmetry gravitini field strength, the SU(2) R-symmetry curvature, and the Lorentz curvature of the superconformal geometry, whose bosonic part reduces to the Weyl tensor 
$C_{ab}{}^{cd}=\frac{1}{4}(\g_{ab})_\g{}^\a(\g^{cd})_\d{}^\b C_{\a\b}{}^{\g\d}$ 
written in spinor notation.

Having clarified the notation, we describe the practicalities of how we reduced the \((1,0)\) Lagrangians to all components that we need for the analysis in our paper. First, the expressions for $F$, $\Omega^\prime{}_\a{}^{ai}$, $\hat{F}$, $\hat{\Omega}_\a{}^{ai}$, $C_{ab}$, $\Lambda_{\b c}{}^i$, and $\rho^{a \a i}$ defined in Appendix~\ref{app:c3fermquadlag} and Appendix~\ref{app:cbcfermquadlag} were obtained by restricting their full expressions in the supplementary file\footnote{The supplementary file of \cite{Butter:2017jqu} can be downloaded from the following link: \href{https://arxiv.org/src/1701.08163}{\tt{arxiv.org/src/1701.08163}}. 
} of \cite{Butter:2017jqu} to quadratic order in all fields but the vielbein --- pure vielbein terms arise \textit{only} from \(\mathcal{Y}_{\a\b}{}^{\g\d}\), which has bosonic part given by the Weyl tensor, as commented above. Second, we extract all the implicit fermions within the superconformal covariant derivatives. Practically, this means that we degauge the \(Q,S\)-connections from superconformal covariant derivatives \(\hat{\nabla}_a\) as they contain implicit gravitino contributions, i.e.\ the \(Q\)-connection is the gravitino and the \(S\)-connection \(\hat{\phi}_{a \a}{}^i\) is composite in terms of the gravitino. Moreover, we degauge parts of the $M$ and $K$ connections which have implicit gravitini, but also $\chi$ spinors as well.\footnote{A technicality of this is that \(Q\)-supersymmetry actions on \((T_{abc}, \chi^{\a i}, D, \mathcal{X}_{\a}^i{}^{\b\g}, \mathcal{Y}_\a{}^\b{}^{kl}, \mathcal{Y}_{\a\b}{}^{\g\d})\) often produce \(\hat{\nabla}_a\) acting on another field. So, degauging the \(Q\)-connection is an iterative process. All other degauging of \(M,K,S\)-connections are simpler than this and do not require iteration.} Third, we extract the fermions within the component expansions of the composite fields \(\mathcal{X}_\a^i{}^{\b\g}, \mathcal{Y}_\a^{\b kl}, \mathcal{Y}_{\a\b}{}^{\g\d}\) --- this mostly amounts to substitution rules using formulae from \cite{Butter:2017jqu}, which we list in Appendix~\ref{app:l10}, and discarding any resulting higher-order terms. Fourth, we gauge-fix dilatation \(b_a=0\) and degauge the remaining bosonic part of the \(K\)-connection \(\hat{\mathfrak{f}}_{ab} = - \frac{1}{2} S_{ab}\). As we are only interested in pure gravitational backgrounds, the R-symmetry connection \(\mathcal{V}_a{}^{ij}\) in the covariant derivatives is discarded and this leaves us with \textit{Levi-Civita} covariant derivatives. Fifth, we integrate by parts to isolate the differential operators acting on fields of the Weyl multiplet, then use symmetries like Bianchi identities to simplify.

Though the algorithm described above to obtain the required component Lagrangians is conceptually straightforward, applying it in practice is daunting. In fact, the expansion of the gravitini alone includes thousands of terms in intermediate steps, and the symmetries provided by Bianchi identities, integration by parts and fermion bilinears can be highly non-trivial. To assist in this process, we have largely employed algorithms developed in the computer algebra software \emph{Cadabra} \cite{Peeters:2007,Peeters:2018no1,Peeters:2018no2}. Due to such complexity, a complete analysis of the 3-form and gravitini fields will be given elsewhere. Now we move to present the results we obtained so far and used in our paper.

\subsubsection{Scalar}
Reading directly off the bosonic \((1,0)\) Lagrangians in $\Lagr_{(2,0)\to(1,0)}$ in eq.~\eqref{l20} above, we obtain
\begin{equation}
\Lagr_D = \frac{4}{45}\, D \check{\nabla}^2 D
\,.
\end{equation}
Degauging the bosonic \(K\)-connection with \(f_{ab} = - \frac{1}{2} S_{ab}\) following \eqref{kdeg} and discarding the R-symmetry connection \(\mathcal{V}_a{}^{ij}\) leads to
\begin{equation}\label{dl}
\Lagr_D = \frac{4}{45}\, D \Big( 
\mathcal{D}^2 + \frac{1}{5} \mathcal{R}
\Big) D
\,,
\end{equation}
where \(\mathcal{R}\) is the scalar curvature and we used that \(D\) is a conformal primary with weight two. Up to an overall factor, the result is simply the expected Lagrangian for a 2-derivative conformally coupled scalar in six dimensions.

\subsubsection{Vector}
Again, we use the sum of the \((1,0)\) Lagrangians in eq.~\eqref{l20} to obtain the \((2,0)\) truncation, but this time for the vector \(\mathcal{V}_a{}^{ij}\). However, now we keep contributions to \textit{all} orders in order to check that, with the appropriate rewriting, the \(\mathrm{SU}(2)\) R-symmetry curvature-cubed term cancels. Indeed, this term is compatible with Weyl symmetry, but is incompatible with supersymmetry --- we will elaborate more on this shortly.
We refer to Appendix~\ref{app:bosonic10lag} for the full bosonic \((1,0)\) Lagrangians. We find
\begin{align}\label{vecl}
\Lagr_{\mathcal{V}}
    &=
-\mathcal{R}_{ab}{}^{ij} \check{\nabla}^2 \mathcal{R}^{ab}{}_{ij}
	-4\mathcal{R}_{ab}{}_i{}^j \mathcal{R}^{ac}{}_j{}^k \mathcal{R}^b{}_c{}_k{}^i
	+C^{abcd} \mathcal{R}_{ab}{}^{ij} \mathcal{R}_{cd}{}_{ij}
\nonumber\\
&\quad\,
	+4 S_a{}^b \mathcal{R}_{bc}{}^{ij} \mathcal{R}^{ac}{}_{ij}
	-2 S \mathcal{R}_{ab}{}^{ij} \mathcal{R}^{ab}{}_{ij}\,,
\end{align}
where \(S = S_a{}^a=\frac{1}{10}\mathcal{R}\) is the trace of the Schouten tensor and 
\begin{equation}
\mathcal{R}_{ab}{}^{ij} = e_a{}^m e_b{}^n \Big( 2\partial_{[m} \mathcal{V}_{n]}{}^{ij} + 2 \mathcal{V}_{[m}{}^{k(i} \mathcal{V}_{n]k}{}^{j)} \Big)
\end{equation}
is the \(\mathrm{SU}(2)\) curvature.
We can also see that \eqref{vecl} matches exactly with the bosonic \((2,0)\) Lagrangian \eqref{bl20} as the \(\mathrm{USp}(4)\) curvature
\begin{equation}
R(V)_{ab}{}^{\underline{ij}} = e_a{}^m e_b{}^n \Big( 2\partial_{[m} V_{n]}{}^{\underline{ij}} + 2 V_{[m}{}^{\uk(\ui} V_{n]\uk}{}^{\uj)} \Big)
\end{equation}
truncates (ignoring the vector and gravitini multiplets) to exactly the \(\mathrm{SU}(2)\) curvature. 

As we are interested in cancelling the R-symmetry curvature-cubed term, we need to keep the R-symmetry connections so we introduce the covariant derivative
\begin{equation}
\check{\mathcal{D}}_a = e_a - \frac{1}{2}\omega(e)_a{}^{bc} M_{bc} - \mathcal{V}_a{}^{ij} J_{ij} 
\end{equation}
consisting of just the Levi-Civita and R-symmetry connections. Degauging the \(K\)-connection and gauge-fixing dilatation to zero gives
\begin{align}\label{deg1}
\check{\nabla}^2 \mathcal{R}_{ab}{}^{ij} 
= \check{\mathcal{D}}^2  \mathcal{R}_{ab}{}^{ij} + \frac{1}{5} \mathcal{R} \mathcal{R}_{ab}{}^{ij}\,.
\end{align}
We can further manipulate equation \eqref{deg1} to extract a new \(\mathrm{SU}(2)\) curvature-cubed term by making use of\footnote{Using \(\check{\mathcal{D}}_{[a} \mathcal{R}_{bc]}{}^{ij}=0\), which follows from the \(\mathfrak{so}(1,5) \oplus \mathfrak{su}(2)\) curvature's Bianchi identity \newline \(\check{\mathcal{D}}_{[a}(\frac{1}{2}\mathcal{R}_{bc]}{}^{de} M_{de} + \mathcal{R}_{bc]}{}^{ij} J_{ij}) = 0\), we obtain
\begin{equation*}
\mathcal{R}^{ab}{}_{ij} \check{\mathcal{D}}^2 \mathcal{R}_{ab}{}^{ij}
	=
-2 \mathcal{R}^{ab}{}_{ij} \check{\mathcal{D}}_b \check{\mathcal{D}}_c \mathcal{R}^c{}_a{}^{ij}
+ C_{abcd} \mathcal{R}^{ab}{}_{ij}  \mathcal{R}^{cd}{}^{ij}
- \mathcal{R}_b{}^d \mathcal{R}^{ba}{}_{ij}  \mathcal{R}_{da}{}^{ij}
-\frac{1}{10} \mathcal{R} \mathcal{R}^{ab}{}_{ij} \mathcal{R}_{ab}{}^{ij}
+4\mathcal{R}^{ab}{}_{ij} \mathcal{R}_b{}^c{}^{il}  \mathcal{R}_{ca}{}_l{}^j\,.
\end{equation*}
}
\begin{equation}
[\check{\mathcal{D}}_a, \check{\mathcal{D}}_b] = - \frac{1}{2} \mathcal{R}_{ab}{}^{cd} M_{cd} - \mathcal{R}_{ab}{}^{ij} J_{ij}
\,,
\end{equation}
where \(\mathcal{R}_{abcd}\) is the usual Riemann curvature.
In the end, the Lagrangian reduces to
\begin{equation}\label{laa}
\Lagr_{\mathcal{V}} 
	=
2 \mathcal{R}^{ab}{}_{ij} \check{\mathcal{D}}_b \check{\mathcal{D}}_c \mathcal{R}^c{}_a{}^{ij}
	+ 2\Big(\mathcal{R}_b{}^d - \frac{1}{5} \d_b^d \mathcal{R} \Big)\mathcal{R}^{ba}{}_{ij}  \mathcal{R}_{da}{}^{ij}\,,
\end{equation}
which coincides with the action that already appeared in \cite{Beccaria:2017dmw} (with Levi-Civita connection only).

We can clearly see that in \eqref{laa} the cubic term in $\mathcal{R}_{ab}{}^{ij}$ cancels and also that the Weyl tensor terms cancel out. The Weyl tensor terms need to vanish if there exists a factorisation of differential operators on Einstein backgrounds. We discuss this in more detail in the next section.
The cancellation of the cubic term  is consistent with the fact that the truncation to flat space should precisely coincide with the bosonic Lagrangian for a \((1,0)\) supersymmetric extension of a $F\Box F$ term that was first constructed in \cite{Ivanov:2005qf} and agrees with \eqref{laa} upon taking a flat limit and replacing $\mathcal{R}_{ab}{}^{ij}$ with $F_{ab}{}^{ij}$. Since an independent supersymmetric extension of a $F^3$ term only constructed out of the fields of a vector multiplet in six dimensions is not known, and it is not expected to exist, it is consistent to observe that such cubic term does not appear in \eqref{laa}.

\subsubsection{Spinor}
Using the quadratic \((1,0)\) Lagrangians in Appendix~\ref{app:l10}, and considering the combination in eq.~\eqref{l20}, we obtain
\begin{align}\label{chi}
\Lagr_\chi
	&=
\frac{4\mathrm{i}}{225} \Big(  4\hat{\nabla}_{\a\b}{\chi^{\alpha i}} \hat{\nabla}^2{\chi^{\beta}{}_i}
+16  \hat{\nabla}_{b}{\chi^{\alpha i}} \hat{\nabla}^{b}{\hat{\nabla}_{\a\b}{\chi^{\beta}{}_i}}
\nonumber\\
&\qquad\quad\;\;
- 5\mathcal{R} \chi^{\alpha i}  \hat{\nabla}_{\a\b}{\chi^{\beta}{}_i}
+6 \mathcal{R}_{a b} \chi^{\alpha i} (\gamma_{a})_{\alpha \beta}  \hat{\nabla}_{b}{\chi^{\beta}{}_i} \Big)
\,,
\end{align}
where the full background gravitational and R-symmetry couplings are present, and the hidden fermions in \(Q,S,M,K\)-connections can be ignored since we are working at quadratic order. Considering only gravitational coupling, we need only degauge the \(K\)-connection, gauge-fix dilatation to zero and discard everything else. Thus, we can follow \eqref{kdeg} and discard its R-symmetry terms to obtain
\begin{subequations}
\begin{align}
\hat{\nabla}_a \chi^{\a i} 
&= \mathcal{D}_a \chi^{\a i}
\,, \\
\hat{\nabla}_a \hat{\nabla}_b \chi^{\a i} 
&= \mathcal{D}_a \mathcal{D}_b \chi^{\a i} + \frac{3}{2} S_{ab} \chi^{\a i} - \frac{1}{2} (\g_{bc})_\b{}^\a S_{a}{}^c \chi^{\b i}
\,.
\end{align}
\end{subequations}
Integration by parts then yields the expression\footnote{We denote \(\slashed{\cd }^3 = \g_a \tilde{\g}_b \gamma_c \mathcal{D}^a \mathcal{D}^b \mathcal{D}^c\) and also use that \(\chi_i \g_a \chi^i = 0\).}
\begin{align}\label{chil}
\Lagr_\chi &=
\chi_i \left[ \slashed{\mathcal D}^3 + 2S^{ab} \g_a \mathcal{D}_b \right] \chi^i
\,.
\end{align}
This Lagrangian has already appeared  in \cite{Beccaria:2017dmw} on a generic geometric background.

We briefly mention a reformulation of the spinor's pure gravitational quadratic Lagrangian in terms of a \(K\)-invariant operator.\footnote{We thank Arkady Tseytlin and Matteo Beccaria for private communication on this reformulation.} By construction, such an operator guarantees Weyl covariance when the \(K\)-connection is degauged from the covariant derivatives and dilatation is gauge-fixed to zero.\footnote{We discuss this formalism briefly earlier in this section, but  more details can be found in \cite{Butter:2016qkx,Freedman:2012zz,Lauria:2020rhc,Kuzenko:2022skv,Kuzenko:2022ajd}.}  
To this end, we introduce a conformal covariant derivative without R-symmetry
\begin{equation}
\tilde{\nabla}_a = e_a - \frac{1}{2}\omega(e,b)_a{}^{bc}M_{bc} - b_a \mathbb{D} - f_a{}_b K^b
\end{equation}
which consists of just the connection \(\omega(e,b)\) of eq.~\eqref{oeb}, the dilatation connection and the \(K\)-connection which reduces to the Schouten tensor, \(f_{ab}= -\frac{1}{2}S_{ab}\), upon setting $b_a=0$. One can show that
\bsubeq 
\begin{align}
K_{d} \left[ (\g_a \tilde{\g}_b \g_c -\g_{abc})\tilde{\nabla}^a \tilde{\nabla}^b \tilde{\nabla}^c \chi^i\right] &= 0 
\,, \\
K_{d} \left[ \chi_i \g_{abc}\tilde{\nabla}^a \tilde{\nabla}^b \tilde{\nabla}^c \chi^i\right] &= 0\,,
\end{align}
\esubeq
where the second equation holds only if one contracts with a second \(\chi\) on the left as well. Moreover, one can show that
\begin{equation}
\chi_i \g_{abc} \tilde{\nabla}^a \tilde{\nabla}^b \tilde{\nabla}^c \chi^i = 0
\end{equation}
vanishes identically, but would \textit{not}, and would not be \(K\)-invariant, if background R-symmetry was included in \(\tilde{\nabla}\). 
Hence, a manifestly conformal Lagrangian is given by 
\begin{align}\label{chil2}
\Lagr_\chi &=
\chi_i\,\tilde{\slashed{\nabla}}{}^3\,\chi^i
\,.
\end{align}
Degauging the \(K\)-connection to return to Levi-Civita covariant derivatives \(\mathcal{D} = d - \omega(e)\) reproduces \eqref{chil} obtained from the truncated \((2,0)\) Lagrangian.

\subsubsection{Gravitino}
We will not give the gravitino's quadratic Lagrangian on a general gravitational background as it is too unwieldy.  
We also focus on the physical transverse \((\mathcal{D}_a \psi^{a i \a} = 0)\) and gamma-traceless \((\gamma_a\psi^{a i }=0)\) components, which we denote \(\psi^{\perp a i \a}\). These conditions arise from \(Q\)- and \(S\)-supersymmetry gauge-fixing respectively. Moreover, we substantially simplify the gravitational background, by restricting to spherical and Ricci-flat, parallel curvature backgrounds. 

It is useful to note that the gamma-traceless condition, implies the following relations 
\begin{equation}
\begin{aligned}
\g_{a} \psi^{\perp a}{}_{i} &= 0 
\,,&
\tilde{\g}_{a b} \psi^{\perp b}{}_{i} &= \psi^\perp_{a i}
\,,&
\g_{a b c} \psi^{\perp c}{}_{i} &= 2\g_{[a} \psi^\perp_{b] i}
\,, \\
\tilde{\g}_{a b c d} \psi^{\perp d}{}_{i} &= 3 \tilde{\g}_{[a b} \psi^\perp_{c] i}
\,,\quad &
\g_{a b c d e} \psi^{\perp e}{}_{i} &= 4\g_{[a b c} \psi^\perp_{d] i}
\,,\quad &
\tilde{\g}_{a b c d e f} \psi^{\perp f}{}_{i} &= 5\tilde{\g}_{[a b c d} \psi^\perp_{e] i}
\,.
\end{aligned}
\end{equation}
As a result of gamma-tracelessness and the transverse condition, the quadratic Lagrangian arising from \eqref{l20} on a spherical background takes the simple form
\begin{equation}\label{gs}
\Lagr_\psi =
\mathrm{i} \psi^\perp_{a i} \mathcal{D}^4\slashed{\mathcal{D} }\psi^{\perp a i}    
+ \frac{2\mathrm{i}}{5} \psi^\perp_{a i} \mathcal{R} \mathcal{D}^2 \slashed{\mathcal{D}} \psi^{\perp a i}      
+ \frac{3\mathrm{i}}{80} \psi^\perp_{a i} \mathcal{R}^2 \slashed{\mathcal{D}} \psi^{\perp a i}
\,,
\end{equation}
where \(\mathcal{D}^4 = \mathcal{D}^2 \mathcal{D}^2\).
On a Ricci-flat, parallel curvature background we have evaluated the 5- and 3-derivative terms as
\begin{equation}\label{grp}
\Lagr_\psi = 
\mathrm{i} \psi^\perp_{a i} \mathcal{D}^4\slashed{\mathcal{D} }\psi^{\perp a i} 
+ \mathrm{i} C^{abcd} \psi^\perp_{a i} \g_{cde} \mathcal{D}^2 \mathcal{D}^e  \psi^\perp_{b}{}^i
+ \text{lower derivatives}
\,.
\end{equation}
We do not evaluate the lower-derivative terms, since the analysis is considerably complicated due to the interplay between the symmetries of the fermion bilinears, the transverse and gamma-traceless conditions, and the parallel curvature condition.\footnote{The parallel curvature condition, i.e.\ the requirement that the Riemann tensor be covariantly constant, implies
\begin{equation*} \label{Pcurv}
0 
	= [\mathcal{D}_{a}, \mathcal{D}_{b}]\mathcal{R}_{cdef} 
	=  \mathcal{R}_{abc}{}^{g} \mathcal{R}_{dgef} - \mathcal{R}_{abd}{}^{g} \mathcal{R}_{cgef} + \mathcal{R}_{abe}{}^{g} \mathcal{R}_{cdfg} - \mathcal{R}_{abf}{}^{g} \mathcal{R}_{cdeg} 
\,.
\end{equation*}
It is interesting to note that the tensor \(T_{abcdef} := \mathcal{R}_{abc}{}^{g} \mathcal{R}_{gdef}\), when considered with the constraint from parallel curvature, does not live in a single Young tableau with \(T_{ab[cdef]}=0\) as a Garnir symmetry. Tensors living in a Young tableau naturally have multi-term symmetries where they vanish if antisymmetrised over some subsets of indices, e.g.\ \(\mathcal{R}_{[a b c] d} = 0\) for the Riemann tensor. Such symmetries are called Garnir symmetries \cite{Peeters:2011,Fischbacher:2005fy}.
}

Before moving to the 3-form field, it is important to stress that we have explicitly checked that mixed terms between spinor \(\chi\) and gravitino \(\psi_a{}^i\)  (without restricting to the transverse and gamma-traceless component) at quadratic level vanish in generic gravitational backgrounds.\footnote{This requires integration by parts and the use of involved relations that come from Bianchi identities, symmetries of the curvature tensor, and symmetries of fermion bilinears.
In particular, we used
\begin{align*}
0
    &=
    2 C_{a b c d} C^{a b c e} \chi^{i} \gamma^{d} \psi_{e i}
-  C_{a b c d} C^{a b c d} \chi^{i} \gamma^{e} \psi_{e i} 
+  C_{a b c d} C^{a b e f} \chi^{i} \gamma^{c d}{}_{e} \psi_{f i} 
- \frac{1}{2}C_{a b}{}^{c d} C^{a b e f} \chi^{i} \gamma_{c d e f h} \psi^h{}_i
\,.
\end{align*}
}
This is a reassuring consistency check on the fermionic Lagrangian since an off-diagonal term would not be allowed by \(S\)-supersymmetry.

\subsubsection{3-form}

The quadratic Lagrangian for the 3-form can be read from the bosonic \((1,0)\) Lagrangian in the combination of eq.~\eqref{l20} as
\begin{align}
\Lagr_T
	&=
\frac{32}{3}C^{a b c d} T_{a b}{}^{e} \check{\nabla}_{e}{\check{\nabla}^{f}{T_{c d f}}}
+\frac{16}{3}C^{a b c d} T_{a b}{}^{e} \check{\nabla}^{f}{\check{\nabla}_{e}{T_{c d f}}} 
- \frac{32}{3}C^{a b c d} T_{a}{}^{e f} \check{\nabla}_{b}{\check{\nabla}_{c}{T_{d e f}}}
\nonumber\\
&\quad\,
+8C^{a b c d} \check{\nabla}_{a}{T_{b}{}^{e f}} \check{\nabla}_{c}{T_{d e f}}
+12C^{a b c d} \check{\nabla}^{e}{T_{a b}{}^{f}} \check{\nabla}_{f}{T_{c d e}} 
- \frac{16}{3}C^{a b e f} C^{c d}{}_{e f} T_{a b}{}^{g} T_{c d g}
\nonumber\\
&\quad\,
-4T^{abc} \Delta^4 T_{abc}
\,,
\end{align}
where \(T^{abc} \Delta^4 T_{abc}\) retains the same form as \eqref{3fop}. Gauge-fixing dilatation and degauging the \(K\)-connection, \(f_{ab} = - \frac{1}{2} S_{ab}\), gives the gravitational background result
\begin{align}
\Lagr_T
	&=
\frac{16}{3}C^{a b c d} T_{a b}\,^{e} \mathcal{D}^{f}{\mathcal{D}_{e}{T_{c d f}}}
+\frac{32}{3}C^{a b c d} T_{a b}\,^{e} \mathcal{D}_{e}{\mathcal{D}^{f}{T_{c d f}}} 
- \frac{32}{3}C^{a b c d} T_{a}\,^{e f} \mathcal{D}_{b}{\mathcal{D}_{c}{T_{d e f}}}
\nonumber\\
&\quad\,
+12C^{a b c d} \mathcal{D}^{f}{T_{a b}\,^{e}} \mathcal{D}_{e}{T_{c d f}}
+8C^{a b c d} \mathcal{D}_{a}{T_{b}\,^{e f}} \mathcal{D}_{c}{T_{d e f}} 
- \frac{16}{3}C^{a b c d} C_{a b}\,^{e f} T_{c d}\,^{g} T_{e f g}
\nonumber\\
&\quad\,
+\frac{16}{15}C^{a b c d} T_{a b}\,^{e} T_{c d e} \mathcal{R} 
- \frac{16}{3}C^{a b c d} T_{a b}\,^{e} T_{c d}\,^{f} \mathcal{R}_{e f}
-4 T^{abc} \Delta^4 T_{abc}
\,,
\end{align}
where
\begin{align}
T^{abc} \Delta^4 T_{abc}
	&=
T^{a b c} \mathcal{D}^2 \mathcal{D}_{a}{\mathcal{D}^{d}{T_{b c d}}}
+\frac{1}{3}T^{a b c} \mathcal{D}_{a}{\mathcal{D}^2{\mathcal{D}^{d}{T_{b c d}}}} 
- \frac{4}{3}T^{a b c} \mathcal{D}^{e}{\mathcal{D}_{a}{\mathcal{D}^{d}{\mathcal{D}_{e}{T_{b c d}}}}}
\nonumber\\
&\quad\,
+T^{a b c} \mathcal{D}_{a}{\mathcal{D}^{d}{\mathcal{D}^2{T_{b c d}}}} 
- \frac{1}{12}T^{a b c} T_{a b}\,^{d} \mathcal{D}_{d}{\mathcal{D}_{c}{\mathcal{R}}} 
- \frac{1}{12}T^{a b c} T_{a b}\,^{d} \mathcal{D}_{c}{\mathcal{D}_{d}{\mathcal{R}}} 
\nonumber\\
&\quad\,
- \frac{1}{2}T^{a b c} T_{a b}\,^{d} \mathcal{D}^2{\mathcal{R}_{c d}}
+\frac{2}{3}T^{a b c} T_{a}\,^{d e} \mathcal{D}_{d}{\mathcal{D}_{b}{\mathcal{R}_{c e}}}
+\frac{1}{5}T^{a b c} \mathcal{D}^{d}{T_{a b d}} \mathcal{D}_{c}{\mathcal{R}} 
\nonumber\\
&\quad\,
- \frac{1}{3}T^{a b c} \mathcal{D}^{e}{T_{a b}\,^{d}} \mathcal{D}_{e}{\mathcal{R}_{c d}} 
- \frac{4}{3}T^{a b c} \mathcal{D}^{e}{T_{a b}\,^{d}} \mathcal{D}_{c}{\mathcal{R}_{e d}}
-T^{a b c} \mathcal{D}^{d}{T_{a d}\,^{e}} \mathcal{D}_{b}{\mathcal{R}_{c e}} 
\nonumber\\
&\quad\,
- \frac{11}{30}T^{a b c} \mathcal{D}_{a}{T_{b c}\,^{d}} \mathcal{D}_{d}{\mathcal{R}} 
- \frac{2}{3}T^{a b c} \mathcal{D}_{a}{T_{b}\,^{d e}} \mathcal{D}_{d}{\mathcal{R}_{c e}} 
- \frac{1}{6}T^{a b c} \mathcal{D}^{e}{\mathcal{D}^{d}{T_{a b d}}} \mathcal{R}_{c e} 
\nonumber\\
&\quad\,
- \frac{1}{2}T^{a b c} \mathcal{D}^{d}{\mathcal{D}^{e}{T_{a b d}}} \mathcal{R}_{c e} 
- \frac{1}{2}T^{a b c} \mathcal{D}^2{T_{a b}\,^{d}} \mathcal{R}_{c d} 
- \frac{17}{60}T^{a b c} \mathcal{D}^{d}{\mathcal{D}_{a}{T_{b c d}}} \mathcal{R}
\nonumber\\
&\quad\,
-T^{a b c} \mathcal{D}^{e}{\mathcal{D}_{a}{T_{b c}\,^{d}}} \mathcal{R}_{e d}
+\frac{37}{60}T^{a b c} \mathcal{D}_{a}{\mathcal{D}^{d}{T_{b c d}}} \mathcal{R} 
- \frac{2}{3}T^{a b c} \mathcal{D}_{a}{\mathcal{D}^{e}{T_{b c}\,^{d}}} \mathcal{R}_{e d} 
\nonumber\\
&\quad\,
- \frac{5}{3}T^{a b c} \mathcal{D}^{d}{\mathcal{D}_{a}{T_{b d}\,^{e}}} \mathcal{R}_{c e}
+\frac{4}{3}T^{a b c} \mathcal{D}_{a}{\mathcal{D}^{d}{T_{b d}\,^{e}}} \mathcal{R}_{c e} 
- \frac{1}{12}T^{a b c} T_{a b}\,^{d} \mathcal{R} \mathcal{R}_{c d}
\nonumber\\
&\quad\,
+\frac{1}{12}T^{a b c} T_{a b}\,^{d} \mathcal{R}_{c}\,^{e} \mathcal{R}_{d e}+\frac{1}{6}T^{a b c} T_{a}\,^{d e} \mathcal{R}_{b d} \mathcal{R}_{c e}
\,.
\end{align}

For completeness, we also provide the explicit \(K\)-connection degauging formulae which have been used (we discard the R-symmetry connection)\footnote{Note that \(K_{a}T_{bcd} = 0\) being primary implies \(\check{\nabla}_a T_{bcd} = \mathcal{D}_a T_{bcd}\). Also, these are consistent with the similar formulae in \cite{Butter:2017jqu}.} 
\begin{subequations}
\begin{align}
\check{\nabla}_{a} T_{def} 
	&= 
\mathcal{D}_{a} T_{def}
\,,\\
\check{\nabla}_{b}{\check{\nabla}_{c}{T_{d e f}}}
	&=
\mathcal{D}_{b}{\check{\nabla}_{c}{T_{d e f}}}
-2f_{b c} T_{d e f}
-6f_{b [d} T_{e f] c}
+6f_{b}{}^{g} \eta_{c [d} T_{e f] g}
\,,\\
\check{\nabla}_{a}{\check{\nabla}_{b}{\check{\nabla}_{c}{T_{d e f}}}}
	&=
\mathcal{D}_{a}{\check{\nabla}_{b}{\check{\nabla}_{c}{T_{d e f}}}}
-8f_{a (b} \check{\nabla}_{c)}{T_{d e f}}
-6f_{a [d} \check{\nabla}_{|c|}{T_{e f] b}}
-6f_{a [d} \check{\nabla}_{|b|}{T_{e f] c}}
\nonumber\\
&\quad\,
+2f_{a}{}^{h} \eta_{b c} \check{\nabla}_{h}{T_{d e f}}
+6f_{a}{}^{h} \delta_{b [d} \check{\nabla}_{|c|}{T_{e f] h}}
+6f_{a}{}^{h} \delta_{c [d} \check{\nabla}_{|b|}{T_{e f] h}}
\,,\\
\check{\nabla}_{a}{\check{\nabla}_{b}{\check{\nabla}_{c}{\check{\nabla}_{d}{T_{e f g}}}}}
	&=
\mathcal{D}_{a}{\check{\nabla}_{b}{\check{\nabla}_{c}{\check{\nabla}_{d}{T_{e f g}}}}}
-2\check{\nabla}_{c}{\check{\nabla}_{b}{T_{e f g}}} f_{a d}
-4\check{\nabla}_{b}{\check{\nabla}_{c}{T_{e f g}}} f_{a d}
-6\check{\nabla}_{b}{\check{\nabla}_{d}{T_{e f g}}} f_{a c}
\nonumber\\
&\quad\,
-6\check{\nabla}_{c}{\check{\nabla}_{d}{T_{e f g}}} f_{a b}
-6\check{\nabla}_{c}{\check{\nabla}_{d}{T_{b [e f}}} f_{|a| g]}
-6\check{\nabla}_{b}{\check{\nabla}_{d}{T_{c [e f}}} f_{|a| g]}
-6\check{\nabla}_{b}{\check{\nabla}_{c}{T_{d [e f}}} f_{|a| g]}
\nonumber\\
&\quad\,
+2\eta_{b c} \check{\nabla}_{h}{\check{\nabla}_{d}{T_{e f g}}} f_{a}{}^h
+2\eta_{b d} \check{\nabla}_{c}{\check{\nabla}_{h}{T_{e f g}}} f_{a}{}^h
+2\eta_{c d} \check{\nabla}_{b}{\check{\nabla}_{h}{T_{e f g}}} f_{a}{}^h
\nonumber\\
&\quad\,
+6\eta_{b [e} \check{\nabla}_{|c}{\check{\nabla}_{d|}{T_{f g] h}}} f_{a}{}^h
+6\eta_{c [e} \check{\nabla}_{|b}{\check{\nabla}_{d|}{T_{f g] h}}} f_{a}{}^h
+6\eta_{d [e} \check{\nabla}_{|b}{\check{\nabla}_{c|}{T_{f g] h}}} f_{a}{}^h
\,.
\end{align}
\end{subequations}
When the 3-form's quadratic Lagrangian is restricted to the sphere \(S^6\), we obtain
\begin{equation}\label{3fs}
\Lagr_T = T_{abc} \mathcal{D}^a \mathcal{D}^2 \mathcal{D}_d T^{bcd} + \frac{1}{5} \mathcal{R} T_{abc} \mathcal{D}^a \mathcal{D}_d T^{bcd}
\,.
\end{equation}

We now turn to using these results to determine the same expressions in the Euclidean case. These will then be used for the heat kernel calculation of the conformal anomalies.

\section{Quadratic Lagrangians} \label{sect:quadrlagra}
In this section we  present the expressions for the quadratic Lagrangians in Euclidean space used in the derivation of the conformal anomalies and discuss the factorisation of the operators that we employ. We often drop overall factors in analysis of the actions, as they are irrelevant for our purposes.
 
We emphasize that from this section on we switch to Euclidean notation to have a formally convergent functional integral and we use Euclidean Majorana spinors. This is not without subtleties but is inconsequential for conformal anomalies (cf.\ footnote~\ref{foot:ferm}).
The notation is presented in Appendix \ref{app:notQFT}.

For the graviton case we perform a general calculation clarifying some aspects of the previous literature.
For the gravitino and the 3-form we do not have a full calculation but we have sound Ansatz and consistency check. We will discuss this in detail.

We use the symbol \(\perp\) to indicate the irreducible representations given by covariantly transverse and traceless (or gamma-traceless in the case of the gravitino) component.
We introduce several second-order differential operators denoted as \(\Delta_ f\), where \(f\) indicates the field they act upon, hopefully in a self-explanatory notation.\footnote{This notation differs from the one used elsewhere in the literature, where the subscript refers to the order of the differential operator.}
For reference, the operators are also reported in \eqref{ope0}-\eqref{ope32}.

\subsection{Graviton}
We consider now the particular Weyl-invariant gravitational theory that admits the \(\mathcal{N}=(2,0)\) supersymmetric extension, which is \(W^{(2,0)}\) defined in \eqref{ww2}.\footnote{
As mentioned in the introduction, this analysis was  independently done in \cite{Aros:2019tjw}.
}
It has the peculiar property that, by adding a term proportional to  \(\mathbb E _6\) that does not change the classical and one-loop structure, it can be cast in the form (see also \cite{Metsaev:2010kp,Beccaria:2017lcz})
\begin{equation}\label{efl}
S = \int \! d{^6x} \sqrt g \left[
	R^{mn} \cd^2 R_{mn} 
	- \frac3{10} R \cd^2 R
	- 2 R^{mnrs} R_{nr} R_{ms}
	- R^{mn} R_{mn} R
	+ \frac 3 {25} R^3
\right]\,.
\end{equation}
This is remarkable in that it is  at most linear in the uncontracted Riemann tensor, thus vanishing on Ricci-flat geometries, and admits Einstein-type solutions. This form of the action is significantly simpler to manipulate than \eqref{ww2}, as  the application of Bianchi identities is more straightforward. As mentioned in the introduction, the integrand of \eqref{efl} is exactly the \(\mathscr Q\)-curvature \(\mathscr Q_6\) given in \eqref{Q6} and this fact has a four-dimensional analogue (see footnote~\ref{foot:Q4}).

We then need the quadratic action on a geometric background, so we expand \(g_{mn} \to g_{mn} + h_{mn}\), where the new metric \(g_{mn}\) is taken to be Einstein. We further decompose \(h_{mn}\) in its irreducible components via \eqref{jkf2}; only the transverse traceless mode \(h_{mn}^\perp\) appears then in the action \eqref{efl} due to diffeomorphism and conformal invariance. 
It is most convenient to present the result of the expansion in the form of the differential operators  acting on symmetric rank-2 tensors
\begin{equation}\label{ibd}
\begin{aligned}
\Delta_2[r]  h_{mn} = 
- \cd^2 h_{mn}
 - 2  R_{mrns}h^{ rs} 
 + r R h_{mn} 
 \,,
\end{aligned}
\end{equation}
which are deformations of the Lichnerowicz operator \cite{Pang:2012rd,Beccaria:2017lcz}.
A direct calculation shows that the  quadratic expansion of the action \eqref{efl} can be rewritten in the form
\begin{equation}\label{oaa}
S_h = \int 
 \! \sqrt{g} \  
 h_{mn}^\perp \ \Delta_2 [\tfrac{2}{15}] \ \Delta_2 [\tfrac{1}{5}]\  \Delta_2 [0]\	h_{mn}^\perp
\,.
\end{equation}
We notice here a first difference in the structure of the differential operators between the graviton and lower-rank fields as e.g.\ those considered in \cite{Casarin:2023ifl}, namely possible couplings with the Riemann tensors due to the higher number of Lorentz indices. In fact, this very coupling plays an important role, because it decouples the (nonphysical) longitudinal mode, since
\begin{equation}\label{fey}
\cd^m \Delta_2[r]  h_{mn} 
 = 
- \cd^2 \cd^m h_{mn} 
 + \left(r- \frac1d\right) R \cd^m h_{mn}\,,
\end{equation}
which follows from
 \(\cd^m  \cd^2 h_{mn} 
 =   2R_{namc}\cd^c h^{  am } \),
and therefore \(\cd^m \Delta_2[r]  h_{mn} \) is transverse if and only if \(h_{mn}\) is. The scalar mode corresponding to the trace decouples as well, since clearly  \(g^{mn} \Delta_2[r]  h_{mn} =0 \) if and only if \(g^{mn}h_{mn} =0 \). Furthermore, these two decouplings take place independently of each other. The structure of the operator \eqref{ibd} matches the expressions for those appearing in \(4d\) conformal supergravity, see \cite{Fradkin:1985am,Tseytlin:2013jya}, as well as those employed to study conformal higher spin theories on  \(S^6\) in \cite{Tseytlin:2013fca}.

The expression \eqref{oaa} was first given in  \cite{Pang:2012rd}, where it was derived with the additional parallel (covariantly constant) curvature assumption \(\cd_a R_{mnrs}=0\).
However the factorisation of \eqref{oaa}  relies solely on the Einstein condition, without any other assumption on the background geometry, as already emphasized in \cite{Aros:2019tjw}. This  therefore extended the result	 of  \cite{Pang:2012rd} and  a posteriori justified the subsequent analysis of \cite[Appendix D]{Beccaria:2017lcz}.

\subsection{Gravitino and 3-form}
Here we present the analogous calculation for the gravitino and 3-form. It is somewhat more conjectural than the one performed for the graviton, however given the amounts or cross-checks discussed in the introduction we are confident in the results.

The gravitino, which in Euclidean signature we describe as a \(8\) component Majorana spinor and denote  \(\psi_m\), is discussed in full analogy to the graviton, i.e.\ in terms of its irreducible transverse and gamma-traceless component \(\psi^\perp_m\) and  has a 5-derivative kinetic term.

The 3-form field \(T_{mnr}\) has a more complicated structure.  Our analysis is heavily inspired by the one  done for the 2-form of \(4d\) conformal supergravity in \cite{Fradkin:1981jc,Fradkin:1985am,Fradkin:1982xc}, with the additional complication of the (anti)-self-duality constraint.
The quadratic term has the form
\begin{equation}\label{fre}
\Lagr _T \simeq \cd^m T_{mac} \cd^2 \cd_n T^{nac} + \text{lower derivative terms},
\end{equation}
where the lower derivative terms involve couplings with the curvature tensors. The way of writing \eqref{fre} is not unique due to (anti)-self-duality, however there is a feature of the action \eqref{fre} that cannot be avoided: the kinetic operator is non-minimal, because the highest derivative term is not a power of the Laplacian. In fact, such a term is algebraically zero, \(T^{mnr}\cd^4 T_{mnr} =0\), as a consequence of self-duality.  To overcome this difficulty we use the Hodge decomposition of the type \eqref{jkf4}.  The requirement of (anti)-self-duality forces the 2-form \(W_{mn}\) to be proportional to \(V_{mn}\). The Lagrangian \eqref{fre} gives then a six-derivative operator for \(V\), which if restricted to the  transverse mode is minimal.

We therefore adopt the decomposition
\begin{equation}\label{dfn}
T_{mnr} = 
\cd_{  [m } V^\perp_{ nr]  } \pm \frac {\rm i} 6 \varepsilon_{mnrace} \cd^{ a }   V^{\perp\,   ce }
\,,
\end{equation}
where the sign  in front of \(\rm i\) depends on the details of the Euclidean rotation\footnote{In six Lorentzian dimensions one has real (anti)-self-dual forms obeying \(3!\,T^{\pm}_{mnr} =\pm \varepsilon_{mnracs} T^{\pm acs}\); in six Euclidean dimensions this condition becomes \(3!\,T^{\pm}_{mnr} =\pm {\rm i} \varepsilon_{mnracs} T^{\pm acs}\).} (see e.g.~\cite{Bilal:2003es}) but it does not matter in our calculation, as it only enters squared. We use real \(V_{mn}\) so as to preserve the number of independent degrees of freedom.
In writing \eqref{dfn} we furthermore restricted \(V_{mn}\) to be transverse.   The   two-form \(V_{mn}\)  thus introduced has  a 6-derivative kinetic term, which is by construction a minimal one in the transverse mode  \(V_{mn}^\perp\).

In redefining the fields as described we ignore the contribution of harmonic forms. These are known to influence the  topological (type-A) anomaly, i.e.\ the \(a\)-coefficient \cite{Duff:1980qv,Fradkin:1985am,Fradkin:1982xc}. Since the \(a\)-anomaly for the 3-form is already known from \cite{Beccaria:2015uta} and the focus of our work is essentially the \(c_i\) coefficients, here we do not dwell on this aspect any further.

\subsubsection{Structure of the quadratic  Lagrangians}

Given the structure of \eqref{ibd}, we expect factorisations of the form 
\bsubeq
\begin{align}\label{fdk}
\mathcal L_\psi & \simeq
\bar \psi^\perp _m (-\cd^2 + \cdots) (-\cd^2 + \cdots) \slashed \cd   \psi^{\perp m}  \,, \\
\label{fdh}
\mathcal L_T  &   \simeq V^\perp_{mn} (-\cd^2 + \cdots) (-\cd^2 + \cdots) (-\cd^2 + \cdots)V^{\perp mn}.
\end{align}
\esubeq
We  introduce the following parametric families of differential operators\footnote{In \(\Delta_{2\text f}\) we use the Weyl tensor instead of the Riemann tensor for convenience, but this choice is of course arbitrary.
}
\bsubeq\label{eff2}
\begin{align}\label{eff}
\Delta_{\frac32}[r] \psi_m &= 
- \cd^2 \psi_m -\frac12 R_{mnrs} \Gamma^{rs} \psi^n + r R \psi_m
\,,
\\* \label{efg}
\Delta_{2\text f}[s] V_{mn}  &= 
- \cd^2 V_{mn} - C_{mnrs}V^{rs} + s   R  V_{mn}\,.
\end{align}
\esubeq
Notice that, when restricted to transverse and (gamma-)traceless components as in \eqref{fdk} and \eqref{fdh}, no other structure can appear, because e.g.\ \(\Gamma^{mn} \psi^\perp_{n} \propto \psi^{\perp m}\) as a consequence of the Clifford algebra and of \(\Gamma^m\psi^\perp_m =0\).
We have chosen the numerical coefficients of the curvature tensors because they are the only ones that preserve transversality condition mimicking   the graviton case, \eqref{fey}, and they allow for natural properties to be satisfied, as we are about to discuss. Furthermore, in analogy with the graviton operator \eqref{ibd}, the gravitino operators  \(\Delta_{\frac32}\)   as defined in \eqref{eff} match those of  \(4d\) conformal supergravity, cf.\ \cite{Fradkin:1985am,Tseytlin:2013jya}, as well as those used for conformal higher spin theories on  \(S^6\) in \cite{Tseytlin:2013fca}.

We propose a factorisation on physical fields of the type
\bsubeq
\begin{align}\label{ife}
S_\psi & =  \int \sqrt{g} \ \bar \psi^\perp_m  \ \Delta_{\frac32}[r_1] \ \Delta_{\frac32}[r_2] \ {\rm i} \slashed \cd \  \psi^{\perp m} \,,
\\
\label{ife2}
S_T & =  \int \sqrt{g} \   V^\perp_{mn}  \ \Delta_{2\text f}[s_1]  \ \Delta_{2\text f}[s_2] \
\Delta_{2\text f}[s_3] \ V^{\perp mn}
\,,
\end{align}
\esubeq
where we ignored irrelevant normalisation factors for the fields. Before discussing explicit values, we motivate our Ansatz  \eqref{ife}.

Let us start with the   the 2-form operator. First, having the same coupling to the Weyl tensor in all three factors ensures that the operators commute, consistently with self-adjointness. Then, the specific coefficient \(-1\) ensures that  \(\Delta_{2 \text f}\) preserves transversalitly as a  consequence of
\begin{equation}\label{few}
\cd^m \cd^2 V_{mn} =
 \cd^2 \cd^m V_{mn }
+ R_{acrs}\cd^c V^{rs}
+\frac16 R \cd^m V_{mn}
\,.
\end{equation}

For the gravitino we observe  the  relations
\bsubeq\label{plj}
\begin{align}
\Gamma^m \Delta_{\frac{3}{2}}[r] \psi_m
	& = - \cd^2(\Gamma^m \psi_m) + \left(r - \frac 1d \right)R \,  \Gamma^m \psi_m\,,
\\ 
\cd^m  \Delta_{\frac{3}{2}}[r] \psi_m
	& =  - \cd^2 (\cd^m \psi_m)  +\left(r - \frac 1 d\right) R\, \cd^m \psi_m \,.
\end{align}
\esubeq
As a consequence,  \(\Delta_{\frac32}\) preserves transversality and gamma-tracelessness \emph{separately}, and this fact plays a key role in the evaluation of its determinant, cf.\ Appendix~\ref{app:detp}. This is not the case for the Dirac operator, since
\begin{equation}\label{dop}
\Gamma^m \slashed \cd \psi_m 
	 = - \slashed \cd ( \Gamma^m \psi_m ) + 2 \cd^m \psi_m\,,
	\qquad
\cd^m \slashed \cd \psi_m 
	 = \slashed \cd \cd^m \psi_m + \frac1{2d} R \Gamma^m \psi_m\,.
\end{equation}
However,   the Dirac operator commutes\footnote{This follows from
\(
[\slashed  \cd , -\cd^2] \psi_m
	 = - 2
		\Gamma^r R_{rams} \cd^a \psi^s
\)
and \eqref{qab}. This depends on the value of the coupling with the curvature tensor.}  with  \(\Delta_{\frac32}[r]\) for all \(r\) and the square of the Dirac operator on the gravitino has precisely the form \(\Delta_{\frac32}\),
\begin{equation}
\label{pljj}
({\rm i} \slashed \cd_\frac32)^2  
	= \Delta_{\frac32}[\tfrac14]   \,.
\end{equation}
Once again, the fact that the coupling to the Riemann tensor is the same in the two-derivative factors in \eqref{ife} guarantees their commutation, hence the three operators in the factorisation totally commute as required by (anti)-self-adjointness.

For these reasons factorisations of the form \eqref{ife} and \eqref{ife2} appear to be the natural expectation in analogy to the graviton case. 
We have not explicitly verified in full generality yet that on Einstein backgrounds the Lagrangian density does factorize in terms of the operators \eqref{eff} and \eqref{efg}. However in the next subsection we gather the evidence that we have found and we fix the values of the remaining coefficients. A complete analysis is left for future work.

\subsubsection{Comparison with explicit evaluations}
Focusing on the physical components (transverse and gamma-traceless), we have evaluated the conformal supergravity action on the six sphere \(S^6\), whose Riemann tensor in our notation is  \eqref{s6}. In such a background the factorisation of the operators is automatic and allows us to fix the values of the coefficients \(r_i\)'s and \(s_i\)'s. 

An additional analysis of the 3-form action based on Weyl-symmetry alone is reported in Appendix~\ref{app:T3}. In particular, it confirms the \(S^6\) case and shows that a factorised action  is necessarily of the form \eqref{ife2}.

Furthermore, \eqref{grp} shows that the factorisation \eqref{ife} is consistent in the gravitino case in the 5- and 3-derivative for a Ricci-flat background obeying the parallel curvature condition.

The quadratic Lagrangians for the gravitino and 3-form on \(S^6\) have been evaluated in Section~\ref{sect:confsugra}.
Translating to the Euclidean QFT notation the gravitino Lagrangian \eqref{gs}, we obtain the action
\begin{equation}\label{sph}
S_{\psi} =\int \! \sqrt{g}\
\bar \psi^\perp_m \left[
	- \cd^4 + \frac25 R \cd^2 - \frac3 {80}R^2
\right]{\rm i} \slashed \cd \,\psi^\perp_m \,,
\end{equation} 
Comparing \eqref{sph} with \eqref{ife} gives 
\begin{equation}\label{pya}
r_1 =\frac {7} {60}
\,,
\qquad\qquad
r_2 = \frac {13} {60} 
\,.
\end{equation}
The 3-form Lagrangian on \(S^6\) was evaluated in \eqref{3fs}. Switching to Euclidean and expressing the 3-form \(T\) in terms of \(V\) as in \eqref{dfn} yields the action
\begin{align}\label{tub}
S_{T} =\int \! \sqrt{g}\ 
\left[
V^{ac}   \cd^6V_{ac} 
- \frac {11}{15 }R V^{ac} \cd^4 V_{ac}
+\frac 8 {45} R^2 V^{ac} \cd^2V_{ac}
-\frac{16}{1125}  R^3 \,   V^{ac}  V_{ac}	
\right].
\end{align}
Comparison with \eqref{ife2} gives
\begin{equation}\label{alvs}
s_1= \frac 4 {15}=s_2\,,\qquad\qquad
s_ 3= \frac 15\,.
\end{equation}
The two pieces of information \eqref{pya} and \eqref{alvs}
completely fix the factorised action with our Ansatz and the  numerical values are in line with those obtained for other fields in similar calculations, which is quite reassuring.

\subsection{Vector, spinor and scalar}

The low-spin fields that are relevant for the \((2,0)\) conformal supergravity  have  been reported elsewhere in the literature, but we summarize them here in Einstein backgrounds for completeness and clarity of exposition. 

The scalars are real and appear with a standard 2-derivative conformally coupled kinetic term as shown in \eqref{dl}, 
\begin{equation}\label{dcs}
S_D = \frac12 \int \! \sqrt{g} \ D \, \Delta_{0}[\tfrac15]\,  D\,,
\qquad\qquad
\Delta_{0}[r]	
	 =  -  \cd^2  +    r    R  \,,
\end{equation}
where \(\frac{d-2}{4(d-1)} \to \frac15\) is the conformal-coupling parameter in  \(d=6\).

We denote the \(8\)-component Majorana spinor field as \(\chi\). We obtained the Lagrangian in \eqref{chil}; converting it to the present notation the action reads
\begin{equation}\label{dcr}
S_\chi = {\rm i}\int \! \sqrt{g}  \ \bar \chi\left[
\slashed \cd^3 +\frac1{30} R\, \slashed \cd
\right] \chi\,.
\end{equation}
Notice that the operator factorises as 
\begin{equation}\label{dcr2}
S_\chi = -\int \! \sqrt{g}  \ \bar \chi
\, \Delta_{\frac12}[\tfrac{13}{60}]	\, {\rm i} \slashed \cd \,  \chi\,,
\qquad\quad
\Delta_{\frac12}[r]	
	 = 
			- \mathbbm 1_8 \cd^2  +    r  \mathbbm 1_8  R  
   \,,
\end{equation}
where we used \(\slashed \cd^2 = \cd^2 - \frac14 R\).

Finally we have the 4-derivative R-symmetry gauge vector, which for simplicity we denote  \(A_m\). The action, from \eqref{laa} is 
\begin{equation}\label{dcd}
\begin{aligned}
S_A & = \int \! \sqrt{g}  \ \left[
\cd_r F^{rm} \cd^n F_{nm} + \frac1{30} R F^{mn} F_{mn}
\right] \,,
\qquad \qquad 
F_{mn} = \cd_{ m } A_{ n }-\cd_{ n } A_{ m }\,,
\\
& =
 \int \! \sqrt{g}  \  A^{\perp m} \Big[
   			\cd^4 
			 - \frac{2}{5}   R\cd^2  
			 +  \frac{7}{180}   R^2
  \Big] A^\perp_{m}\,,
\end{aligned}
\end{equation}
where we have dropped R-symmetry indices and the terms that lead in the action to orders higher than quadratic in $A_m$, and we have also used that, owing to gauge invariance, only the transverse part  \(A^\perp_m\) actually  appears in the action. 
We observe also in this case that the kinetic operator factorizes in terms of minimal 2-derivative operators as
\begin{equation}\label{dcd2} 
S_A  =
  \int \! \sqrt{g}  \  A^{\perp m}
  	\Delta_{1}[\tfrac16] \ 
  	  	\Delta_{1}[\tfrac7{30}] \ 
   A^\perp_{m}\,, 
   \qquad\quad
   \Delta_1[r]	_{mn}
   	 = -g_{mn} \cd^2 + r  R g_{mn} \,.
\end{equation}
As mentioned around \eqref{laa}, there is another  Weyl- and gauge-invariant term that could, in principle, appear in the action (when written in the form \eqref{dcd}), namely   \(C_{mnrs} F^{mn} F^{rs} \). In fact, this term is present in the \((1,0)\) conformal supergravity Lagrangians, but it remarkably disappears in the \((2,0)\) combination. Such a contribution  prevents  a factorisation, e.g.\ because it produces 2-derivative terms \( R^{manc} A^\perp_m  \cd_a \cd_c A^\perp_n   \), cf. \cite{Casarin:2023ifl}. 

\subsection{Summary}
Given the complexity of the Lagrangian for conformal supergravity, the expression of the Lagrangian for quadratic fluctuations of physical fields on an Einstein background is by itself a remarkable result. Our expressions come from combining  explicit evaluations with educated guess. Let us present the results of this section in one place and summarise explicitly the evidence in favour of our Ansatz. The Lagrangians for the physical (transverse traceless) fields are 
\bsubeq\label{summ}
\begin{align}
    L_h &= 
        h_{mn}^\perp \ \Delta_2 [\tfrac{2}{15}] \ \Delta_2 [\tfrac{1}{5}]\  \Delta_2 [0]\	h_{mn}^\perp
   \,,\\
    L_T &= V^\perp_{mn}  \ \Delta_{2\text f}[\tfrac 4 {15}]  \ \Delta_{2\text f}[\tfrac 4 {15}] \
\Delta_{2\text f}[\tfrac  1 {5}] \ V^{\perp mn}
   \,,\\
    L_\psi &= \bar \psi^\perp_m  \ \Delta_{\frac32}[\tfrac 7 {60}] \ \Delta_{\frac32}[\tfrac{13}{60}] \ {\rm i} \slashed \cd \  \psi^{\perp m} 
   \,,\\
    L_A &= A^{\perp m}
  	\Delta_{1}[\tfrac16] \ 
  	  	\Delta_{1}[\tfrac7{30}] \ 
   A^\perp_{m}
   \,,\\
    L_\chi &= \bar \chi
\, \Delta_{\frac12}[\tfrac{13}{60}]	\, {\rm i} \slashed \cd \,  \chi
   \,,\\
    L_D &= D \, \Delta_{0}[\tfrac15]\,  D
\end{align}
\esubeq
where the operators are defined above and collected in \eqref{opeapp}. \(V_{mn}\) is the potential for \(T_{mnr}\) as in \eqref{dfn}.

We have explicitly evaluated in full detail the actions for graviton \(h\), vector \(A\), spinor \(\chi\) and scalar \(D\), so that the associated Lagrangians displayed above are exact. The graviton case, also studied in \cite{Aros:2019tjw} with the same results, is particularly insightful, in that the index structure of the field is rich enough to allow for a coupling with the Riemann tensor. The particular coupling we find is such that the action of the operator preserves transversality, cf.\ \eqref{fey}. We observe the same feature also in the 4d conformal supergravity case \cite{Fradkin:1985am} and in the 4d conformal higher spin context \cite{Tseytlin:2013jya}. It is also natural for 6d conformal higher spin theory \cite{Tseytlin:2013fca}. In fact, this particular curvature coupling allows one to decouple the physical from the gauge modes, cf.\ Appendix~\ref{app:detp}.

The factorisation of the operators associated to the quadratic flucutations on particular backgrounds has been observed to extend to all the fields in the supermultiplet in 4d conformal supergravity  \cite{Fradkin:1985am,Tseytlin:2013jya} (on Einstein background) and also for the 4-derivative flat-space 6d \(F \Box F\) theory \cite{Casarin:2019aqw} (on a \(D^m F_{mn}=0\) background). It was also instrumental in the already mentioned conformal-higher-spin analyses \cite{Tseytlin:2013jya,Tseytlin:2013fca}.
That the factorisation of the operators works also for the gravitino and the three-form (or rather its potential \(V\)) seems thus plausible and is our working assumption, which we plan to prove explicitly in future work.

The operators that we propose for \(V_{mn}^\perp\) and \(\psi_m^\perp\) involve nontrivial couplings with the Riemann tensor. In analogy with the cases just discussed we assume that these couplings preserve transversality. This uniquely fixes the curvature couplings as given in \eqref{eff2}. As a bonus, we observe that the gravitino operator also maps physical fields into physical fields since it preserves gamma-tracelessness as well, cf.\ \eqref{plj}. Gamma-tracelessness and transversality are also preserved  by the standard Dirac operator, cf.\ \eqref{plj}.  For these reasons our Ansatz seem rather natural and the only remaining unknowns are the couplings with the Ricci scalar. These are fixed by comparing with the explicit evaluation of the action on the sphere \(S^6\), where factorisation is automatic.

To check the solidity of the curvature coupling in our Ansatz we inspected the actions. Working on a Ricci-flat background with parallel curvature, we observe that the gravitino Lagrangian \eqref{grp} displays exactly the 3-derivative coupling expected by our Ansatz.
For the three-form, we performed a tedious but thorough analysis in Appendix~\ref{app:T3}. We studied in detail the structure of a generic Weyl invariant action. A cross-check of the analyses is that we confirm the value of the action on the sphere. We then focus on the Ricci-flat parallel curvature case and show that a factorised action for the 2-form transverse potential is possible and \emph{necessarily} of the form given in our Ansatz \eqref{efg}.

Furthermore, following the discussion about the anomalies of maximal CSG and its \((1,0)\) truncation given in the Introduction, our results appear nicely consistent with very different analyses in the literature. This in turn provides additional indirect evidence for the expressions \eqref{summ}. However, given the complexity of the Lagrangian and the potential future applications we believe a more complete analysis is deserved and worthy of a separate publication.

\section{Partition functions and conformal anomalies}\label{sec:partf}
In this section we use the quadratic operators to construct the partition functions and  evaluate the conformal anomalies.
We also verify the number of dynamical degrees of freedom \(\nu\)  propagated by each field.\footnote{%
In this context the number of on-shell degrees of freedom is given by the partition function in flat background as \(Z[\text{flat}] = [\det(-\partial^2)]^{-\nu/2} \), so that \(\nu=1\) for a real 2-derivative scalar. Notice that \(\nu>0\) for bosonic and \(\nu <0\) for fermionic fields.
}

The technical details concerning the functional determinants and the heat-kernel coefficients are relegated to Appendices~\ref{app:HK} and \ref{app:jac}.

\subsection{Scalar, spinor and vector}
The scalar field with action \eqref{dcs} has a partition function of the form 
\begin{equation}\label{gfd}
Z_D = \int \DD{D} e^{-S_D} =   \textstyle{ \Big[ \det_0 \Delta_0 [\tfrac15] \Big]^{-\frac12}   }\,,
\end{equation}
which clearly propagates one degree of freedom and immediately gives the conformal anomaly
\begin{equation}\label{rwr}
\mathscr A_D 
= b_6(\Delta_0[\tfrac15]) 
=  \frac{1}{7!}\left(   
	- \frac5{72} \mathbb E_6 
	- \frac {28}3 I_1
	+ \frac 53 I_2
	+ 2 I_3\,,
   \right)
\end{equation}
as quoted in the introduction. 

The $\chi$ spinor can be treated in full analogy to obtain
\begin{equation}\label{ggd}
Z_\chi = \int \DD{D} e^{-S_\chi} =   \textstyle{
 \Big[ \det_{\frac12} \Delta_{\frac12} [\tfrac{13}{60}] \Big]^{\frac12}   
  \Big[ \det_{\frac12}  {\rm i} \slashed \cd \Big]^{\frac12}    
 }\,,
\end{equation}
where the exponent \(\frac12\) comes from using Majorana spinors in   Dirac gamma-matrix notation.
 To evaluate the determinant of the Dirac operator we use the standard procedure of relating it to its square, \(-\slashed \cd^2 =   \Delta_{\frac12}[\frac14]\), so that we obtain
\begin{equation}\label{gqd}
Z_\chi  
 =
 \textstyle{
  \Big[ \det_{\frac12} \Delta_{\frac12} [\tfrac{13}{60}] \Big]^{\frac12} 
    \Big[ \det_{\frac12}  \Delta_{\frac12}[\tfrac14]   \Big]^{\frac14}    
 }\,.
\end{equation}
From \eqref{gqd}, we can also verify the number of dynamical degrees of freedom \(\nu\) of a \(3\)-derivative Majorana or Weyl spinor, 
\begin{equation}\label{gid}
Z_\chi[\text{flat}] =     \textstyle{
 \Big[ \det_{\frac12} [-\mathbb I_8 \partial^2  ]\Big]^{\frac32}    = 
 \Big[\det_0 [-\partial^2]\Big]^{12}
 }\,,
 \qquad
 \nu(\chi)=-12,
\end{equation}
as reported in the table in the introduction.

From \eqref{gqd} we immediately obtain the anomaly via the evaluation of the heat-kernel coefficients
\begin{equation}\label{rwo}
\mathscr A_\chi 
= - b_6(\Delta_{\frac12} [\tfrac{13}{60}])  - \frac 12 b_6  (    \Delta_{\frac12}[\tfrac14]  ) 
=  \frac{1}{7!}\left(   
	\frac{39}{16} \mathbb E_6 
	+ \frac {896}3 I_1
	+ \frac {220}3 I_2
	- 24 I_3
   \right)\,,
\end{equation}
as mentioned in the introduction and originally computed by \cite{Beccaria:2017dmw,Casarin:2021fgd}.

The vector field requires some additional care due to gauge invariance, which we deal with by restricting to  transverse fields. In turn, this entails additional Jacobian factors in the functional integral and determinants restricted to transverse modes. These aspects are reviewed in Appendix~\ref{app:jac}. In particular, in decomposing \eqref{dcd} in terms of the transverse component and the longitudinal (unphysical) mode as in \eqref{jkf1}, the functional integral picks up the Jacobian factor \(J_1\)  explicitly given in \eqref{jac1}. 
As a result, the partition function reads
\begin{equation}\label{fpo}
Z_A = \int \DD{A_1} e^{- S_A} 
= J_1 \Big[ \textstyle \det_{1\perp} \Delta_1 [\tfrac16] \  \textstyle \det_{1\perp} \Delta_1 [\tfrac7{30}] \Big]^{-\frac12}
\,,
\end{equation}
where the determinant indicates explicitly that it is restricted to transverse modes. 

The transversality restriction can be lifted as described in Appendix~\ref{app:detp}, in particular using \eqref{pr1}.
As a result, we arrive at the partition function\footnote{
The standard Faddeev-Popov procedure with gauge  \(\mathscr G = \mathscr G[A]\) produces  (see e.g.\ \cite{Avramidi:2000bm,Casarin:2021fgd,Casarin:2023ifl})  the family of partition functions (on Einstein background geometry) 
\begin{gather*} 
Z_{A,\text{FP}} = 
{\textstyle{ 
	\big[   \det_0 \mathscr G'  \big]  
		\big[   \det_0  \Delta_{\mathscr G} \,  \big]  ^{\frac12}
	}}
	\int \DD{A} e^{- S_{A,\text {FP}}}\,,
		\qquad
S_{A,\text {FP}}= \int	\!	  A^m \Delta_A A_m 
					+ (\cd\cdot A) (-\cd^2)(\cd \cdot  A)
					- \mathscr G  \, \Delta_{\mathscr G}  \mathscr G
	\,,
\\
\mathscr G' = -\cd_m \frac{ \delta \mathscr G[A ]}{\delta  A_m } 
\,,
\qquad 
	 \Delta_{A} A_{m} 
	 =
	  \cd^4 A_m 
	  - \frac25 R \,\cd^2 \! A_m 
	  + \frac1{15} R \, \cd_m \! \cd^r \! A_r
	  + \frac 7 {180}  R^2A_m
\,,
\end{gather*}
where \(\det_0\mathscr G'\) is the Faddeev-Popov determinant and \(\Delta_\mathscr G \) is an \(A\)-independent Gaussian weight with some degree of arbitrariness.
For the Feynman-Landau-DeWitt-type gauge \(\mathscr G = \cd^m A_m\) one has \(\mathscr G' = \Delta_0[0]\), but there are still several possible choices for \(\Delta_\mathscr G = -\cd^2 + \ldots\). With the simplest \(\Delta_\mathscr G = \Delta_0[0]\) one gets 
\begin{equation*}
Z_{A,\text{FP}} = 
\textstyle{ 
	\big[   \det_0 \Delta_{0}[   0  ]   \big]^\frac32 
	\big[	\det_{1} \Delta_{A} \big]^{-\frac12} 
	} 
\,,
\end{equation*}
which is the one used in \cite{Casarin:2023ifl}. The choice \( \Delta_\mathscr G = \Delta_0[\tfrac1{15}]\) produces \eqref{fpu}.
In either case the total \(b_6\) coefficient  governing the one loop divergences and the conformal anomaly  (cf.\ \eqref{iab}) is the same, since it is physical. }
\begin{equation}\label{fpu}
Z_{A} = 
\left[
\frac{ 
	\big[\det_0 \Delta_{0}[0]\big]^2 
	\det_0 \Delta_{0}[\frac1{15}] 
}{ 
	\det_{1} \Delta_{1}[\frac16] \
	\det_{1} \Delta_{1}[\frac7{30}] 
	} \right]^{\frac12}\,,
\end{equation}
where now all determinants are unconstrained and explicit. Specialising \eqref{fpu} to flat space we can verify the number of dynamical degrees of freedom,
\begin{equation}\label{ifh}
Z_A[\text{flat}]   =  
 				\Big[ \textstyle \det _{1} [ - \mathbb  I_{6} \partial^2 ]  \Big]^{-1}
				\Big[ \textstyle \det _{0} [ -  \partial^2 ]  \Big]^{\frac32} 
 = 
 				\Big[ \textstyle \det _{0} [ -  \partial^2 ]  \Big]^{-\frac92} 
 \,,
 \qquad
 \nu(A) = 9
 \,, 
\end{equation}
matching the entry in Table~\ref{tab} in the introduction.

Then from \eqref{fpu} one can obtain the conformal anomaly via the heat-kernel coefficients,
\begin{equation}\label{ioe}
\begin{aligned}
 \mathscr A_A  &=
     b_6( \Delta_1[ \tfrac16 ] ) 
     +    b_6( \Delta_1[ \tfrac7{30} ] )
      - 2    b_6( \Delta_0  [  0   ] ) 
        -    b_6( \Delta_0  [  \tfrac1{15}   ] ) 
 \\
 &= 
\frac{275}{8 \cdot 7!} \mathbb E_6 
+ \frac{97}{180} I_1
+ \frac{911}{5040} I_2
- \frac{5}{168}  I_3
\,,
\end{aligned}
\end{equation}
which are the values quoted in the introduction.
In relation to the comments around \eqref{laa} and \eqref{dcd2}, if we include the contribution \(C_{mnrs} F^{mn}F^{rs}\) in the analysis, the \(c_i\) coefficients get modified according to its coupling, see \cite{Casarin:2023ifl}.

\subsection{Graviton}
The partition function corresponding to the classical action \eqref{oaa} reads
\begin{equation}\label{ffe}
Z_h = \int \DD {h_2} 
		e^{-S_h  }
	= J_2 \, J_1^{-1}
	\left[
			\textstyle \det_{2\perp}\Delta[\tfrac2{15} ]  \ 
			\textstyle \det_{2\perp} \Delta[\tfrac15 ]  \ 
			\textstyle \det_{2\perp}   \Delta[0]
	\right]^{-\frac12}
	\,,
\end{equation}
where we have introduced the Jacobian  \(J_2\) associated to the decomposition into irreducible representations given in \eqref{jac2}. 
The factor \(J_1^{-1}\) is due to the fact that the partition function is normalised to un-contrained fields.

We then relax the transversality condition following the discussion in Appendix~\ref{app:detp} and obtain
\begin{equation}\label{kif}
Z_h =
\left[
\frac{
	\det_1 \Delta_{1}[-\frac16]
	\det_1 \Delta_{1}[\frac1{30}]
	\det_1 \Delta_{1}[-\frac1{30}]
	\det_1 \Delta_{1}[-\frac16]
	\det_0 \Delta_{0}[-\frac15]
}{
	\det_0 \Delta_{0}[-\frac13]
	\det_{2^0} \Delta_{2}[\frac2{15}]
	\det_{2^0} \Delta_{2}[\frac15]
	\det_{2^0} \Delta_{2}[0] 
	} \right]^{\frac12}\,,
\end{equation}
where the superscript \(0\) in \(2^0\) indicates that the determinant is taken on \emph{traceless} symmetric rank-2 tensors.
We can then verify that \eqref{kif} reproduces the expected number of degrees of freedom \(\nu\),
\begin{equation}\label{pth} 
Z_h[\text{flat}]   = 
 				\Big[ \textstyle \det _{2^0} [ - \mathbb  I_{20} \partial^2 ]  \Big]^{- \frac32}
 				\Big[ \textstyle \det _{1} [ - \mathbb  I_{6} \partial^2 ]  \Big]^{2}
 = 
 				\Big[ \textstyle \det _{0} [ -  \partial^2 ]  \Big]^{-18} 
 \,,
 \qquad
 \nu(h) = 36
 \,, 
\end{equation}
where \( 20 \) is the dimension of the space of symmetric traceless tensors (cf.\ \eqref{p20}) and the number of degrees of freedom, \(36\), matches the one given in  Table~\ref{tab} in the introduction.

The anomaly can then be computed from \eqref{kif} by direct evaluation of the heat-kernel coefficients (see appendix~\ref{app:HK} for formulae) and the result is 
\begin{equation}\label{ioh}
\begin{aligned}
 	\mathscr  A_h   
& = 
		b_6 (    \Delta_{0}[-\tfrac13]   )
		+ b_6 (  \Delta_{2^0}[-\tfrac2{15}]   )
		+ b_6 (  \Delta_{2^0}[-\tfrac15]   )
		+ b_6 (  \Delta_{2^0}[-\tfrac13]   )
\\
&
\qquad		- b_6 (  \Delta_{1}[- \tfrac{1}6]  )
		- b_6 (  \Delta_{1}[\tfrac1{30}]   )
		- b_6 (  \Delta_{1}[-\tfrac{1}{30}]   )
		- b_6 (  \Delta_{1}[ -\tfrac{1}6]   )
		- b_6 (  \Delta_{0}[ -\tfrac{1}5]   )
\\ 
& = 
	\frac{601}{2016 } \mathbb E_6 
	+ \frac{1507}{45 } I_1
	+  \frac{635}{126 }   I_2
	-  \frac{1639}{420 } I_3\,,
\end{aligned}
\end{equation}
matching what was anticipated Table~\ref{tab} in the introduction.

\subsection{Gravitino}

We can evaluate the 1-loop partition function from the action \eqref{ife} as
\begin{equation}\label{pta}
\begin{aligned}
Z_\psi &  = 
		\int \DD {\psi_\frac32} 
			e^{ - S_\psi  }
=
		J_\frac{3}{2}  
		    \left[ \textstyle \det_{\frac32 \perp} {\rm i} \slashed   \cd 
		     \ 
   		     \textstyle \det_{\frac32 \perp} \Delta_\frac32[\tfrac 7 {60}]
   		     \
  		     \textstyle \det_{\frac32 \perp} \Delta_\frac32[\tfrac {13} {60} ]
  		     \right]^{\frac12}\,,
\end{aligned}
\end{equation}
where we have included the Jacobian factor \eqref{jac5}.  Then we relax the transverse gamma-traceless condition and obtain 
\begin{equation}\label{iue}
Z_\psi  
 =
 \left[
 \frac{
   \big[ \det_{\frac32} \Delta[\frac14]\big]^{\frac12}  \
 	\det_{\frac32} \Delta[\frac 7 {60}]   \
 	\det_{\frac32} \Delta[\frac {13} {60}] 
 }{ 
    \det_{\frac12} \Delta[\frac1{12} ]  \
 \big[ \det_{\frac12} \Delta[\frac1{20}]  \big]^{2} \
 \big[\det_{\frac12} \Delta[-\frac1{20}] \big]^{3} 
 	} \right]^{\frac12}\,.
\end{equation}
We can reproduce the expected number of degrees of freedom from   \eqref{pta} via
\begin{equation}\label{pt}
\begin{aligned} 
Z_\psi[\text{flat}]   = 
 				\left[ \textstyle \det _{\frac12} [ - \mathbb  I_{8} \partial^2 ]  \right]^{- 3}
 				\left[ {\det }_{\frac32} [ - \mathbb  I_{8\otimes6} \partial^2] \right]^{\frac54} 
 = 
 				\left[ \textstyle \det _{0} [ -  \partial^2 ]  \right]^{36} 
 \,,
 \qquad
 \nu(\psi) = -72\,,
\end{aligned}
\end{equation}
where we used that all operators in flat space are the scalar Laplacian times the identity in spinor and vector-spinor space (we are using Dirac gamma-matrix notation). As a result, \(Z_\psi\) describes  \(72\) fermionic degrees of freedom, which is the number quoted in Table~\ref{tab}.

Finally, the anomaly is given by the combination of \(b_6\) coefficients determined by \eqref{iue},
\begin{equation}\label{iot}
\begin{aligned}
\mathscr A_\psi  & = 
- \frac12 b_6(  \Delta_{\frac32} [ \tfrac 14] )
- b_6(  \Delta_{\frac32} [ \tfrac {7} {60}] )
- b_6(  \Delta_{\frac32} [ \tfrac {13} {60}] )   
\\
&  \qquad  
+  b_6(  \Delta_{\frac12} [  \tfrac1{12}  ] )  
+  2 b_6(  \Delta_{\frac12} [ \tfrac1{20}   ] )  
+  3 b_6(  \Delta_{\frac12} [ -\tfrac1{20}   ] )  
\\
& 
= - \frac{  4643  }{ 8  \cdot 7!} \mathbb E_6 
- \frac{644}{45} I_1
- \frac{110}{63}  I_2
+ \frac{379}{210}  I_3\,,
\end{aligned}
\end{equation}
which are the values given in Table~\ref{tab}.

\subsection{3-form}

The construction of the partition function for the 3-form requires some specific discussion. The Jacobian associated to the field redefinition \eqref{dfn} cannot be computed with the standard techniques employed for the other fields and discussed in Appendix~\ref{app:jac} due to the self-duality constraint, which forces \(T_{mnr} T^{mnr} =0\) algebraically. In \eqref{jac4} we have computed the Jacobian for the transformation \(T_{mnr}\to (V_{mn} , W_{mn})\) given in \eqref{jkf4}, namely without imposing self-duality; in this case  the two components \(V\) and \(W\) contribute independently and with the same contribution, cf.~\eqref{fds}.  Since \eqref{dfn} is the same type of transformation but with \(W\) given in terms of \(V\),  we postulate that the associated Jacobian is the square root \(\sqrt{J_{3\text f}}\) of \eqref{jac4}.\footnote{
We emphasize that our treatment of the 3-form field is somewhat speculative and stems from the idea that the two chiralities should contribute symmetrically to a parity-even observable such as the conformal anomaly. Given the consistency of the results that we find, we believe our results to be correct.}  As a result,
\begin{equation}\label{guf}
Z_T = \int \DD{T_{3\text f}} e^{- S_T }
= \sqrt{J_{3\text f}} \int \DD{V^\perp_{2\text f}}e^{-S_T}
= 
\textstyle{ 
		\Big[ \det_{2 \text f \perp} \Delta[\tfrac4{15} ]   \Big]^{-\frac12}
		 \Big[ \det_{2 \text f \perp} \Delta[\tfrac1{5} ]   \Big]^{-\frac12}	
		 }
\,,
\end{equation}
where  one of the determinants from \eqref{ife2},\eqref{alvs} is cancelled by the Jacobian factor.

Eliminating the transversality constraint with the techniques described in Appendix~\ref{app:detp} we then obtain
\begin{equation}\label{fke}
Z_T = 
\left[
\frac{
	\det_1 \Delta_{1}[   \tfrac16   ] \ 
	\det_1 \Delta_{1}[  \tfrac 1 {10}  ]  
}{
	\det_{2\text f } \Delta_{2}[\frac4{15}] \
	\det_{2\text f} \Delta_{2}[\frac15]  \
		\det_0 \Delta_{0}[  - \tfrac 1 {15}     ] \
		\det_0 \Delta_{0}[ 0   ] 
	} \right]^{\frac12}\,,
\end{equation}
where now all determinants are explicit and unconstrained.
As a check of \eqref{fke} we count the number \(\nu\) of degrees of freedom by computing
\begin{equation}\label{fkde} 
 Z_T[\text{flat}] = 
 \textstyle{
\Big[ \det_{2 \text f}[-\mathbb I_{15} \partial^2] \Big]^{ -1 }
\Big[ \det_1[-\mathbb I_6 \partial^2] \Big]^{ 1 }
\Big[ \det_0[- \partial^2] \Big]^{ -1 }
=\Big[\det_0 - \partial^2\Big]^{   -10  }
}
\,,
\qquad
\nu(T) = 20\,,
\end{equation}
where \(15 \) is the dimension of the space of 2-forms as in \(\eqref{p2f}\) and we obtain the   result \(\nu = 20\) anticipated in the introduction.

All operators in \eqref{fke} are second order differential operators and the anomaly can thus be evaluated with the usual techniques. We obtain
\begin{equation}\label{puo} 
\begin{aligned}
\mathscr A_T & = 
  b_6    (\Delta_{2\text f }[\tfrac4{15} ]  )
+    b_6    ( \Delta_{2\text f }[\tfrac1{5} ]  )
+    b_6    ( \Delta_{0}[-\tfrac1{15}] )
+    b_6    ( \Delta_{0}[0])
\\
&  \qquad 
- b_6 (\Delta_1 [\tfrac 1 {6}])
-b_6(\Delta_1[\tfrac 1 {10}])
- a _\text h  \mathbb E_6
\\
&=- \frac{1}{7!}   \Big(\frac{779 }{9} + a _\text h\Big) \mathbb E_6-\frac{11 }{27} I_1
-\frac{11}{42}  I_2
- \frac{5}{252}  I_3
\,.
\end{aligned}
\end{equation}
We have symbolically included the contribution of the harmonic forms in the coefficient \(a_\text h\).

The \(a\) coefficient for the 3-form was computed in \cite{Beccaria:2015uta} via AdS$_7$/CFT$_6$ correspondence without invoking the Hodge decomposition.  In particular, their calculation also captures the contribution of the harmonic forms that we spuriously introduce with the change of variables \eqref{dfn}. Their result  \(a=-\frac{166}{9 \cdot 7!  }\) (which corresponds to    \(a_\text h= - 105\)) is the one used in the introduction. The \(c_i\) coefficients are an original result of this paper.

\subsection*{Acknowledgments}
We are grateful to Arkady Tseytlin for several discussions, enlightening comments, and for reading the draft of this paper. 
It is also a pleasure to acknowledge discussions with
 Fiorenzo Bastianelli, Franz Ciceri, Gregory Gold, Saurish Khandelwal, Sergei Kuzenko, Olaf Lechtenfeld, Ruben Minasian, and Stefan Theisen.
 LC thanks the Department of Mathematical Sciences of
Durham University for hospitality during the final stages of this project.
CK and GT-M acknowledge the kind hospitality and financial support extended to them at the MATRIX Program ``New Deformations of Quantum Field and Gravity Theories,'' between 22 Jan and 2 Feb 2024.
CK is supported by a postgraduate scholarship at the University of Queensland.
GT-M has been supported by the Australian Research Council (ARC) Future Fellowship FT180100353, ARC Discovery Project DP240101409, and a Capacity Building Package of the University of Queensland.
Many of the calculations presented in this paper have been carried out with the aid of the software Wolfram Mathematica and the xAct package suite
\cite{xAct,Nutma:2013zea,Martin-Garcia:2008ysv}
as well as the software \emph{Cadabra} \cite{Peeters:2007,Peeters:2018no1,Peeters:2018no2}.

\appendix

\section{Notation and conventions}
\label{app:not}

In this paper we use two different sets of conventions to connect with the notation used in the two different subfields we consider in this paper.  
For the  (classical) conformal supergravity derivation of the quadratic Lagrangians, we follow the notation established in the respective literature \cite{Butter:2016qkx,Butter:2017jqu}. In particular, the signature is Lorentzian. This applies only to Section~\ref{sect:confsugra} and Appendix~\ref{app:l10}.  
In the introduction, Sections~\ref{sect:quadrlagra} and \ref{sec:partf} as well as appendices~\ref{app:HK}, \ref{app:jac} and \ref{app:T3}, which are  devoted to the evaluation of the  anomaly, we adhere to the QFT notation of e.g.\ \cite{Beccaria:2015uta,Beccaria:2015ypa,Beccaria:2017dmw,Beccaria:2017lcz,Casarin:2023ifl}. In particular, it is customary to use  Euclidean signature to have a formally convergent functional integral.
The following two subsections discuss the details of the notation employed in the paper.

\subsection{Lorentzian Supergravity notation}
\label{app:notSugra}

We give a brief overview of our Lorentzian notation. For further information about the Lorentzian supergravity conventions we refer to Appendix A of \cite{Butter:2016qkx}. Our index notation is as follows 
\begin{table}[H]
\centering
\begin{tabular}{cccccccc}
{indices} & {usage} & ranges \\[0.5em]
\(m,n,r,s,p,q\) & curved vector & 0,1,2,3,4,5 \\[0.25em]
\(a,b,c,d,e,f,g,h\) & flat vector & 0,1,2,3,4,5  \\[0.25em]
\(\a,\b,\g,\d,\epsilon,\varepsilon,\zeta,\eta,\theta\)    & Weyl spinor \(\mathrm{SU}^*(4)\)  & 1,2,3,4 \\[0.25em]
\(i,j,k,l,i',j',k',l'\) & \((1,0)\) R-symmetry \(\mathrm{SU}(2)\)  & 1,2 \\[0.25em]
\(\ui,\uj,\uk,\ul\) & \((2,0)\) R-symmetry \(\mathrm{USp}(4)\)  & 1,2,3,4 \\[0.75em]
\end{tabular}
\label{tab3}
\caption{Index notations used in \((1,0)\) and \((2,0)\) conformal supergravity. The Weyl spinor representations of \(\mathrm{Spin}(1,5)\simeq \mathrm{SU}^*(4)\) are dual to one another, so we need only one set of spinor indices.
}
\end{table}

In an orthonormal frame given by the vielbein \(e_m{}^a\), our metric is \(\eta_{ab} =\mathrm{diag}(-1,1,1,1,1,1)\) and our Lorentzian Levi-Civita symbols are \(\varepsilon_{012345} = -\varepsilon^{012345} = 1\). Our \(\mathrm{SU}(2)\) invariant tensors are normalised as \(\varepsilon^{12}=-\varepsilon_{12} =1\) and analogously for the primed case. We take the standard definitions for the other Levi-Civita symbols, i.e.\ \(\varepsilon_{1234} = \varepsilon^{1234}=1\) for \(\mathrm{SU}^*(4)\) and \(\mathrm{USp}(4)\). We take our symplectic form for \(\mathrm{USp}(4)\) as
\begin{equation}
    \Omega^{\underline{ij}} =
    \begin{pmatrix}
        \varepsilon^{ij} & 0 \\
        0 & \varepsilon^{i'j'}
    \end{pmatrix}
    \,,\qquad
    \Omega_{\underline{ij}} =
    \begin{pmatrix}
        \varepsilon_{ij} & 0 \\
        0 & \varepsilon_{i'j'}
    \end{pmatrix}
    \,.
\end{equation}

Our Levi-Civita covariant derivative is denoted \(\mathcal{D} = d - \omega(e)\) where the 1-form \(\omega(e)= \frac{1}{2}\omega(e)^{bc} M_{bc}\), with \(M_{ab}\) the Lorentz generators, is related to the usual Christoffel symbols by the vielbein-generated \(\mathrm{GL}(6)\)  gauge transformation
\bsubeq
\begin{align}
\omega(e)_{abc}&=-\frac{1}{2}(\mathcal{C}_{abc}+\mathcal{C}_{cab}-\mathcal{C}_{bca})
~,~~~~~~
C_{mn}{}^a:=2\partial_{[m}e_{n]}{}^a
~,\\
\Gamma^n{}_{pm} &= e_a{}^n \Big( - \omega(e)_p{}^a{}_b \Big) e^b{}_m  + e_a{}^n \partial_p e^a{}_m
= \frac{1}{2} g^{nq} \Big( \partial_m g_{pq} + \partial_p g_{mq} - \partial_q g_{pm} \Big)
\,.
\end{align}
\esubeq
Explicitly, one can define the Riemann curvature in our Lorentzian conventions by
\begin{equation}
[\mathcal{D}_a, \mathcal{D}_b] V_c = - \frac{1}{2} \mathcal{R}_{ab}{}^{ef} M_{ef} V_c = - \mathcal{R}_{abcd} V^d
\,.
\end{equation}
Notice that the Riemann, Ricci, scalar, Weyl, Schouten curvature tensors are the minus those in the Euclidean conventions described below. 
Even though we will comment more in Appendix \ref{app:l10}, it is important to note here that the \((1,0)\) superconformal geometry described by the covariant derivative $\hat{\nabla}_a$ of equation \eqref{sccov} introduces a more general Lorentz connection
\begin{subequations}\label{Lconn}
\begin{align}
\hat{\omega}_{a}{}_{bc}
	&=
\omega(e,b)_{a}{}_{bc}
-\frac{\mathrm{i}}{4}\psi_b{}^k \g_a \psi_c{}_k
-\frac{\mathrm{i}}{2}\psi_a{}^k \g_{[b} \psi_{c]}{}_k
\,, \\
\omega(e,b)_{a}{}_{bc} 
    &=
\omega(e)_{abc}
-2\eta_{a[b} b_{c]}
\,,
\end{align}
\end{subequations}
where \(b_a\) is the dilatation connection and \(\psi_{a i}{}^{\a}\) is the gravitino. The bosonic connection \(\omega(e,b)\) of \eqref{Lconn} is a feature of conformal symmetry only, so it holds equally as well in the \((2,0)\) case, but the gravitino contributions will be different. The associated Lorentz curvatures are defined by Cartan's structure equations as
\begin{subequations}\label{Lcurv}
\begin{align}
\mathcal{R}_{ab}{}^{cd}(\hat{\omega})
	&:=
	e_a{}^me_b{}^n\Big(
	2\partial_{[m}\hat{\omega}_{n]}{}^{cd}
	-2\hat{\omega}_{[m}{}^{ce} \hat{\omega}_{n]}{}_e{}^d
	\Big) 
\,, \\
\mathcal{R}_{a}{}^{b}(\hat{\omega})
	&:= 
	\mathcal{R}_{ac}{}^{b c}(\hat{\omega}) 
\,, \\ 
\mathcal{R}(\hat{\omega}) 
	&:= \mathcal{R}_a{}^a(\hat{\omega})
\,, \\
C_{ab}{}^{cd}(\hat{\omega})
    &:= 
\mathcal{R}_{ab}{}^{cd}(\hat{\omega})
- \d_{[a}^{[c} \mathcal{R}_{b]}{}^{d]}(\hat{\omega}) + \frac{1}{10} \d_{[a}^{[c} \d_{b]}^{d]} \mathcal{R}(\hat{\omega})
\,,
\end{align}
\end{subequations}
where, in the bosonic case, one can replace the \(\hat{\omega}\) in the formulae above with \(\omega(e,b)\). A key point is that, since the dilatation curvature vanishes and \(C_{ab}{}^{cd}(\omega(e,b))\) is a conformal primary, then \(C_{ab}{}^{cd}(\omega(e,b)) = C_{ab}{}^{cd}(\omega(e))\) is independent of \(b_a\). That is, it is the usual Weyl tensor of (pseudo-)Riemannian geometry and this justifies why we always denote it as \(C_{ab}{}^{cd}\). The same does not hold for the non-primary \(\mathcal{R}_{ab}{}^{cd}(\omega(e,b))\) or its traces, but one can check that \(\mathcal{R}_{ab}{}^{cd}(\omega(e)) = \mathcal{R}_{ab}{}^{cd}\) is the usual Riemann tensor --- note we are in an orthonormal frame \(e_a{}^m\) here, not the coordinate frame. More details can be found in \cite{Butter:2016qkx,Butter:2017jqu}.

In regards to the \((2,0)\) \(\mathrm{USp}(4)\) and \((1,0)\) \(\mathrm{SU}(2)\) R-symmetry curvatures, we define them via Cartan's structure equations as
\begin{subequations}\label{Rcurv}
\begin{align}
\mathcal{R}_{ab}{}^{ij} &= e_a{}^m e_b{}^n \Big( 2\partial_{[m} \mathcal{V}_{n]}{}^{ij} + 2 \mathcal{V}_{[m}{}^{k(i} \mathcal{V}_{n]k}{}^{j)} \Big)
\,, \\
R(V)_{ab}{}^{\underline{ij}} &= e_a{}^m e_b{}^n \Big( 2\partial_{[m} V_{n]}{}^{\underline{ij}} + 2 V_{[m}{}^{\uk(\ui} V_{n]\uk}{}^{\uj)} \Big)
\,,
\end{align}
\end{subequations}
where \(V_a{}^{\underline{ij}}\) and \(\mathcal{V}_a{}^{ij}\) are the \(\mathrm{USp}(4)\) and \(\mathrm{SU}(2)\) connections, respectively.

Our Clifford algebra conventions with Weyl gamma matrices \((\g_a)_{\a\b},(\tilde{\g}_a)^{\a\b}\) are
\begin{equation}\label{gammas}
(\gamma_a)_{\a\g}(\tilde\gamma_b)^{\g\b}+(\gamma_b)_{\a\g}(\tilde\gamma_a)^{\g\b} = -2 \eta_{ab} \d_\a^\b
\,,
\qquad\quad
(\tilde{\gamma}_a)^{\a\b} = \frac 12 \varepsilon^{\a \b\g\d}(\gamma_a)_{\g\d}
\,.
\end{equation}
Given any Weyl spinor \(\chi^{\a}\) we can enforce a Symplectic-Majorana-Weyl (SMW) condition by writing \(\chi^{\a i} = (\chi^\a, \bar{\chi}^\a )\) using \(\mathrm{SU}(2)\) and using the symplectic form \(\varepsilon_{ij}\) to write
\begin{equation}
\overline{(\chi^{\a i})} = \chi^\a{}_i = \varepsilon_{ij}\chi^{\a j}
\,,
\end{equation}
where \(\bar{\chi} = \chi^c\) is the charge conjugate, and charge conjugation satisfies \((\chi^c)^c = -\chi\) in \(d=1+5\) dimensions. The same principle holds with \(\mathrm{USp}(4)\) but now
\(\chi^{\a \ui}~=~(\chi^{\a 1}, \bar{\chi}^{\a 1}, \chi^{\a 3}, \bar{\chi}^{\a 3} )\). More details on charge conjugation and gamma matrices in the Weyl representation can be found in Appendix A of \cite{Butter:2016qkx}.  

As we have already mentioned in Section~\ref{sect:confsugra}, the fundamental isospinor representations of SU(2) and USp(4) are complex and act on \(\mathbb{C}^2\) \((v^i)\) and \(\mathbb{C}^4\) \((v^{\ui})\) respectively. Hence, like fermions, bosons also require reality constraints, e.g.\ for \(B^{ij}\)
\begin{equation}
\overline{(B^{ij})} = B_{ij} = \varepsilon_{ik} \varepsilon_{jl} B^{kl}
\,,
\end{equation}
where the same principle holds in the USp(4) case. A key point is that bosons may only have an even number of (SU(2) or USp(4)) indices, whilst fermions must have an odd number of indices. This is a result of complex conjugation squaring to unity for bosons, but charge conjugation squaring to minus unity for fermions. For bosons, one can interpret this as real subrepresentations only existing for \((\mathbb{C}^2)^{\otimes 2n}\) for \(n\geq 0\) for SU(2), and analogously for USp(4). In more precise terms, for bosons, the complex representations are of real type and, for fermions, the complex representations are of quaternionic type.

\subsection{Euclidean  QFT notation}
\label{app:notQFT}

We work in 6 Euclidean dimensions and indicate spacetime indices with Latin lowercase letters. There is no distinction between flat and curved indices as we work always in the latter case. However, we will give many  formulae for general \(d\)
for completeness.

The metric is \(g_{mn}\) and we   write the covariant derivative with Levi-Civita (spin) connection as
\(\cd  \). 
The curvature tensors are 
\begin{equation}\label{opa}
\begin{gathered}
\relax
[\cd_m,\cd_n]V_a = R_{mnac}V^c
\,,
\qquad 
R_{mn} = R\indices{^a_{man}}\,,
\qquad 
R = R^m_m\,,
\end{gathered}
\end{equation}
where \(V^m\) is a   vector.
The Weyl tensor reads
\begin{equation}\label{ugya}
C_{mnrs}
=
 R_{mnrs}
 + \frac 2 {d-2}   g_{r [n }   R_{   m ]s}   
- \frac 2 {d-2}   g_{ s [n }  R _{  m ]r}   
 + \frac{2}{(d-1)(d-2)} g_{m [ r} g_{s ] n} R \,.
\end{equation}
We stress that the notation here differs from the supergravity one discussed in Appendix~\ref{app:notSugra}  in that all curvatures get a minus sign.

The basis of the conformal anomaly that we adopt throughout the paper is
\begin{equation}\label{krt}
\begin{aligned}
\mathbb E_6  & 
= - \varepsilon_{mnrspq} \varepsilon_{abcdef}  R^{mnab}   R^{rscd}   R^{pqef}
=   -32 R\indices{^a^ c  _m _n} R\indices{^m^ n _r_ s} R\indices{^r^ s _a _c}	+\ldots\,,
  \\
I_1   &  = C\indices{^a_m_n^c} C\indices{^m_r_s^n} C\indices{^r_a_c^s}
\,,
\\
I_2  &   = 
C\indices{^a^ c  _m _n} 
C\indices{^m^ n _r_ s} 
C\indices{^r^ s _a _c}\,,
  \\
I_3   &  = 
C^{amnr}\Big[    
g_a^c\cd^2
+ 4 R_a^c 
- \frac{6}{5} g_a^c R  
\Big] C_{c mnr}\,.
\end{aligned}
\end{equation}

The  geometric background is always taken to be Einstein,
\begin{equation}\label{ekl}
R_{mn} = \frac1d R g_{mn} \,,
\qquad\qquad
\cd_m R =0\,,
\end{equation}
where the second relation follows from the Bianchi identity.
We  note the identity (on Einstein \eqref{ekl})
\begin{equation}\label{fkfr}
\cd^2 R_{mnrs} =
- R_{acrs} R\indices{^{ac}_{mn}} 
+ \frac{2 }{d} R R_{mnrs}
+ 2 R\indices{_{m}^{a}_{s}^{c}} R_{narc} 
- 2 R\indices{_{m}^{a}_{r}^{c}} R_{nasc}
\,,
\end{equation}
which follows from \(\cd^m \cd_{ [m } R_{ ns]ac } = 0\).
Furthermore, we use that given any   tensor \(V^{ac\cdots}\) such that \(V^{ac\,\cdots}=V^{[ac]\,\cdots}\), from the Bianchi identity it follows that
\begin{equation}\label{ids}
\begin{gathered}
2 V^{ac\,\cdots}R_{macn} = - V^{ac\,\cdots} R_{mnac} \,,
\qquad\qquad
2 V^{ac\,\cdots}C_{macn} = - V^{ac\,\cdots} C_{mnac} \,,
\end{gathered}
\end{equation}
which are used in several places without mentioning.

The metric on the sphere \(S^d\) is
\begin{equation}\label{s6}
    R_{mnrs } = \frac{R}{d(d-1)} 
    (g_{mr}g_{ns} - g_{ms} g_{nr})
    \,.
\end{equation}

Spinors are discussed in terms of Dirac gamma matrices \(\Gamma_m\) obeying the Clifford algebra
\begin{equation}\label{fes}
\{\Gamma_m,\Gamma_n\} = 2 g_{mn}\,,
\end{equation}
where \(\Gamma_m\)'s are \(8\times8\) complex matrices. In Euclidean signature they are all Hermitian and the Dirac conjugate is the Hermitian conjugate. We also have the standard relations
\begin{equation}
\begin{gathered}\label{qaa}
\Gamma^{mn} 
= \Gamma^{[m} \Gamma^{n]}
= \frac12 (  \Gamma^m \Gamma^n - \Gamma^n \Gamma^m )
\,,
\qquad
\qquad
[\Gamma^m, \Gamma^{nr}]
= - 4 \Gamma^{ [n }  g^{ r ] m  }\,,
\\
\Gamma^{mnr} 
= \Gamma^{  [ m   } \Gamma^n \Gamma^{ r ]  }
= 
\frac12
\{
\Gamma^m , \Gamma^{nr}
\}
\,.
\end{gathered}
\end{equation}
Together with the Bianchi identities  we obtain
\begin{equation}\label{qab}
\begin{aligned}
\Gamma^m \Gamma^{nr} R_{mnrs }
&= - 2 \Gamma^{ n }  R_{ns}
\\
& =  - \frac{2}{d}  \Gamma_{ s }  R  \qquad \text{(on Einstein)}\,,
\end{aligned}
\end{equation}
which follows from   \(\Gamma^{mnr}R_{mnrs}=0\) and the Lorentz algebra, and
similarly,
\begin{equation}
\begin{aligned}\label{qac}
\Gamma^m \Gamma^{nr} \cd_mR _{nrac } 
& = 2 \Gamma^{ n } ( \cd_c R_{an} - \cd_{a} R_{cn})
 \\
  & =0 \qquad \text{(on Einstein)\,.}
\end{aligned}
\end{equation}
We can connect with four-component Weyl matrices used in the supergravity section via the representation
\begin{equation}\label{fdg}
\Gamma_m = {\rm i} \begin{pmatrix}
0 & \tilde \gamma^\text E_m    \\
 \gamma^\text E_m   &    0 
\end{pmatrix} \,,
\end{equation}
where  
the additional \(\rm i\) switches the sign of the Clifford algebra to obtain \eqref{fes} and \(\gamma^\text E_m~=~(-{\rm i} \gamma_0,\dots, \gamma_6)\).

We collect here the definitions of the following differential operators used in the text:
\bsubeq\label{opeapp}
\begin{align}
\label{ope0}
\Delta_0[r]	
	& = -\cd^2 + r R
 \,,
\\
\label{ope1}
\Delta_1[r]	_{mn}
	& = -g_{mn} \cd^2 + r  R g_{mn} 
 \,,
\\
\label{ope2}
\Delta_2[r]_{mnrs}	
	& =  -g_{r(m} g_{n)s }\cd^2  -2 R_{mrns}  + r  R g_{r(m} g_{n)s }
 \,,
\\
\label{ope2f}
\Delta_{2 \text f}[r]	_{mnrs}
	& = -g_{r[m} g_{n]s }\cd^2  - C_{mnrs}  + r  R g_{r[m} g_{n]s }
 \,,
\\
\label{ope12}
\Delta_{\frac12}[r]	
	& = 
			- \mathbbm 1_8 \cd^2  +    r  \mathbbm 1_8  R  
   \,,
\\
\label{ope32}
\Delta_{\frac32}[r]_{mn}	
	& = 
			-g_{mn } \mathbbm 1_8 \cd^2  -\frac12 R_{mnrs}\Gamma^{rs} + r   \mathbbm 1_8 R g_{mn}
   \,.
\end{align}
\esubeq
Notice that the subscript \(s\) in \(\Delta_s\) indicates the field representation on which the operator acts and not, as frequently in the literature, the order of the differential operator. Our definition of \(\Delta_2\) is also slightly different from the one of \cite{Pang:2012rd,Beccaria:2017dmw}.
Furthermore, the operator \(\Delta_{2}\) naturally restricts to \emph{traceless} symmetric rank-2 tensors as 
\begin{equation}\label{fin}
P\indices{_{mn}^{rs}}  \ \Delta_2[r]\indices{_{rs} ^{ac}} \  P\indices{ _{ac} ^ {uv}  } 
\,,
\end{equation}
where \(P\indices{ _{ mn  }  ^{ac} } \) are the projectors on symmetric traceless tensors given explicitly in \eqref{p20}.

\section{\texorpdfstring{\((1,0)\) Conformal supergravity Lagrangians}{(1,0) Conformal supergravity Lagrangians}}
\label{app:l10}
Here we give the Lagrangians of the two \((1,0)\) conformal supergravity actions constructed in \cite{Butter:2016qkx,Butter:2017jqu} restricted to quadratic order in all fields except for the purely gravitational ones (vielbein/metric), which is given to all orders as to allow an arbitrary gravitational background. Moreover, we also give the bosonic \((1,0)\) Lagrangians to all orders in Appendix~\ref{app:bosonic10lag}.  Similar to Section~\ref{sect:confsugra}, we follow the notation of the original conformal supergravity papers \cite{Butter:2016qkx,Butter:2017jqu}.

Recall from Section~\ref{sect:confsugra} that, when only the \((1,0)\) Weyl multiplet is considered, the truncation of the \((2,0)\) Lagrangian is given by
\begin{equation}
\Lagr_{(2,0) \to (1,0)} = \Lagr_{C\Box C} + \frac{1}{2} \Lagr_{C^3} \,,
\end{equation}
where \(\Lagr_{(2,0) \to (1,0)}\) stands for the unique gravitational \((1,0)\) Lagrangian which permits an uplift to \((2,0)\) conformal supersymmetry.

The construction of both \((1,0)\) action principles relies  on finding a suitable closed super six-form \(J\) which transforms into an exact form under super-Weyl transformations \cite{Butter:2016qkx}. Closure, $d J=0$, ensures \(J\) transforms into an exact form under super-diffeomorphisms as well. The action is then given by the pullback
\begin{equation}\label{a10}
S = \int \dd{^6x} i^*J = \int \dd{^6x \sqrt{-g}} \, *J\,,
\qquad
*J := \frac{1}{6!} \varepsilon^{abcdef} J_{abcdef}
\,,
\end{equation}
where \(i : \mathcal{M}^6 \to \mathcal{M}^{6|8}\) is the inclusion map of the base manifold into the supermanifold and we have abused notation by treating \(J\) as \(i^*J\) when applying the Hodge star. The Lagrangians \(*J\) are determined in terms of a primary superfield and its descendants, and their expansions in terms of the standard Weyl multiplet are taken from the supplementary file of \cite{Butter:2017jqu}.  We give the explicit  forms of both Lagrangians up to degauging the \(Q,S\)-connections from the covariant derivatives, as this makes things very messy. 
We give again the formulas from Section~\ref{sect:confsugra} for easy reference. The independent descendants of the $(1,0)$ super-Weyl tensor \(W^{\a\b}=W^{(\a\b)}\) take the form
\begin{subequations}
\begin{align}
X^{\a i} 
	&:= 
	-\frac{\mathrm{i}}{10}\hat{\nabla}_\b^i W^{\a\b}
\,,\quad
X_\g^k{}^{\a\b} 
	:= 
	-\frac{\mathrm{i}}{4}\hat{\nabla}_\g^k W^{\a\b} 
	- \d_\g^{(\a} X^{\b)k}
\,, \\
Y
	&:=
	\frac{1}{4}\hat{\nabla}_\g^k X_k^\g 
\,,\quad
Y_\a{}^{\b ij}
	:=
	-\frac{5}{2}\left( \hat{\nabla}_\a^{(i} X^{\b j)}
	-\frac{1}{4}\d_\a^\b \hat{\nabla}_\g^{(i}X^{\g j)} \right)
\,, \\
Y_{\a\b}{}^{\g\d}
	&:=
	\hat{\nabla}_{(\a}^kX_{\b) k}{}^{\g\d}
	-\frac{1}{3}\hat{\nabla}_\rho^k X_{(\a k}{}^{\rho(\g}\d_{\b)}^{\d)}
\,,
\end{align}
\end{subequations}
where \(\hat{\nabla}_\a^i\) is the superspace spinor covariant derivative, which projects onto the base manifold as \(Q_\a^i\)-supersymmetry generators.
The component fields obtained from projecting these onto the bosonic base manifold are defined as
\begin{subequations}
\begin{align}
T_{abc} 
	&:= -2 W_{abc} \vert 
\,,\quad
\chi^{\a i} 
	:= 
	\frac{15}{2}X^{\a i} \vert
\,,\quad
\mathcal{X}_{\a}^i{}^{\b\g}
	:=
	X_{\a}^i{}^{\b\g} \vert
\,, \\
D
	&:=
	\frac{15}{2}Y \vert
\,,\quad
\mathcal{Y}_\a{}^\b{}^{kl}
	:=
	Y_\a{}^\b{}^{kl} \vert 
\,,\quad
\mathcal{Y}_{\a\b}{}^{\g\d}
	:=
	Y_{\a\b}{}^{\g\d} \vert	 
\,,
\end{align}
\end{subequations}
where \(W_{abc} = \frac{1}{8} (\g_{abc})_{\a\b} W^{\a\b}\) and the vertical bar denotes setting all fermionic coordinates \(\theta_\a^i\) to zero. Formulae for \(Q,S\)-supersymmetry actions on these fields can be found in Appendix A.3 of \cite{Butter:2017jqu}. It is also important to note that only the field \(\mathcal{Y}_{\a \b}{}^{\g \d}\) has a pure gravitational part, given by the Weyl tensor, so we keep it to all orders.

Our hatted covariant derivatives \(\hat{\nabla}_a\) are in the traceless frame described in \cite{Butter:2017jqu}. There will be some unavoidable computations involving changing from the original frame of \cite{Butter:2016qkx} to the traceless frame in order to obtain the \(C\Box C\) Lagrangian, which will be discussed in detail below.
The expansions of the various covariant derivatives we use to deal with this change of frame, and inclusion of the fermions are
\begin{subequations}\label{cov2}
\begin{align}
\hat{\nabla}_{a} 
	&= 
e_{a} - \frac{1}{2} \psi_{a i}{}^{\alpha} Q_{\alpha}^{i} - \frac{1}{2} \hat{\omega}_{a}{}^{bc} M_{bc} - \mathcal{V}_{a}{}^{ij} J_{ij} - b_{a} \mathbb{D} - \hat{\mathfrak{f}}_{a}{}^{b} K_b - \frac{1}{2} \hat{\phi}_{a \alpha}{}^{i} S^{\alpha}_{i}
\,, \\
\nabla_a
	&= 
e_{a} - \frac{1}{2} \psi_{a i}{}^{\alpha} Q_{\alpha}^{i} - \frac{1}{2} \omega_{a}{}^{bc} M_{bc} - \mathcal{V}_{a}{}^{ij} J_{ij} - b_{a} \mathbb{D} - \mathfrak{f}_{a}{}^{b} K_b - \frac{1}{2} \phi_{a \alpha}{}^{i} S^{\alpha}_{i}
\,, \\
\hat{\mathcal{D}}_a 
	&:= e_a 
	- \frac{1}{2} \hat{\omega}_a{}^{bc} M_{bc}
	- b_a \mathbb{D} 
	- \mathcal{V}_a{}^{ij} J_{ij}
\,, \\
\mathcal{D}_a 
	&:= e_a 
	- \frac{1}{2} \omega(e)_a{}^{bc} M_{bc}
\,,
\end{align}
\end{subequations}
where \(e_a = e_a{}^m \partial_m\) is the vielbein vector field, \(\psi_{a i}{}^{\a}\) is the gravitino, and we have various connection forms for \(Q\)-supersymmetry \((Q_\a^i)\), Lorentz \((M_{ab})\), \(\mathrm{SU}(2)\) R-symmetry \((J_{ij})\), dilatation \((\mathbb{D})\), special conformal \((K_{a})\) and \(S\)-supersymmetry \((S_i^\a)\). The difference between the traceless and original frame's covariant derivatives is
\begin{equation}
\hat{\nabla}_a - \nabla_a = 
- W_a{}^{bc} M_{bc} 
+ \frac{3\mathrm{i}}{8} (\g_{a})_{\a \b} \chi^{\a j} S_j{}^\b 
-  \frac{1}{60} D K_a
+ \frac{1}{2} \nabla^b W_{abc} K^c
- \frac{1}{2} W_a{}^{ef} W_{efc} K^c
\,
\end{equation}
and this corresponds directly to the difference between the connections \(\omega, \mathfrak{f}, \phi\) and \(\hat{\omega}, \hat{\mathfrak{f}}, \hat{\phi}\), whilst other connections are unaffected.
Consequently, the superspace torsions and curvatures differ between frames, and we will provide explicit formulae when necessary --- the reader can find a more complete discussion of the change of frames in \cite{Butter:2017jqu}. The composite traceless frame connections expand into components as
\begin{subequations}\label{cscconn}
\begin{align}
\hat{\omega}_{a}{}_{bc}
	&=
\omega(e)_{a}{}_{bc}
-2\eta_{a[b} b_{c]}
-\frac{\mathrm{i}}{4}\psi_b{}^k \g_a \psi_c{}_k
-\frac{\mathrm{i}}{2}\psi_a{}^k \g_{[b} \psi_{c]}{}_k
\,, \\
\hat{\phi}_m{}^{k}
	&=
\frac{\mathrm{i}}{16}\Big(
\g^{bc}\g_m -\frac{3}{5}\g_m \tilde{\g}^{bc}
\Big)
\Big(
\hat{\Psi}_{bc}{}^{k}
- \frac{1}{6}W_{def} \tilde{\g}^{def} \g_{[b} \psi_{c]}{}^{k}
\Big)
\,, \\
\hat{\mathfrak{f}}_{a}{}^{b}
	&=
-\frac{1}{8}\mathcal{R}_{a}{}^{b}(\hat{\omega})
+\frac{1}{80}\d_{a}^{b} \mathcal{R}(\hat{\omega})
+\frac{1}{8}\psi_{[a}{}_j \g^{bc} \hat{\phi}_{c]}{}^j
-\frac{1}{80}\d_{a}^{b} \psi_{c}{}_j \g^{cd} \hat{\phi}_{d}{}^j
\nonumber\\
&\quad\,
+\frac{\mathrm{i}}{16}\psi_{c}{}_j \g_{a} \hat{R}(Q)^{bc}{}^j
+\frac{\mathrm{i}}{8}\psi_{c}{}_j \g^{[b} \hat{R}(Q)_{a}{}^{c]}{}^{ j}
+\frac{\mathrm{i}}{60}\psi_{a}{}_j \g^b \chi^j 
\nonumber\\
&\quad\,
-\frac{\mathrm{i}}{8}\psi_{a}{}^{ j}\g_c \psi_{d}{}_j\,W^{bcd} 
+\frac{\mathrm{i}}{80}\d_{a}^{b} \psi_{c}{}^{j} \g_d \psi_{e}{}_j W^{cde} 
\,,
\end{align}
\end{subequations}
where \(\omega(e)_{abc}\) is the Levi-Civita connection and we have the following useful formulae
\begin{subequations}
\begin{align}
\hat{\Psi}_{ab}{}^\g_k
	&=
	2e_a{}^m e_b{}^n \hat{\mathcal{D}}_{[m}\psi_{n]}{}^\g_k  
\,, \\
\mathcal{R}_{ab}{}^{cd}(\hat{\omega})
	&=
	e_a{}^me_b{}^n\Big(
	2\partial_{[m}\hat{\omega}_{n]}{}^{cd}
	-2\hat{\omega}_{[m}{}^{ce} \hat{\omega}_{n]}{}_e{}^d
	\Big) 
\,, \\
\mathcal{R}_{a}{}^{b}(\hat{\omega})
	&= 
	\mathcal{R}_{ac}{}^{b c}(\hat{\omega}) 
\,, \\ 
\mathcal{R}(\hat{\omega}) 
	&= \mathcal{R}_a{}^a(\hat{\omega})
\,, \\
\mathcal{R}_{ab}{}^{ij} 
    &= e_a{}^m e_b{}^n \Big( 2\partial_{[m} \mathcal{V}_{n]}{}^{ij} + 2 \mathcal{V}_{[m}{}^{k(i} \mathcal{V}_{n]k}{}^{j)} \Big)
\,.
\end{align}
\end{subequations}
As mentioned in Appendix~\ref{app:notSugra}, one can verify that the curvature \(\mathcal{R}_{ab}{}^{cd}(\omega(e)) = \mathcal{R}_{ab}{}^{cd}\) is the usual Riemann curvature in the orthonormal frame provided by the vielbein \(e_m{}^a\). As a consequence (when ignoring fermionic and \(b_a\) contributions), the pure bosonic part of the \(K\)-connection is \(\hat{\mathfrak{f}}_{ab} = - \frac{1}{2} S_{ab}\) with \(S_{ab} = \frac{1}{4}(\mathcal{R}_{ab} - \frac{1}{10}\eta_{ab} \mathcal{R})\) the Schouten tensor. 
For completeness, we also provide the projections of the superspace torsion \(\hat{R}(Q)\)
and curvatures \( \hat{R}(M), \hat{R}(J)\) and their relation to the standard Weyl multiplet. We present them in their constrained form\footnote{In practice, one first projects the superspace curvatures and uses the supergeometry to constrain them, which yields the expansions of the connections \eqref{cscconn}, but this does not concern us here. We also ignore the \(\hat{R}(K), \hat{R}(S)\) curvatures as they are cumbersome and are not useful to us here.}
\bsubeq
\begin{align}
\hat{R}(Q)_{ab}{}_k
&=
\frac{1}{2} {\hat{\Psi}}_{ab}{}_k
+\mathrm{i} \tilde{\g}_{[a} \hat{\phi}_{b]}{}_{k}
-\frac{1}{12}W_{cde}\tilde{\g}^{cde}\g_{[a}\psi_{b]}{}_k
\,, \\
\hat{R}(J)_{ab}{}^{ij}
&=
\mathcal{R}_{ab}{}^{ij}
+4\psi_{[a}{}^{ (i}  \hat{\phi}_{b]}{}^{j)}
+\frac{4\mathrm{i}}{15} \psi_{[a}{}^{ (i}\g_{b]} \chi^{ j)}
\,, \\
\hat{R}(M)_{ab}{}^{cd}
&=
\mathcal{R}_{ab}{}^{cd}(\hat{\omega})
+8 \d_{[a}^{[c}  \hat{\mathfrak{f}}_{b]}{}^{d]}
+\mathrm{i}\psi_{[a}{}_j\g_{b]} \hat{R}(Q)^{cd}{}^j
+2\mathrm{i}\psi_{[a}{}_j\g^{[c} \hat{R}(Q)_{b]}{}^{d]}{}^{ j}
\nonumber\\
&\quad\,
-\psi_{[a}{}_j  \g^{cd} \hat{\phi}_{b]}{}^j
-\frac{2\mathrm{i}}{15} \d_{[a}^{[c}\psi_{b]}{}_j \g^{d]}\chi^{j}
-\mathrm{i} \psi_{[a}{}^{ j}\g^e\psi_{b]}{}_j\,W_{e}{}^{cd}
\,,
\end{align}
\esubeq
and their relations to the descendant of the super-Weyl tensor of the standard Weyl multiplet are given by
\bsubeq
\begin{align}
\mathcal{X}_{\a i}{}^{\b \g} &= -\frac{1}{8} (\g^{ab})_{\a}{}^{\b} \hat{R}(Q)_{ab}{}^{\g}{}_i
\,, \\
\mathcal{Y}_{\a}{}^{\b ij} &= - \frac{1}{4} (\g^{ab})_{\a}{}^{\b}\hat{R}(J)_{ab}{}^{kl}
\,, \\
\mathcal{Y}_{\a \b}{}^{\g \d} &= \frac{1}{16} (\g^{ab})_{\a}{}^{\g} (\g^{cd})_{\b}{}^{\d}\hat{R}(M)_{ab}{}^{cd}
\,.
\end{align}
\esubeq
We did not provide the expression of
\(\hat{R}(P),\,\hat{R}(\mathbb{D})\) since they are identically zero in the traceless frame, \(\hat{R}(P)_{ab}{}^c=0,\,\hat{R}(\mathbb{D})_{ab}=0\).
Using the above, one can then readily expand the fields of the projected standard Weyl multiplet into components. One mild complication is that the curvature \(\mathcal{R}_{ab}{}^{cd}(\hat{\omega})\), which appears inside \(\hat{\mathfrak{f}}_{ab}\) and \(\hat{R}(M)_{ab}{}^{cd}\), needs to expanded and rewritten in a covariant form beforehand. Up to quadratic order in fermions this gives
\begin{align}\label{hRiem2}
\mathcal{R}_{ab}{}^{cd}(\hat{\omega})
	&=
	\mathcal{R}_{ab}{}^{cd}(\omega(e,b))
	-\mathrm{i} e_{[a}{}^{m} e_{b]}{}^{n}
	\Big(
	\hat{\mathcal{D}}_{m} \psi^{[c}{}^{k} \gamma_{n} \psi^{d]}{}_{k}
	+ \hat{\mathcal{D}}_{m} \psi_{n}{}^{k} \gamma^{[c} \psi^{d]}{}_{k}
	+\psi_{n}{}^{k} \gamma^{[c} \hat{\mathcal{D}}_{m} \psi^{d]}{}_{k}
	\Big)
\,.
\end{align}

Another complication is given by extracting the hidden fermions within multiple $\hat{\nabla}_a$ derivatives acting on some covariant field in the standard Weyl multiplet. With our end goal being a Lagrangian with Levi-Civita covariant derivatives, one clearly needs to algorithmically strip off $Q$ and $S$ supersymmetry generators according to \eqref{cov2}. To achieve this, one has to carefully employ covariance and the detailed structure of the soft superconformal algebra in the traceless frame. This is a conceptually straightforward and algorithmic procedure that can be implemented by using \emph{Cadabra}. We do not provide all of the explicit, cumbersome, results here, but refer the reader to \cite{Butter:2017jqu} for all the relevant building blocks to evaluate expressions such as 
$\hat{\nabla}_a W^{\a\b}$, 
$\hat{\nabla}_a D$, 
$\hat{\nabla}_a \mathcal{Y}_{\a\b}{}^{\g\d}$, 
$\hat{\nabla}_a \mathcal{Y}_{\a}{}^{\b}{}^{ij}$, 
$\hat{\nabla}_a \mathcal{X}_{\a i}{}^{\g\d}$, 
$\hat{\nabla}_a \chi^{\a i}$, 
$\hat{\nabla}_a \hat{\nabla}_b W^{\a\b}$, 
$\hat{\nabla}_a \hat{\nabla}_b \mathcal{Y}_{\a\b}{}^{\g\d}$, 
and so on. We just give the two most used results
\begin{subequations}
\begin{align}
\hat{\nabla}_{a}\mathcal{X}_{\a i}{}^{\g \d}
&=
\check{\nabla}_{a}\mathcal{X}_{\a i}{}^{\d \g} - \frac{1}{4}\mathcal{Y}_{\a \b}{}^{\d \g} \psi_{a i}{}^{\b}
\,, \\
\hat{\nabla}_{a}\hat{\nabla}_{b}\mathcal{X}_{\a i}{}^{\g \d}
&=
\check{\nabla}_{a}\check{\nabla}_{b}\mathcal{X}_{\a i}{}^{\d \g}
- \frac{1}{4}\mathcal{Y}_{\a \b}{}^{\d \g} \check{\nabla}_{a}{\psi_{b i}{}^{\b}} 
- \frac{1}{4}\psi_{a i}{}^{\b} \check{\nabla}_{b}{\mathcal{Y}_{\a \b}{}^{\d \g}} 
- \frac{1}{4}\psi_{b i}{}^{\b} \check{\nabla}_{a}{\mathcal{Y}_{\a \b}{}^{\d \g}}
\nonumber\\
&\quad\,
+\frac{\mathrm{i}}{4}\mathcal{Y}_{\a \b}{}^{\d \g} (\tilde{\g}_{b})^{\b \epsilon} \hat{\phi}_{a i \epsilon}
\,.
\end{align}
\end{subequations}
which involve the purely bosonic conformal covariant derivatives with R-symmetry \eqref{10rccd} and are only valid up to linear order in fields other than \(\mathcal{Y}_{\a\b}{}^{\g\d}\).

\subsection{\texorpdfstring{\(C^3\) Quadratic Lagrangian}{C3 Quadratic Lagrangian}}
\label{app:c3fermquadlag}
The Lagrangian has contributions at quadratic order from two terms, each of which can be obtained by directly truncating the expressions in the supplementary file of \cite{Butter:2017jqu}, explicitly
\begin{equation}\label{lc3}
\Lagr_{C^3}^{(2)} = F - \frac{\mathrm{i}}{4} \psi_{a i}{}^{\a} \Omega'{}_\a{}^{a i} \,,
\end{equation}
and they have quadratic order truncations given by
\bsubeq
\begin{align}
F 
	&=
- \frac{16}{15}D \mathcal{Y}_{\alpha \beta}{}^{\gamma \delta} \mathcal{Y}_{\gamma \delta}{}^{\alpha \beta} 
- \frac{32}{3}\mathcal{Y}_{\alpha \beta}{}^{\gamma \delta} \mathcal{Y}_{\gamma \delta}{}^{\epsilon \varepsilon} \mathcal{Y}_{\epsilon \varepsilon}{}^{\alpha \beta} 
- \frac{32}{3}\mathcal{Y}_{\alpha \beta}{}^{\gamma \delta} \mathcal{Y}_{\gamma \epsilon}{}^{\alpha \varepsilon} \mathcal{Y}_{\delta \varepsilon}{}^{\beta \epsilon}
+\frac{32}{3} W^{\alpha \beta} \mathcal{Y}_{\alpha \gamma}{}^{\delta \epsilon}  \hat{\nabla}_{\beta \varepsilon}{\mathcal{Y}_{\delta \epsilon}{}^{\gamma \varepsilon}}
\nonumber\\
&\quad\,
- \frac{32}{3} W^{\alpha \beta} \mathcal{Y}_{\alpha \gamma}{}^{\delta \epsilon}  \hat{\nabla}_{\delta \varepsilon}{\mathcal{Y}_{\beta \epsilon}{}^{\gamma \varepsilon}}
+16 \mathcal{Y}_{\gamma \alpha}{}^{\delta \epsilon} \mathcal{Y}_{\delta \epsilon}{}^{\gamma \varepsilon}  \hat{\nabla}_{\varepsilon \beta}{W^{\alpha \beta}}
+8 W^{\alpha \beta} \mathcal{Y}_{\alpha \beta}{}^{\epsilon \varepsilon}  \hat{\nabla}_{\epsilon \gamma}{\hat{\nabla}_{\varepsilon \delta}{W^{\gamma \delta}}}
\nonumber\\
&\quad\,
-8 W^{\alpha \beta} \mathcal{Y}_{\alpha \gamma}{}^{\epsilon \varepsilon}  \hat{\nabla}_{\beta \epsilon}{\hat{\nabla}_{\varepsilon \delta}{W^{\gamma \delta}}} 
- \frac{16}{3} W^{\alpha \beta}  \hat{\nabla}_{\alpha \epsilon}{W^{\gamma \delta}} \hat{\nabla}_{\beta \varepsilon}{\mathcal{Y}_{\gamma \delta}{}^{\epsilon \varepsilon}}
+\frac{16}{3} W^{\alpha \beta}  \hat{\nabla}_{\alpha \epsilon}{W^{\gamma \delta}} \hat{\nabla}_{\gamma \varepsilon}{\mathcal{Y}_{\beta \delta}{}^{\epsilon \varepsilon}}
\nonumber\\
&\quad\,
+8 W^{\alpha \beta}  \hat{\nabla}_{\gamma \varepsilon}{W^{\gamma \delta}} \hat{\nabla}_{\alpha \epsilon}{\mathcal{Y}_{\beta \delta}{}^{\epsilon \varepsilon}}
-8 W^{\alpha \beta}  \hat{\nabla}_{\gamma \epsilon}{W^{\gamma \delta}} \hat{\nabla}_{\delta \varepsilon}{\mathcal{Y}_{\alpha \beta}{}^{\epsilon \varepsilon}} 
- \frac{64\mathrm{i}}{15} \mathcal{X}_{\alpha i}{}^{\beta \gamma} \mathcal{Y}_{\beta \delta}{}^{\alpha \epsilon}  \hat{\nabla}_{\gamma \epsilon}{\chi^{\delta i}}  
\nonumber\\
&\quad\,
- \frac{64\mathrm{i}}{5} \mathcal{X}_{\alpha i}{}^{\beta \gamma} \mathcal{Y}_{\beta \gamma}{}^{\alpha \delta}  \hat{\nabla}_{\delta \epsilon}{\chi^{\epsilon i}} 
+\frac{64\mathrm{i}}{9} \mathcal{X}_{\alpha i}{}^{\beta \gamma} \chi^{\delta i}  \hat{\nabla}_{\beta \epsilon}{\mathcal{Y}_{\gamma \delta}{}^{\alpha \epsilon}}  
- \frac{64\mathrm{i}}{9} \mathcal{X}_{\alpha i}{}^{\beta \gamma} \chi^{\epsilon i}  \hat{\nabla}_{\epsilon \delta}{\mathcal{Y}_{\beta \gamma}{}^{\alpha \delta}} 
\nonumber\\
&\quad\,
-8\mathcal{Y}_{\alpha}{}^{\beta i j} \mathcal{Y}_{\gamma}{}^{\delta}{}_{i j} \mathcal{Y}_{\beta \delta}{}^{\alpha \gamma}
-4 \mathcal{Y}_{\alpha \gamma}{}^{\epsilon \varepsilon}  \hat{\nabla}_{\epsilon \beta}{W^{\alpha \beta}} \hat{\nabla}_{\varepsilon \delta}{W^{\gamma \delta}}
+8 \mathcal{Y}_{\alpha \gamma}{}^{\epsilon \varepsilon}  \hat{\nabla}_{\epsilon \delta}{W^{\alpha \beta}} \hat{\nabla}_{\varepsilon \beta}{W^{\gamma \delta}}
\nonumber\\
&\quad\,
- \frac{64\mathrm{i}}{15} \mathcal{Y}_{\beta \gamma}{}^{\delta \alpha} \chi^{\epsilon i}  \hat{\nabla}_{\delta \epsilon}{\mathcal{X}_{\alpha i}{}^{\beta \gamma}}  
- \frac{64\mathrm{i}}{3} \mathcal{Y}_{\delta \beta}{}^{\epsilon \alpha} \chi^{\delta i}  \hat{\nabla}_{\epsilon \gamma}{\mathcal{X}_{\alpha i}{}^{\beta \gamma}} 
+\frac{128\mathrm{i}}{3} \mathcal{X}_{\alpha i}{}^{\beta \gamma} \mathcal{X}_{\beta}{}^{i \delta \epsilon}  \hat{\nabla}_{\delta \varepsilon}{\mathcal{Y}_{\gamma \epsilon}{}^{\alpha \varepsilon}}   
\nonumber\\
&\quad\,
- \frac{128\mathrm{i}}{3} \mathcal{X}_{\alpha i}{}^{\beta \gamma} \mathcal{X}_{\beta}{}^{i \delta \epsilon}  \hat{\nabla}_{\gamma \varepsilon}{\mathcal{Y}_{\delta \epsilon}{}^{\alpha \varepsilon}} 
-128\mathrm{i} \mathcal{X}_{\alpha i}{}^{\beta \gamma} \mathcal{Y}_{\beta \delta}{}^{\alpha \varepsilon}  \hat{\nabla}_{\varepsilon \epsilon}{\mathcal{X}_{\gamma}{}^{i \delta \epsilon}} 
-256\mathrm{i} \mathcal{X}_{\alpha i}{}^{\beta \gamma} \mathcal{Y}_{\beta \epsilon}{}^{\alpha \delta}  \hat{\nabla}_{\gamma \varepsilon}{\mathcal{X}_{\delta}{}^{i \epsilon \varepsilon}} 
\nonumber\\
&\quad\,
+128\mathrm{i} \mathcal{X}_{\alpha i}{}^{\beta \gamma} \mathcal{Y}_{\beta \epsilon}{}^{\varepsilon \delta}  \hat{\nabla}_{\gamma \varepsilon}{\mathcal{X}_{\delta}{}^{i \alpha \epsilon}}  
-384\mathrm{i} \mathcal{X}_{\alpha i}{}^{\beta \gamma} \mathcal{Y}_{\beta \gamma}{}^{\varepsilon \delta}  \hat{\nabla}_{\varepsilon \epsilon}{\mathcal{X}_{\delta}{}^{i \alpha \epsilon}}  
+128\mathrm{i} \mathcal{X}_{\alpha i}{}^{\beta \gamma} \mathcal{Y}_{\delta \epsilon}{}^{\alpha \varepsilon}  \hat{\nabla}_{\beta \varepsilon}{\mathcal{X}_{\gamma}{}^{i \delta \epsilon}}
\,,
\end{align}
\begin{align}
\Omega^\prime{}_{\a a}{}^i
	&=
\frac{32}{15}(\gamma_{a})_{\epsilon \varepsilon} \mathcal{Y}_{\alpha \beta}{}^{\gamma \delta} \mathcal{Y}_{\gamma \delta}{}^{\beta \epsilon} \chi^{\varepsilon i} 
+\frac{32}{15}(\gamma_{a})_{\delta \varepsilon}  \mathcal{Y}_{\alpha \beta}{}^{\gamma \delta} \mathcal{Y}_{\gamma \epsilon}{}^{\beta \varepsilon} \chi^{\epsilon i} 
- \frac{32}{15}(\gamma_{a})_{\alpha \varepsilon} \mathcal{Y}_{\beta \gamma}{}^{\delta \epsilon} \mathcal{Y}_{\delta \epsilon}{}^{\beta \gamma} \chi^{\varepsilon i}
\nonumber\\
&\quad\,
+\frac{64}{15}(\gamma_{a})_{\alpha \varepsilon} \mathcal{Y}_{\beta \gamma}{}^{\delta \epsilon} \mathcal{Y}_{\delta \epsilon}{}^{\beta \varepsilon} \chi^{\gamma i} 
-64(\gamma_{a})_{\varepsilon \zeta} \mathcal{X}_{\beta}{}^{i \gamma \delta} \mathcal{Y}_{\alpha \epsilon}{}^{\beta \varepsilon} \mathcal{Y}_{\gamma \delta}{}^{\epsilon \zeta}  
+64(\gamma_{a})_{\delta \zeta} \mathcal{X}_{\beta}{}^{i \gamma \delta} \mathcal{Y}_{\alpha \epsilon}{}^{\beta \varepsilon} \mathcal{Y}_{\gamma \varepsilon}{}^{\epsilon \zeta}
\nonumber\\
&\quad\,
-64(\gamma_{a})_{\varepsilon \zeta} \mathcal{X}_{\beta}{}^{i \gamma \delta} \mathcal{Y}_{\alpha \gamma}{}^{\epsilon \varepsilon} \mathcal{Y}_{\delta \epsilon}{}^{\beta \zeta} 
+64(\gamma_{a})_{\delta \zeta} \mathcal{X}_{\beta}{}^{i \gamma \delta} \mathcal{Y}_{\alpha \gamma}{}^{\epsilon \varepsilon} \mathcal{Y}_{\epsilon \varepsilon}{}^{\beta \zeta}
-128(\gamma_{a})_{\alpha \zeta} \mathcal{X}_{\beta}{}^{i \gamma \delta} \mathcal{Y}_{\gamma \delta}{}^{\epsilon \varepsilon} \mathcal{Y}_{\epsilon \varepsilon}{}^{\beta \zeta} 
\nonumber\\
&\quad\,
-128(\gamma_{a})_{\alpha \zeta} \mathcal{X}_{\beta}{}^{i \gamma \delta} \mathcal{Y}_{\gamma \epsilon}{}^{\beta \varepsilon} \mathcal{Y}_{\delta \varepsilon}{}^{\epsilon \zeta} 
+128(\gamma_{a})_{\alpha \delta} \mathcal{X}_{\beta}{}^{i \gamma \delta} \mathcal{Y}_{\gamma \epsilon}{}^{\varepsilon \zeta} \mathcal{Y}_{\varepsilon \zeta}{}^{\beta \epsilon}
\,,
\end{align}
\esubeq
where \(\hat{\nabla}_{\a\b} = (\g^{a})_{\a\b} \hat{\nabla}_a\), \(\hat{\nabla}^{\a\b} = (\tilde{\g}^{a})^{\a\b} \hat{\nabla}_a\) and we adhere to the supplementary file's notation denoting the component field \(W^{\a\b} \vert\) as \(W^{\a\b}\). Many of the fields hidden in the superconformal covariant derivatives \(\hat{\nabla}\) drop out and do not contribute at quadratic order. In particular, the fermion bilinears hidden in \(\hat{\omega},\hat{\mathfrak{f}}\) make no contribution to the above \(C^3\) Lagrangian. However, in the \(C\Box C\) Lagrangian below, the fermion bilinears in both \(\hat{\omega},\hat{\mathfrak{f}}\) make contributions. It is worth noting that the fermion bilinears in \(\hat{\mathfrak{f}}\) only make contributions in the \(C\Box C\) Lagrangian due to the \(\hat{\mathfrak{f}}^{ab} C_{ab}\) term, but never from the covariant derivatives \(\hat{\nabla}\).

\subsection{\texorpdfstring{\(C\Box C\) Quadratic Lagrangian}{C C Quadratic Lagrangian}}
\label{app:cbcfermquadlag}
The Lagrangian has quadratic order contributions given by the following terms\footnote{The result in equations (4.2) and (4.5) in arXiv v3 of \cite{Butter:2017jqu} and prior contains some typos. One needs to put a factor of \(\frac{\mathrm{i}}{2}\) in front of the \(\psi_{mi} \g^m \Omega'{}^i\) and \(\psi_{ai} \hat{\Omega}^{ai}\) terms. Moreover, one needs to correct (4.6b) by replacing the \(\mathrm{i}\) coefficient of the second term by \(2\) and by replacing the \(-\frac{\mathrm{i}}{30}\) coefficient of the third term by \(-\frac{1}{15}\). This makes everything consistent with equations (5.20-5.23) in arXiv v2 of \cite{Butter:2016qkx}.
}
\begin{equation}\label{lcbc}
\Lagr_{C\Box C}^{(2)} = \hat{F} 
+ \frac{\mathrm{i}}{2}\psi_{a i}{}^\a \hat{\Omega}_{\a}{}^{a i} 
- 16\hat{\mathfrak{f}}^{ab} C_{ab} 
- 8\mathrm{i}\psi_{ai}{}^{\a} \g^{ab}{}_{\a}{}^{\b} \Lambda_{\b c}{}^i \hat{\mathfrak{f}}_b{}^c
+ 2\hat{\phi}_{a i \a} \rho^{a \a i} \,.
\end{equation}
Here we also witness the appearance of the \(K,S\)-connections \(\hat{\mathfrak{f}}_{ab}, \hat{\phi}_{a \a}{}^i\) as a consequence of the super 6-form defining the \(C \Box C\) action being non-primary (transforming into an exact form instead of vanishing under special conformal $K$ and $S$-susy transformations). 

A complication of \(C\Box C\) is that the supplementary file of \cite{Butter:2017jqu} provides an expression for \(F\), not \(\hat{F}\) and we have to use formulae from the paper to convert it to \(\hat{F}\). The same goes for \(\Omega^{\a i}, \hat{\Omega}_{\a}{}^{ai}\), but at quadratic order we find \(\hat{\Omega}_{\a a}{}^{i} = (\g_{a})_{\a \b} \Omega^{\b i}\). In the case of \(F,\hat{F}\) at quadratic order, we instead find non-trivial contributions given by
\begin{align}
\hat{F}
	&=
F
- \frac{16\mathrm{i}}{3} \Lambda_{\alpha b}{}^{i} \nabla_c R(Q)^{b c\, \alpha}{}_{i}
+ \frac{2}{3} C_{a b\, i j} R(J)^{a b\, i j}
\nonumber\\
&\quad\,
+ 2 \, C^{a b} \Big(
\frac{2}{15} \eta_{a b} D
-  2 \nabla^c T_{c a b}
+  T_a{}^{c d} T_{b c d}
\Big)
+ \frac{\mathrm{i}}{5} \chi^{\a}{}_j (\gamma_a)_{\a \b}  \rho^{a \b j} \,,
\end{align}
where \(\nabla_a\) here represents the covariant derivative in the original frame and \(R(J)_{ab}{}^{ij},R(Q)_{ab}{}^\a{}_i\) are curvatures in the original frame. We need to convert the covariant derivatives to the traceless frame to be consistent, as all the covariant derivatives in the supplementary file are in the traceless frame. Moreover, it is helpful to also rewrite the curvatures \(R(J)_{ab}{}^{ij},R(Q)_{ab}{}^\a{}_i\) in terms of the projections of the standard Weyl multiplet \(\mathcal{Y}_{\a}{}^{\b i j}, \mathcal{X}_{\a i}{}^{\b \g}\). Following \cite{Butter:2017jqu}, we find that
\begin{subequations}
\begin{align}
R(J)_{ab}{}^{ij} &= \hat{R}(J)_{ab}{}^{ij} = \frac{1}{2} (\g_{ab})_\b{}^\a \mathcal{Y}_{\a}{}^{\b i j}
\,, \\
R(Q)_{ab}{}^{\a i} &=  (\g_{ab})_{\b}{}^{\g} \mathcal{X}_{\g}{}^{i \b \a} + \frac{1}{10} (\tilde{\g}_{ab})^{\a}{}_{\b} \chi^{\b i}
\,.
\end{align}
\end{subequations}
The original frame covariant derivative becomes (recalling \(T_{abc}=-2W_{abc}\))
\begin{subequations}
\begin{align}
\nabla_a W_{bcd} &= \hat{\nabla}_a W_{bcd} \label{covW}
\,, \\
\nabla_{a} \chi^{\a i} &= \hat{\nabla}_{a} \chi^{\a i} - 4(\g_{a})_{\b \g} W^{\a \b} \chi^{\g i}
\,, \\
\nabla_{a} \mathcal{X}_{\a i}{}^{\b \g} &= \hat{\nabla}_{a}{\mathcal{X}_{\alpha i}{}^{\beta \gamma}}
+\frac{1}{2} (\gamma_{a})_{\alpha \epsilon} W^{\delta \epsilon} \mathcal{X}_{\delta i}{}^{\beta \gamma} 
+(\gamma_{a})_{\delta \epsilon} W^{\delta (\beta} \mathcal{X}_{\alpha i}{}^{\gamma) \epsilon}
\nonumber\\
&\quad\,
+\frac{3}{8} (\gamma_{a})_{\alpha \delta} W^{\beta \gamma} \chi^{\delta}{}_i 
- \frac{3}{20}(\gamma_{a})_{\delta \epsilon}  \delta_{\alpha}^{(\beta} W^{\gamma) \delta} \chi^{\epsilon}{}_i
\,,
\end{align}
\end{subequations}
where \eqref{covW} follows from the identity \(W_{a[b}{}^e W_{cd]e} =0 \) implied by anti-self-duality of \(W\). Finally one needs to use the expressions of \(\Lambda_{\a a}{}^i, C_{ab}{}^{ij}, C_{ab}, \rho_{a}{}^{\a i}\) given in the supplementary file of \cite{Butter:2017jqu}. Putting this together, one obtains the quadratic order result 
\begin{align}\label{Fhat0}
\hat{F} 
	&=
F
+\frac{8}{15}D \mathcal{Y}_{\alpha \beta}{}^{\gamma \delta} \mathcal{Y}_{\gamma \delta}{}^{\alpha \beta}
-8 \mathcal{Y}_{\gamma \alpha}{}^{\delta \epsilon} \mathcal{Y}_{\delta \epsilon}{}^{\gamma \varepsilon}  \hat{\nabla}_{\varepsilon \beta}{W^{\alpha \beta}}
-8W^{\alpha \beta} W^{\gamma \delta} \mathcal{Y}_{\alpha \epsilon}{}^{\varepsilon \zeta} \mathcal{Y}_{\gamma \varepsilon}{}^{\epsilon \eta} \varepsilon_{\beta \delta \zeta \eta} 
\nonumber\\
&\quad\,
- \frac{8}{3} W^{\alpha \beta}  \hat{\nabla}_{\gamma \varepsilon}{W^{\gamma \delta}} \hat{\nabla}_{\alpha \epsilon}{\mathcal{Y}_{\beta \delta}{}^{\epsilon \varepsilon}}
+\frac{8}{3} W^{\alpha \beta}  \hat{\nabla}_{\gamma \epsilon}{W^{\gamma \delta}} \hat{\nabla}_{\delta \varepsilon}{\mathcal{Y}_{\alpha \beta}{}^{\epsilon \varepsilon}}
+\frac{64\mathrm{i}}{15} \mathcal{X}_{\alpha i}{}^{\beta \gamma} \mathcal{Y}_{\beta \delta}{}^{\alpha \epsilon}  \hat{\nabla}_{\gamma \epsilon}{\chi^{\delta i}} 
\nonumber\\
&\quad\,
+\frac{128\mathrm{i}}{15} \mathcal{X}_{\alpha i}{}^{\beta \gamma} \mathcal{Y}_{\beta \gamma}{}^{\alpha \delta}  \hat{\nabla}_{\delta \epsilon}{\chi^{\epsilon i}} 
+\frac{32\mathrm{i}}{15} \mathcal{X}_{\alpha i}{}^{\beta \gamma} \chi^{\delta i}  \hat{\nabla}_{\beta \epsilon}{\mathcal{Y}_{\gamma \delta}{}^{\alpha \epsilon}} 
- \frac{32\mathrm{i}}{15} \mathcal{X}_{\alpha i}{}^{\beta \gamma} \chi^{\epsilon i}  \hat{\nabla}_{\epsilon \delta}{\mathcal{Y}_{\beta \gamma}{}^{\alpha \delta}} 
\nonumber\\
&\quad\,
+4 \mathcal{Y}_{\alpha \beta}{}^{\epsilon \varepsilon}  \hat{\nabla}_{\epsilon \gamma}{W^{\alpha \beta}} \hat{\nabla}_{\varepsilon \delta}{W^{\gamma \delta}}
+\frac{16}{3}\mathcal{Y}_{\alpha}{}^{\beta i j} \mathcal{Y}_{\gamma}{}^{\delta}{}_{i j} \mathcal{Y}_{\beta \delta}{}^{\alpha \gamma}
+\frac{32\mathrm{i}}{5} \mathcal{Y}_{\delta \beta}{}^{\epsilon \alpha} \chi^{\delta i}  \hat{\nabla}_{\epsilon \gamma}{\mathcal{X}_{\alpha i}{}^{\beta \gamma}}  
\nonumber\\
&\quad\,
- \frac{256\mathrm{i}}{3} \mathcal{X}_{\alpha i}{}^{\beta \gamma} \mathcal{Y}_{\beta \delta}{}^{\alpha \varepsilon}  \hat{\nabla}_{\varepsilon \epsilon}{\mathcal{X}_{\gamma}{}^{i \delta \epsilon}} 
+\frac{256\mathrm{i}}{3} \mathcal{X}_{\alpha i}{}^{\beta \gamma} \mathcal{Y}_{\beta \epsilon}{}^{\alpha \delta}  \hat{\nabla}_{\gamma \varepsilon}{\mathcal{X}_{\delta}{}^{i \epsilon \varepsilon}} \,.
\end{align}

We may now proceed as in the \(C^3\) case and explicitly truncate the expressions in the supplementary file to quadratic order and obtain the following expressions that should be plugged in \eqref{lcbc} (for $\hat{F}$ one should also use \eqref{Fhat0})
\bsubeq
\begin{align}
F 
	&=
- \frac{4}{45}\hat{\nabla}_{a}{D} \hat{\nabla}^{a}{D} 
- \frac{8}{9}\hat{\nabla}_{a}{\mathcal{Y}_{\alpha \beta}{}^{\gamma \delta}} \hat{\nabla}^{a}{\mathcal{Y}_{\gamma \delta}{}^{\alpha \beta}} 
- \frac{8}{9}(\gamma^{a b})_{\epsilon}{}^{\alpha} \hat{\nabla}_{a}{\mathcal{Y}_{\alpha \beta}{}^{\gamma \delta}} \hat{\nabla}_{b}{\mathcal{Y}_{\gamma \delta}{}^{\epsilon \beta}} 
- \frac{8}{3} W^{\alpha \beta} \mathcal{Y}_{\alpha \gamma}{}^{\delta \epsilon}  \hat{\nabla}_{\beta \varepsilon}{\mathcal{Y}_{\delta \epsilon}{}^{\gamma \varepsilon}}
\nonumber\\
&\quad\,
+\frac{8}{3} W^{\alpha \beta} \mathcal{Y}_{\alpha \gamma}{}^{\delta \epsilon}  \hat{\nabla}_{\delta \varepsilon}{\mathcal{Y}_{\beta \epsilon}{}^{\gamma \varepsilon}}
+\mathcal{Y}_{\alpha}{}^{\beta i j} \hat{\nabla}^2 {\mathcal{Y}_{\beta}{}^{\alpha}{}_{i j}} 
- \hat{\nabla}_{a}{\hat{\nabla}_{\alpha \gamma}{W^{\alpha \beta}}} \hat{\nabla}^{a}{\hat{\nabla}_{\beta \delta}{W^{\gamma \delta}}}
\nonumber\\
&\quad\,
- \hat{\nabla}^2{W^{\alpha \beta}} \hat{\nabla}_{\alpha \gamma}{\hat{\nabla}_{\beta \delta}{W^{\gamma \delta}}}
+\frac{4}{3} (\gamma^{b c})_{\epsilon}{}^{\gamma} \hat{\nabla}_{b}{\mathcal{Y}_{\gamma \beta}{}^{\delta \epsilon}} \hat{\nabla}_{\delta \alpha}{\hat{\nabla}_{c}{W^{\alpha \beta}}} 
- \frac{8}{9} \hat{\nabla}_{\gamma \varepsilon}{\mathcal{Y}_{\alpha \beta}{}^{\gamma \delta}} \hat{\nabla}^{\alpha \epsilon}{\mathcal{Y}_{\epsilon \delta}{}^{\varepsilon \beta}} 
\nonumber\\
&\quad\,
- \frac{5}{9}\hat{\nabla}_{a}{\mathcal{Y}_{\alpha}{}^{\beta i j}} \hat{\nabla}^{a}{\mathcal{Y}_{\beta}{}^{\alpha}{}_{i j}} 
+\frac{8}{5}W^{\alpha \beta} W^{\gamma \delta} \mathcal{Y}_{\alpha \epsilon}{}^{\varepsilon \zeta} \mathcal{Y}_{\gamma \varepsilon}{}^{\epsilon \eta} \varepsilon_{\beta \delta \zeta \eta}
+2 W^{\alpha \beta} \mathcal{Y}_{\alpha \beta}{}^{\epsilon \varepsilon}  \hat{\nabla}_{\epsilon \gamma}{\hat{\nabla}_{\varepsilon \delta}{W^{\gamma \delta}}} 
\nonumber\\
&\quad\,
- \frac{4}{5} W^{\alpha \beta} \mathcal{Y}_{\alpha \gamma}{}^{\epsilon \varepsilon}  \hat{\nabla}_{\varepsilon \delta}{\hat{\nabla}_{\beta \epsilon}{W^{\gamma \delta}}}
+\frac{4}{5} W^{\alpha \beta} \mathcal{Y}_{\alpha \gamma}{}^{\epsilon \varepsilon}  \hat{\nabla}_{\beta \epsilon}{\hat{\nabla}_{\varepsilon \delta}{W^{\gamma \delta}}} 
- \frac{8}{3}W^{\alpha \beta} \hat{\nabla}_{\alpha \epsilon}{W^{\gamma \delta}} \hat{\nabla}_{\beta \varepsilon}{\mathcal{Y}_{\gamma \delta}{}^{\epsilon \varepsilon}}
\nonumber\\
&\quad\,
+\frac{8}{3} W^{\alpha \beta}  \hat{\nabla}_{\alpha \epsilon}{W^{\gamma \delta}} \hat{\nabla}_{\gamma \varepsilon}{\mathcal{Y}_{\beta \delta}{}^{\epsilon \varepsilon}}
+4 W^{\alpha \beta}  \hat{\nabla}_{\gamma \varepsilon}{W^{\gamma \delta}} \hat{\nabla}_{\alpha \epsilon}{\mathcal{Y}_{\beta \delta}{}^{\epsilon \varepsilon}}
-4 W^{\alpha \beta}  \hat{\nabla}_{\gamma \epsilon}{W^{\gamma \delta}} \hat{\nabla}_{\delta \varepsilon}{\mathcal{Y}_{\alpha \beta}{}^{\epsilon \varepsilon}}
\nonumber\\
&\quad\,
- \frac{32\mathrm{i}}{15} \mathcal{X}_{\alpha i}{}^{\beta \gamma} \chi^{\delta i}  \hat{\nabla}_{\beta \epsilon}{\mathcal{Y}_{\gamma \delta}{}^{\alpha \epsilon}} 
+\frac{32\mathrm{i}}{15} \mathcal{X}_{\alpha i}{}^{\beta \gamma} \chi^{\epsilon i}  \hat{\nabla}_{\epsilon \delta}{\mathcal{Y}_{\beta \gamma}{}^{\alpha \delta}} 
+2\mathcal{Y}_{\alpha}{}^{\beta i j} \mathcal{Y}_{\gamma}{}^{\delta}{}_{i j} \mathcal{Y}_{\beta \delta}{}^{\alpha \gamma}
\nonumber\\
&\quad\,
+\frac{32\mathrm{i}}{15}\mathcal{Y}_{\delta \beta}{}^{\epsilon \alpha} \chi^{\delta i}  \hat{\nabla}_{\epsilon \gamma}{\mathcal{X}_{\alpha i}{}^{\beta \gamma}} 
+\frac{16\mathrm{i}}{225} \hat{\nabla}_{\alpha \beta}{\chi^{\alpha i}} \hat{\nabla}^2{\chi^{\beta}{}_i}
+\frac{64\mathrm{i}}{225} \hat{\nabla}_{b}{\chi^{\alpha i}} \hat{\nabla}^{b}{\hat{\nabla}_{\alpha \beta}{\chi^{\beta}{}_i}}   
\nonumber\\
&\quad\,
- \frac{32\mathrm{i}}{45} (\gamma^{b c})_{\gamma}{}^{\alpha} \hat{\nabla}_{b}{\mathcal{X}_{\alpha i}{}^{\beta \gamma}} \hat{\nabla}_{\beta \delta}{\hat{\nabla}_{c}{\chi^{\delta i}}} 
-32\mathrm{i} \hat{\nabla}_{\beta \delta}{\mathcal{X}_{\alpha i}{}^{\beta \gamma}} \hat{\nabla}^2{\mathcal{X}_{\gamma}{}^{i \delta \alpha}}
- \frac{32\mathrm{i}}{9} \hat{\nabla}_{b}{\mathcal{X}_{\alpha i}{}^{\beta \gamma}} \hat{\nabla}^{b}{\hat{\nabla}_{\beta \delta}{\mathcal{X}_{\gamma}{}^{i \delta \alpha}}}  
\nonumber\\
&\quad\,
+\frac{32\mathrm{i}}{45} (\gamma^{b c})_{\gamma}{}^{\alpha} \hat{\nabla}_{b}{\chi^{\delta i}} \hat{\nabla}_{\delta \beta}{\hat{\nabla}_{c}{\mathcal{X}_{\alpha i}{}^{\beta \gamma}}} 
+\frac{1}{9} (\gamma^{a b})_{\gamma}{}^{\alpha} \hat{\nabla}_{a}{\mathcal{Y}_{\alpha}{}^{\beta i j}} \hat{\nabla}_{b}{\mathcal{Y}_{\beta}{}^{\gamma}{}_{i j}}
+64\mathrm{i} \mathcal{X}_{\alpha i}{}^{\beta \gamma} \mathcal{X}_{\beta}{}^{i \delta \epsilon}  \hat{\nabla}_{\delta \varepsilon}{\mathcal{Y}_{\gamma \epsilon}{}^{\alpha \varepsilon}}  
\nonumber\\
&\quad\,
-64\mathrm{i} \mathcal{X}_{\alpha i}{}^{\beta \gamma} \mathcal{X}_{\beta}{}^{i \delta \epsilon}  \hat{\nabla}_{\gamma \varepsilon}{\mathcal{Y}_{\delta \epsilon}{}^{\alpha \varepsilon}} 
+\frac{544\mathrm{i}}{9} \mathcal{X}_{\alpha i}{}^{\beta \gamma} \mathcal{Y}_{\beta \delta}{}^{\alpha \varepsilon}  \hat{\nabla}_{\varepsilon \epsilon}{\mathcal{X}_{\gamma}{}^{i \delta \epsilon}} 
+\frac{544\mathrm{i}}{9} \mathcal{X}_{\alpha i}{}^{\beta \gamma} \mathcal{Y}_{\beta \epsilon}{}^{\alpha \delta}  \hat{\nabla}_{\gamma \varepsilon}{\mathcal{X}_{\delta}{}^{i \epsilon \varepsilon}} 
\nonumber\\
&\quad\,
+\frac{64\mathrm{i} }{9} \mathcal{X}_{\alpha i}{}^{\beta \gamma} \mathcal{Y}_{\beta \epsilon}{}^{\varepsilon \delta}  \hat{\nabla}_{\gamma \varepsilon}{\mathcal{X}_{\delta}{}^{i \alpha \epsilon}} 
- \frac{544\mathrm{i}}{9} \mathcal{X}_{\alpha i}{}^{\beta \gamma} \mathcal{Y}_{\beta \gamma}{}^{\varepsilon \delta}  \hat{\nabla}_{\varepsilon \epsilon}{\mathcal{X}_{\delta}{}^{i \alpha \epsilon}}   
- \frac{32\mathrm{i}}{9} \mathcal{X}_{\alpha i}{}^{\beta \gamma} \mathcal{Y}_{\delta \epsilon}{}^{\alpha \varepsilon}  \hat{\nabla}_{\beta \varepsilon}{\mathcal{X}_{\gamma}{}^{i \delta \epsilon}} 
\nonumber\\
&\quad\,
+\frac{1}{3} \mathcal{Y}_{\alpha}{}^{\beta i j}  \hat{\nabla}_{\beta \delta}{\hat{\nabla}^{\alpha \gamma}{\mathcal{Y}_{\gamma}{}^{\delta}{}_{i j}}} 
- \frac{1}{9} \hat{\nabla}_{\beta \delta}{\mathcal{Y}_{\alpha}{}^{\beta i j}} \hat{\nabla}^{\alpha \gamma}{\mathcal{Y}_{\gamma}{}^{\delta}{}_{i j}}
+\frac{8\mathrm{i}}{81} (\gamma^{b c})_{\epsilon}{}^{\alpha} \hat{\nabla}_{b}{\mathcal{X}_{\alpha i}{}^{\beta \gamma}} \hat{\nabla}_{\beta \delta}{\hat{\nabla}_{c}{\mathcal{X}_{\gamma}{}^{i \delta \epsilon}}} 
\nonumber\\
&\quad\,
+\frac{296\mathrm{i}}{81} (\gamma^{b c})_{\gamma}{}^{\delta} \hat{\nabla}_{b}{\mathcal{X}_{\alpha i}{}^{\beta \gamma}} \hat{\nabla}_{\beta \epsilon}{\hat{\nabla}_{c}{\mathcal{X}_{\delta}{}^{i \epsilon \alpha}}} 
- \frac{296\mathrm{i}}{81} \gamma_{c}{}^{} \hat{\nabla}_{\beta \epsilon}{\mathcal{X}_{\alpha i}{}^{\beta \gamma}} \hat{\nabla}_{\gamma \varepsilon}{\hat{\nabla}^{\alpha \delta}{\mathcal{X}_{\delta}{}^{i \epsilon \varepsilon}}}
\nonumber\\
&\quad\,
+\frac{8\mathrm{i}}{81} \hat{\nabla}^{\alpha \delta}{\mathcal{X}_{\alpha i}{}^{\beta \gamma}} \hat{\nabla}_{\beta \epsilon}{\hat{\nabla}_{\gamma \varepsilon}{\mathcal{X}_{\delta}{}^{i \epsilon \varepsilon}}}
\,,
\end{align}
\begin{align}
\Omega^{\a i} 
	&=
- \frac{16}{45} (\gamma^{a b})_{\delta}{}^{\beta} \hat{\nabla}_{a}{\mathcal{Y}_{\beta \gamma}{}^{\alpha \delta}} \hat{\nabla}_{b}{\chi^{\gamma i}}
+\frac{64}{9}\hat{\nabla}_{a}{\mathcal{X}_{\beta}{}^{i \gamma \delta}} \hat{\nabla}^{a}{\mathcal{Y}_{\gamma \delta}{}^{\alpha \beta}}
+\frac{32}{3} (\gamma^{a b})_{\epsilon}{}^{\alpha} \hat{\nabla}_{a}{\mathcal{X}_{\beta}{}^{i \gamma \delta}} \hat{\nabla}_{b}{\mathcal{Y}_{\gamma \delta}{}^{\epsilon \beta}} 
\nonumber\\
&\quad\,
- \frac{32}{9} (\gamma^{a b})_{\epsilon}{}^{\beta} \hat{\nabla}_{a}{\mathcal{X}_{\beta}{}^{i \gamma \delta}} \hat{\nabla}_{b}{\mathcal{Y}_{\gamma \delta}{}^{\alpha \epsilon}}
- \frac{32}{3} (\gamma^{a b})_{\epsilon}{}^{\delta} \hat{\nabla}_{a}{\mathcal{X}_{\beta}{}^{i \alpha \gamma}} \hat{\nabla}_{b}{\mathcal{Y}_{\delta \gamma}{}^{\epsilon \beta}}
+\frac{32}{3} \hat{\nabla}^{\alpha \epsilon}{\mathcal{X}_{\beta}{}^{i \gamma \delta}} \hat{\nabla}_{\gamma \varepsilon}{\mathcal{Y}_{\epsilon \delta}{}^{\varepsilon \beta}}
\nonumber\\
&\quad\,
- \frac{32}{9} \hat{\nabla}^{\beta \epsilon}{\mathcal{X}_{\beta}{}^{i \gamma \delta}} \hat{\nabla}_{\gamma \varepsilon}{\mathcal{Y}_{\epsilon \delta}{}^{\alpha \varepsilon}}
\,,
\end{align}
\begin{align}
C_{ab} 
	&=
\frac{1}{45}\eta_{a b} D^2 
+\frac{1}{15} (\gamma_{a b c})_{\alpha \beta} D  \hat{\nabla}^{c}{W^{\alpha \beta}}
+\frac{1}{15} (\gamma_{a b c})_{\alpha \beta} W^{\alpha \beta}  \hat{\nabla}^{c}{D} 
- \frac{1}{2} (\gamma_{a})_{\alpha \gamma} (\gamma_{b})_{\beta \delta} W^{\alpha \beta}  \hat{\nabla}^2{W^{\gamma \delta}}
\nonumber\\
&\quad\,
+\frac{1}{2} (\gamma_{a})_{\alpha \gamma} W^{\alpha \beta}  \hat{\nabla}_{b}{\hat{\nabla}_{\beta \delta}{W^{\gamma \delta}}} 
- \frac{2}{3} (\gamma_{b})_{\alpha \delta} (\gamma_{a c})_{\epsilon}{}^{\gamma} W^{\alpha \beta}  \hat{\nabla}^{c}{\mathcal{Y}_{\beta \gamma}{}^{\delta \epsilon}}
+\frac{1}{2} (\gamma_{b})_{\alpha \gamma} W^{\alpha \beta}  \hat{\nabla}_{a}{\hat{\nabla}_{\beta \delta}{W^{\gamma \delta}}} 
\nonumber\\
&\quad\,
- \frac{1}{2} (\gamma_{a b d})_{\beta \delta} W^{\alpha \beta}  \hat{\nabla}_{\alpha \gamma}{\hat{\nabla}^{d}{W^{\gamma \delta}}} 
- \frac{4\mathrm{i}}{15} (\gamma_{a b c})_{\beta \gamma} \mathcal{X}_{\alpha i}{}^{\beta \gamma}  \hat{\nabla}^{c}{\chi^{\alpha i}} 
- (\gamma_{b})_{\delta \varepsilon} (\tilde{\gamma}_{a})^{\beta \epsilon} \mathcal{Y}_{\alpha \beta}{}^{\gamma \delta} \mathcal{Y}_{\gamma \epsilon}{}^{\alpha \varepsilon} 
\nonumber\\
&\quad\,
+ (\gamma_{b})_{\delta \beta} (\gamma_{a c})_{\epsilon}{}^{\gamma} \mathcal{Y}_{\gamma \alpha}{}^{\delta \epsilon}  \hat{\nabla}^{c}{W^{\alpha \beta}} 
- \frac{1}{2} (\gamma_{a b})_{\epsilon}{}^{\gamma} \mathcal{Y}_{\gamma \alpha}{}^{\delta \epsilon}  \hat{\nabla}_{\delta \beta}{W^{\alpha \beta}}
+\frac{16\mathrm{i}}{45} (\gamma_{a b c})_{\beta \gamma} \chi^{\alpha i}  \hat{\nabla}^{c}{\mathcal{X}_{\alpha i}{}^{\beta \gamma}}  
\nonumber\\
&\quad\,
- \frac{5}{8} (\gamma_{a})_{\alpha \gamma} (\gamma_{b})_{\beta \delta} \hat{\nabla}_{c}{W^{\alpha \beta}} \hat{\nabla}^{c}{W^{\gamma \delta}}
+\frac{1}{8} (\gamma_{a})_{\alpha \gamma} \hat{\nabla}_{b}{W^{\alpha \beta}} \hat{\nabla}_{\beta \delta}{W^{\gamma \delta}}
+\frac{1}{8} (\gamma_{b})_{\alpha \gamma}  \hat{\nabla}_{a}{W^{\alpha \beta}} \hat{\nabla}_{\beta \delta}{W^{\gamma \delta}}
\nonumber\\
&\quad\,
+\frac{1}{8} (\gamma_{b})_{\alpha \gamma} (\gamma_{a c d})_{\beta \delta} \hat{\nabla}^{c}{W^{\alpha \beta}} \hat{\nabla}^{d}{W^{\gamma \delta}}
+\frac{3}{8} (\gamma_{a b d})_{\beta \delta} \hat{\nabla}_{\alpha \gamma}{W^{\alpha \beta}} \hat{\nabla}^{d}{W^{\gamma \delta}}
+ (\gamma_{a})_{\gamma \epsilon} (\gamma_{b})_{\delta \varepsilon} W^{\alpha \beta} W^{\gamma \delta} \mathcal{Y}_{\alpha \beta}{}^{\epsilon \varepsilon} 
\nonumber\\
&\quad\,
- \frac{1}{2} \eta_{a b} W^{\alpha \beta}  \hat{\nabla}_{\alpha \gamma}{\hat{\nabla}_{\beta \delta}{W^{\gamma \delta}}}
-4\mathrm{i} (\gamma_{a})_{\beta \delta} \mathcal{X}_{\alpha i}{}^{\beta \gamma}  \hat{\nabla}_{b}{\mathcal{X}_{\gamma}{}^{i \alpha \delta}}  
- \frac{8\mathrm{i}}{15} (\gamma_{b})_{\beta \delta} (\gamma_{a c})_{\gamma}{}^{\alpha} \mathcal{X}_{\alpha i}{}^{\beta \gamma}  \hat{\nabla}^{c}{\chi^{\delta i}} 
\nonumber\\
&\quad\,
+\frac{4\mathrm{i}}{3} (\gamma_{b})_{\beta \delta} \mathcal{X}_{\alpha i}{}^{\beta \gamma}  \hat{\nabla}_{a}{\mathcal{X}_{\gamma}{}^{i \alpha \delta}} 
+\frac{4\mathrm{i}}{15} (\gamma_{a b})_{\gamma}{}^{\alpha} \mathcal{X}_{\alpha i}{}^{\beta \gamma}  \hat{\nabla}_{\beta \delta}{\chi^{\delta i}} 
+4\mathrm{i} (\gamma_{a b c})_{\beta \delta} \mathcal{X}_{\alpha i}{}^{\beta \gamma}  \hat{\nabla}^{c}{\mathcal{X}_{\gamma}{}^{i \alpha \delta}} 
\nonumber\\
&\quad\,
+\frac{5}{12} \eta_{a b} \mathcal{Y}_{\alpha}{}^{\beta i j} \mathcal{Y}_{\beta}{}^{\alpha}{}_{i j} 
+\frac{1}{4} (\gamma_{a b})_{\gamma}{}^{\alpha} \mathcal{Y}_{\alpha}{}^{\beta i j} \mathcal{Y}_{\beta}{}^{\gamma}{}_{i j} 
+\frac{2\mathrm{i}}{75} (\gamma_{a})_{\alpha \beta} \chi^{\alpha i}  \hat{\nabla}_{b}{\chi^{\beta}{}_i} 
\nonumber\\
&\quad\,
+\frac{2\mathrm{i}}{75} (\gamma_{b})_{\alpha \beta} \chi^{\alpha i}  \hat{\nabla}_{a}{\chi^{\beta}{}_i} 
- \frac{14\mathrm{i}}{225} (\gamma_{a b c})_{\alpha \beta} \chi^{\alpha i}  \hat{\nabla}^{c}{\chi^{\beta}{}_i} 
- \frac{16\mathrm{i}}{45} (\gamma_{b})_{\delta \beta} (\gamma_{a c})_{\gamma}{}^{\alpha} \chi^{\delta i}  \hat{\nabla}^{c}{\mathcal{X}_{\alpha i}{}^{\beta \gamma}} 
\nonumber\\
&\quad\,
+\frac{8\mathrm{i}}{45} (\gamma_{a b})_{\gamma}{}^{\alpha}\chi^{\delta i}  \hat{\nabla}_{\delta \beta}{\mathcal{X}_{\alpha i}{}^{\beta \gamma}} 
+\frac{1}{8} \eta_{a b} \hat{\nabla}_{\alpha \gamma}{W^{\alpha \beta}} \hat{\nabla}_{\beta \delta}{W^{\gamma \delta}}
+8\mathrm{i} \eta_{a b} \mathcal{X}_{\alpha i}{}^{\beta \gamma}  \hat{\nabla}_{\beta \delta}{\mathcal{X}_{\gamma}{}^{i \alpha \delta}} 
\nonumber\\
&\quad\,
+\frac{28\mathrm{i}}{3} (\gamma_{b})_{\beta \delta} (\gamma_{a c})_{\epsilon}{}^{\alpha}\mathcal{X}_{\alpha i}{}^{\beta \gamma}  \hat{\nabla}^{c}{\mathcal{X}_{\gamma}{}^{i \delta \epsilon}} 
-4\mathrm{i} (\gamma_{b})_{\beta \epsilon} (\gamma_{a c})_{\gamma}{}^{\delta} \mathcal{X}_{\alpha i}{}^{\beta \gamma}  \hat{\nabla}^{c}{\mathcal{X}_{\delta j}{}^{i \alpha \epsilon}} 
+4\mathrm{i} (\gamma_{a b})_{\gamma}{}^{\delta} \mathcal{X}_{\alpha i}{}^{\beta \gamma}  \hat{\nabla}_{\beta \epsilon}{\mathcal{X}_{\delta}{}^{i \alpha \epsilon}} 
\nonumber\\
&\quad\,
- \frac{1}{4} (\gamma_{b})_{\beta \delta} (\tilde{\gamma}_{a})^{\alpha \gamma} \mathcal{Y}_{\alpha}{}^{\beta i j} \mathcal{Y}_{\gamma}{}^{\delta}{}_{i j} 
- \frac{22\mathrm{i}}{225} \eta_{a b} \chi^{\alpha i}  \hat{\nabla}_{\alpha \beta}{\chi^{\beta}{}_i} 
- \frac{20\mathrm{i}}{3} (\gamma_{b})_{\beta \epsilon} (\tilde{\gamma}_{a})^{\alpha \delta} \mathcal{X}_{\alpha i}{}^{\beta \gamma}  \hat{\nabla}_{\gamma \varepsilon}{\mathcal{X}_{\delta}{}^{i \epsilon \varepsilon}}
\,,
\end{align} 
\begin{align}
\Lambda_{\a a}{}^i
	&=
8 (\g_{a})_{\d \epsilon} \mathcal{X}_{\b}{}^{i \g \d} \mathcal{Y}_{\a \g}{}^{\b \epsilon}
\,,
\end{align}

\begin{align}
\rho_{a}{}^{\g i}
	&=
- \frac{8}{15} (\gamma_{a b})_{\delta}{}^{\alpha} \mathcal{Y}_{\alpha \beta}{}^{\gamma \delta}  \hat{\nabla}^{b}{\chi^{\beta i}}
+\frac{32}{3} \mathcal{Y}_{\beta \delta}{}^{\gamma \alpha} \hat{\nabla}_{a}{\mathcal{X}_{\alpha}{}^{i \beta \delta}}
- \frac{16}{45} (\gamma_{a b})_{\delta}{}^{\beta} \chi^{\alpha i}  \hat{\nabla}^{b}{\mathcal{Y}_{\alpha \beta}{}^{\gamma \delta}} 
\nonumber\\
&\quad\,
- \frac{16}{3} (\gamma_{a b})_{\epsilon}{}^{\delta} \mathcal{X}_{\alpha}{}^{i \gamma \beta}  \hat{\nabla}^{b}{\mathcal{Y}_{\beta \delta}{}^{\alpha \epsilon}}
+\frac{16}{3} (\gamma_{a b})_{\epsilon}{}^{\alpha} \mathcal{Y}_{\beta \delta}{}^{\gamma \epsilon}  \hat{\nabla}^{b}{\mathcal{X}_{\alpha}{}^{i \beta \delta}}
+16 (\gamma_{a b})_{\delta}{}^{\epsilon} \mathcal{Y}_{\epsilon \beta}{}^{\gamma \alpha}  \hat{\nabla}^{b}{\mathcal{X}_{\alpha}{}^{i \beta \delta}}
\nonumber\\
&\quad\,
- \frac{16}{3} (\gamma_{a})_{\varepsilon \delta} \mathcal{Y}_{\epsilon \beta}{}^{\gamma \varepsilon}  \hat{\nabla}^{\epsilon \alpha}{\mathcal{X}_{\alpha}{}^{i \beta \delta}}
\,.
\end{align}
\esubeq

\subsection{\texorpdfstring{Bosonic \((1,0)\) Lagrangians}{Bosonic (1,0) Lagrangians}}
\label{app:bosonic10lag}
We list the full bosonic Lagrangians from \cite{Butter:2017jqu} for completeness, with \(T^-_{abc} \to T_{abc}\). We again give the expression for the conformal covariant derivative with SU(2) R-symmetry \eqref{10rccd} for easy reference
\begin{align*}
\check{\nabla}_a &= e_a - \frac{1}{2} \omega(e,b)_a{}^{bc} M_{bc} - \mathcal{V}_a{}^{ij} J_{ij} - b_a \mathbb{D} - f_a{}^b K_b
\,,
\\
[\check{\nabla}_a ,\check{\nabla}_b]
&=
- \frac{1}{2} C_{ab}{}^{de}M_{de}
-\mathcal{R}_{ab}{}^{ij}J_{ij}
-\frac{1}{6}\check{\nabla}^d C_{abcd}K^d
\,.
\end{align*}
The bosonic \(C^3\)  Lagrangian is given by
\begin{align}
\Lagr_{C^3}^{(\text{b})}
&=
\frac{8}{3}\,{C}_{a bc d} {C}^{a b e f} {C}^{c d}{}_{e f} 
  - \frac{16}{3}\, {C}_{a b c d} {C}^{aecf} {C}^{b}{}_{ e}{}^{d}{}_{f} 
-2\, {C}_{a b c d} {\mathcal{R}}^{a b}\,^{i j} {\mathcal{R}}^{c d}\,_{ij}  
+ 4\, { \mathcal{R}}_{a b}\,^{i j} { \mathcal{R}}^{a c}\,_{i}{}^{k} { \mathcal{R}}^{b}{}_{c}\,_{j k} 
\nonumber\\
&\quad\,
     - \frac{32}{225} D^3
  - \frac{4}{15} \, D {C}_{a b c d} { C}^{a b c d} 
  + \frac{8}{5} D { \mathcal{R}}_{a b}\,^{i j} {\mathcal{R}}^{a b}\,_{ij}
    + \frac{128}{15} \, T_{a b c} {T}^{ - a d e} D {C}^{b}{}_{d}{}^{c}{}_{e} 
\nonumber\\
&\quad\,
     + \frac{64}{15} \, T_{a b c} D {\check{\nabla}}^{a}{{\check{\nabla}}_{d}{T^{b c d}}}
- \frac{4}{5} \, D {\check{\nabla}}^{a}{T_{a b c}}\,  {\check{\nabla}}_{d}{T^{b c d}}
+ \frac{4}{15} \, D {\check{\nabla}}_{a}{T_{b c d}}\,  {\check{\nabla}}^{a}{T^{b c d}}
- \frac{4}{3} \, D {\check{\nabla}}_{a}{T_{b c d}}\,  {\check{\nabla}}^{b}{T^{a c d}}
\nonumber\\
&\quad\,
  - \frac{16}{5} \, T_{a b c} T^{a b d} T^{c e f} {T}_{ d e f} D 
- \frac{32}{3}\, T_{a b c} { C}^{a b d e} {\check{\nabla}}^{f}{{ C}^{c}{}_{d e f}}
   + \frac{16}{3}\, { C}_{a b c d} { C}^{a b e f} {\check{\nabla}}^{c}{T^{d}{}_{e f}}
\nonumber\\
&\quad\,
    - 16\, T_{a b c} {\check{\nabla}}_{d}{T^{a b e}}\,  {\check{\nabla}}_{e}{{\check{\nabla}}_{f}{T^{cd f}}}   
   - 16 \, T_{a b c} {\check{\nabla}}_{d}{T^{a d e}}\,  {\check{\nabla}}_{e}{{\check{\nabla}}_{f}{T^{b c f}}}
  - 48 \, T_{a b c} {\check{\nabla}}_{d}{T^{a d e}}\,  {\check{\nabla}}^{b}{{\check{\nabla}}^{f}{T^{c}{}_{ e f}}}
\nonumber\\
&\quad\,
    + 16 \, {\check{\nabla}}^{e}{T_{e a b}}\,  {\check{\nabla}}^{f}{T_{f c d}}\,  {\check{\nabla}}^{a}{T^{b c d}}\,  
    - 40\, T_{a b e} T^{c d e} {\check{\nabla}}_{f}{T^{f a b}}\,  {\check{\nabla}}^{g}{T_{g c d}}
 + 16 \, T_{a b c} { C}^{a b d e} {\check{\nabla}}^{c}{{\check{\nabla}}^{f}{T_{d e f}} }
\nonumber\\
&\quad\,
 - 16\, T_{a b c} {C}^{a b d e} {\check{\nabla}}_{d}{{\check{\nabla}}^{f}{T^{c}{}_{e f}}}
 - 4\, {C}_{a b c d} {\check{\nabla}}_{e}{T^{a b e}} {\check{\nabla}}_{f}{T^{c d f}}
 +8\, { C}_{a b c d} {\check{\nabla}}_{e}{T^{a b f}}\,  {\check{\nabla}}_{f}{T^{c d e}}
\nonumber\\
&\quad\,
- \frac{64}{3}\, T^{f b}{}_{ d} \,{\check{\nabla}}^{e}{{C}_{e a b c}}\,  {\check{\nabla}}_{f}{T^{a c d}}
+32\, T^{a b}{}_{ d} \,  {\check{\nabla}}^{e}{{ C}_{e a b c}}\,  {\check{\nabla}}_{f}{T^{f c d}}
    - 32\,  T_{f g c} T^{f g d} {\check{\nabla}}^{c}{T_{d a b}}\, {\check{\nabla}}_{e}{T^{e a b}}
\nonumber\\
&\quad\,
- 8\, {\check{\nabla}}_{e}{T_{b a d}}\,  {\check{\nabla}}^{e}{T^{c a d}}\,  T^{f g b} T_{f g c} 
- 8\, T_{a b c} T^{a b d} { C}^{c e f g} {\check{\nabla}}_{e}{T_{d f g}}
- \frac{8}{3}\, T_{a b c} { \mathcal{R}}^{a b}\,^{i j} {\check{\nabla}}_{d}{{ \mathcal{R}}^{c d}\,_{ij}}
\nonumber\\
&\quad\,
+ \frac{28}{3}\, T_{a b c} { \mathcal{R}}^{a d}\,^{i j} {\check{\nabla}}_{d}{{ \mathcal{R}}^{b c}\,_{ij}}
- \frac{32}{9}\, {\mathcal{R}}_{a b}\,^{i j} {\mathcal{R}}_{c d}\,_{ij} {\check{\nabla}}^{a}{T^{b c d}}\, 
+ 4 \,  T_{e f b} T^{e f c} T_{g h a} T^{g h}{}_{ c}\, {\check{\nabla}}_{d}{T^{d a b}}
\nonumber\\
&\quad\,
-8\,  T_{a b c} T^{a b d} T_{e f g} T^{e f h} { C}^{c g}{}_{ d h} 
+12\, T_{a b c} T^{a d e} {\mathcal{R}}^{b c}\,^{i j} {\mathcal{R}}_{d e}\,_{ij} 
\,.
\end{align}
The bosonic \(C\Box C\) Lagrangian is given by
\begin{align}
\Lagr_{C\Box C}^{(\text{b})}
	&=
\frac{1}{3}\, C_{a b c d} \check{\nabla}^2 C^{abcd}
- \frac{1}{3} C_{a b}{}^{c d} C_{c d}{}^{e f} C_{e f}{}^{a b}
- \frac{4}{3} C_{a b c d} C^{a e c f} C^b{}_e{}^d{}_f 
\nonumber\\
&\quad\,
- \mathcal{R}_{ab}\,^{i j} \check{\nabla}^2 \mathcal{R}^{ab}\,_{i j}
- 2\, \mathcal{R}_{a}{}^{b}\,_{i}{}^{j} \mathcal{R}^a{}_{c}\,_{j}{}^{k} \mathcal{R}_{b}{}^{c}\,_{k}{}^{i}
+ 2 \, C^{a b c d} \,\mathcal{R}_{a b}\,^{i j} \mathcal{R}_{c d}\,_{ij}
\nonumber\\
&\quad\,
+ \hat{\mathfrak f}_{a}{}^b (
	\frac{32}{3}\, C^{acde} C_{b c d e} 
	- 8\, \mathcal{R}_{bc}\,^{i j} \mathcal{R}^{ac}\,_{ij}
	) 
- 4\, \hat {\mathfrak f}_{a}{}^a (C_{bcde} C^{bcde} - \mathcal{R}_{bc}\,^{i j} \mathcal{R}^{bc}\,_{ij})
\nonumber\\
&\quad\,
+ \frac{4}{45}\, D \check{\nabla}^2 D
+ \frac{8}{225} \, D^3
+ \frac{2}{15} D\, C_{abcd} C^{abcd}
- \frac{14}{15} D\, \mathcal{R}_{a b}\,^{i j} \mathcal{R}^{a b}\,_{ij}
\nonumber\\
&\quad\,
+ \frac{20}{3} T^{a b e} C_{a b}{}^{c d} \check{\nabla}^f C_{f e c d}
+ 4  \,T^{abe} \check{\nabla}^f C_{ab}{}^{cd} C_{fecd}
\nonumber\\
&\quad\,
+ 2 \, T_{a b c} \check{\nabla}_{d}{\mathcal{R}^{a b}\,^{i j}}\,  \mathcal{R}^{c d}\,_{ij}
+ 4 \, T_{a b c} \check{\nabla}_{d}{\mathcal{R}^{a d}\,^{i j}}\,  \mathcal{R}^{b c}\,_{ij}
\nonumber\\
&\quad\,
- 4 \, T_{a b c} \Delta^4 T^{a b c}
- \frac{16}{3}  C_{a b c d} T^{a b e} \check{\nabla}_{e}{\check{\nabla}_{f}{T^{c d f}}\, }\,  
- \frac{8}{3}  C_{a b c d} T^{a b e} \check{\nabla}_{f}{\check{\nabla}_{e}{T^{c d f}}\, }\,  
\nonumber\\
&\quad\,
+ \frac{16}{3}  C_{a b}{}^{c d} T^{a e f} \check{\nabla}^{b}{\check{\nabla}_{c}{T_{d e f}}\, }\,
- 4  \, C_{a b}{}^{c d} \check{\nabla}^{a}{T^{b e f}}\,  \check{\nabla}_{c}{T_{d e f}}\, 
- 6 \, C_{a b c d} \check{\nabla}_{e}{T^{a b f}}\,  \check{\nabla}_{f}{T^{c d e}}\,  
\nonumber\\
&\quad\,
- \frac{16}{15} \, D\, T_{a b c} \check{\nabla}^{a}{\check{\nabla}_{d}{T^{b c d}}\, }
+ \frac{8}{15} D\, \check{\nabla}^{a}{T_{a b c}}\,  \check{\nabla}_{d}{T^{b c d}}\, 
\nonumber\\
&\quad\,
- 2 \, T_{a b c} T^{a d e} \mathcal{R}^{b c}\,_{i j} \mathcal{R}_{d e}\,^{ij}
- \frac{4}{3} \, C_{a b e f} C^{c d e f} T^{a b g} T_{c d g}
\nonumber\\
&\quad\,
- \frac{1}{2}\,\check{\nabla}^{a_1}{T_{a_1 ab}}\,  \check{\nabla}^{a_2}{T_{a_2 cd}}\,  
\check{\nabla}^{a_3}{T_{a_3 ef}}\,  \varepsilon^{abcdef} 
- 6\, T_{ab}{}^g \check{\nabla}^{a_1}{T_{a_1 g c}}\,  \check{\nabla}_{d}{\check{\nabla}^{a_2}{T_{a_2 e f}}\, }\,  \varepsilon^{abcdef} 
\nonumber\\
&\quad\,
+ 8 \,C_{a b c d} T^{e c d} T_{e f g} \check{\nabla}^{a}{T^{b f g}}\,  
+ \frac{10}{3}\, T_{a b c} T^{a e d} T^{b f}{}_d \check{\nabla}^2 T^{c}{}_{e f}
\,
- 2\, T_{a b c} T^{a b d} \check{\nabla}^{c}{T_{d e f}}\,  \check{\nabla}_{g}{T^{e f g}}\,  
\nonumber\\
&\quad\,
+ 4 \, T_{a b c} T^{a}{}_{d e} \check{\nabla}^{f}{T^{b d g}}\,  \check{\nabla}_{f}{T^{c e}{}_{g}}\,  
+ 2 \, C_{a b c d} T^{a b e} T^{c f g} T^{d}{}_{f h} T_{eg}{}^h 
+ \frac{8}{15} D \,T_{a b c} T^{a b d} T^{c e f} T_{d e f}
\,,
\end{align}
where \(T^{abc} \Delta^4 T_{abc}\) is defined as
\begin{align}
T^{abc} \Delta^{4} T_{a b c} &:= T^{abc}
\Big(\check{\nabla}_{a} \check{\nabla}^{d} \check{\nabla}^2 T_{b c d}
+ \check{\nabla}^2 \check{\nabla}_{a} \check{\nabla}^{d} T_{b c d}
\nonumber\\
&\quad\,
+ \tfrac{1}{3}\, \check{\nabla}_{a} \check{\nabla}^2 \check{\nabla}^{d} T_{b c d}
- \tfrac{4}{3}\, \check{\nabla}_{e} \check{\nabla}_{a} \check{\nabla}^{d} \check{\nabla}^{e} T_{b c d}\Big)
\,.
\end{align}

\section{Relevant facts about the heat kernel}
\label{app:HK}

A standard representation for the determinant of a second-order differential operator  \(\Delta\) is 
\begin{equation}\label{waa}
\log \det \Delta
= - \int  \dd{^dx} \sqrt{g} \intl_\varepsilon^\infty \frac{dt}{t}\,  \tr \braket{x |  e^{-t \Delta}   |x }\, , 
\end{equation}
where tr is the trace over internal indices of the operator and $\varepsilon = \Lambda^{-2}$ is a UV cutoff. The matrix element in the integrand is the heat kernel. It  has  an  asymptotic expansion for $t \to  0^+$    that allows us to write  (see e.g.~\cite{Barvinsky:1985an,Vassilevich:2003xt,Avramidi:2000bm,Casarin:2021fgd})
\begin{equation}\label{kag}
\begin{split}
&{(\log \det \Delta)_\infty }
 =
    - \frac{2}{(4\pi)^{d/2}}
    \log \frac{\Lambda}{\mu}  \int \dd{^d x \sqrt{g}   } b_d(\Delta)\,, 
 \end{split}
 \end{equation}
where \(\mu \)  is a renormalization scale.
In \eqref{kag} we only focused on the logarithmic divergence, responsible for the conformal anomaly, dropping power-law divergences.
The density  \(b_d(\Delta)\) is defined modulo total derivatives and has the form of a trace over internal indices of covariant quantities depending on the differential operator and on the geometric background.
In dimensional regularisation one has  \eqref{kag} with the formal substitution \(\log \frac{\Lambda}{\mu} \to \frac{1}{n-d} \), where \(n\) is the original integer number of dimensions. 
 
Consider the most general minimal second-order differential operator with generically an internal (gauge) and a spacetime connection
\begin{align}
\label{kad}
\Delta_2 &= - \cd^2 + X\,,
\end{align}
where
\(X\) is a   covariant coefficient function,   in general matrix-valued, and the covariant derivative contains both gauge and spacetime connections with curvatures \(\WW_{mn} = [\cd_m,\cd_n]\) and \(R_{mnrs}\) (as in \eqref{opa}) for internal and spacetime indices respectively.
The operator \(\Delta_2\) in
\eqref{kad}  is minimal in the sense that the highest-derivative term is proportional to the squared Laplacian.

For second-order differential operators of the form \eqref{kad} we have, following
 \cite{Gilkey:1975iq,Avramidi:1990je,Avramidi:2000bm},
\begin{align} \label{app2}
\begin{aligned}
	b_6(\Delta_2)
=& \tr\Big[  	 
		- \frac{1}{60} \left( \cd_m \WW_{mn}  \right)^2
		+ \frac{1}{90} \WW_{mn} \WW_{nr} \WW_{rm}
		- \frac{1}{12} X \WW_{mn} \WW_{mn}
		+ \frac{1}{12} X \cd^2 X
		- \frac{1}{6} X^3  
				+\frac1 {12} R X^2 
\\
& \qquad
		- \frac{7}{540} R^2 X  
		- \frac{1}{180} X R_{mnrs}R_{mnrs} 
					+\frac{17}{1080} R \WW _{mn} \WW^{mn} 
+ \frac{1}{180} R_{mnrs} \WW_{mn} \WW_{rs}
	+\mathbbm 1  \,  \mathfrak E     \, 
		\Big] ,
\end{aligned}
\end{align}
where \(\1\) is the identity in the internal space, where \(\tr\) acts, and 
\begin{equation}\label{geom}
\begin{aligned}
\mathfrak E 
={}
& 
		 \frac{17}{45360} R\indices{_{mn}^{pq}}  R\indices{_{pq}^{rs}}  R\indices{_{rs}^{mn}} 
		- \frac{1}{1620} R\indices{_m^p_n^q}  R\indices{_p^r_q^s}  R\indices{_r^m_s^n}   
		+ \frac{31}{51030} R^3
		+ \frac{43}{45360} R R_{mnrs} R^{mnrs} 
\end{aligned}
\end{equation}
is a purely geometrical contribution.

In this paper we are interested in a geometric background only. In this case the formulae above apply with the following values:
\begin{equation}\label{dfsf}
\begin{aligned}
\text{symmetric 2-tensor} \quad & {2} & &: \quad & &   [\WW_{mn}]\indices{_{ac}^{rs}} = 
2 g^{   (  r   }_{  (  c }     R\indices{ _{  a) } ^{  s) }   _{ m n } }
  \,,\quad & & \tr \mathbbm 1 = 21\,,
\\
\text{traceless symm.\ 2-tensor} \quad 
& {2^0} & &: \quad & &   [\WW_{mn}]\indices{_{ac}^{rs}} = 
2 g^{   (  r   }_{  (  c }     R\indices{ _{  a) } ^{  s) }   _{ m n } }
  \,,\quad & & \tr \mathbbm 1 = 20\,,
\\
\text{2-form}\quad 
&2{\text f}& &:  \quad & & [\WW_{mn}]\indices{_{ac}^{rs}} = g_{  [a }^{[r } R\indices{_ {c]} ^{s]}  _{ mn}}\,,\quad & & \tr \mathbbm 1 = 15\,,
\\
\text{vector}\quad 
&1& &:      & & [\WW_{mn}]\indices{_a^c} =   R\indices{_a^c _{mn}}\,,\quad & & \tr \mathbbm 1 =  6\,,
\\
\text{scalar}\quad 
&0 & & :      & & \WW_{mn} = 0\,,\quad & & \tr \mathbbm 1 = 1\,,
\\
\text{vector-spinor}\quad 
&{\tfrac32} & & :      & & [\WW_{mn}]\indices{_a^c} = \frac14 g_{a}^c R_{mnrs}\Gamma^{rs} + R\indices{_a^c _{mn}} \, \mathbb I_8\,,\quad & & \tr \mathbbm 1 = 48 \,,
\\
\text{spinor}\quad 
&{\tfrac12} & & :      & & \WW_{mn} =  \frac14 R_{mnrs}\Gamma^{rs}\,,\quad & & \tr \mathbbm 1 = 8\,.
\end{aligned}
\end{equation} 
Some  explanation is in order.
The identity  for symmetric tensors \(2\) is
\begin{equation}\label{fsf}
S_{mn}^{rs} =  g_{(m}^{r}g_{n)}^s \qquad\quad
\tr \mathbbm 1 = S_{mn}^{mn} = \frac{ d(d+1)}2\,.
\end{equation}
For the symmetric traceless tensors \(2^0\), the identity   is the projector on traceless symmetric 2-tensors
\begin{equation}\label{p20}
P_{mn}^{rs} =    g_{(m}^{r}g_{n)}^s   - \frac 1d g_{mn}g^{rs}  ,
\qquad\quad
\tr \mathbbm 1 = P_{mn}^{mn} =  S_{mn}^{mn} -1= \frac{ (d-1)(d+2)}2\,.
\end{equation} 
The curvature \(\WW_{mn}\) is insensitive to the tracelessness condition, i.e.\ it is the same for \(2\) and \(2^0\), because \([\cd_m,\cd_n]\) commutes with the contraction with the metric.
For rank-2 antisymmetric tensors (\(2\text f\)) the identity is 
\begin{equation}\label{p2f}
A_{mn}^{rs} = g^{  r }_{  [m  }  g^{  s }_{  n] }\,,
\qquad\qquad
\tr \mathbbm 1 = A_{mn}^{mn} = \frac{ d (d-1)}2\,.
\end{equation}

\section{Additional details on gauge fixing and determinants}
\label{app:jac}
In this appendix we give more technical details and explicit expressions for the Jacobians and the calculation of constrained determinants. We keep the dimension \(d\) generic.

\subsection{Jacobian factors}
In the evaluation of the functional integrals and of the determinants we made use of the following change of variables
\bsubeq
\begin{align}\label{jkf1}
A_m &= A_m^\perp + \cd_m \sigma 
\,, \\ \label{jkf2}
h_{mn} & = h_{mn}^\perp 
		+\cd_m A_n^\perp
		+\cd_n A_m^\perp
		+ \cd_m \cd_n \sigma
		- \frac 1d g_{mn} \cd^2 \sigma
		+ \frac 1 d g_{mn} h
\,, \\ \label{jkf20}
h^0_{mn} & = h_{mn}^\perp 
		+\cd_m A_n^\perp
		+\cd_n A_m^\perp
		+ \cd_m \cd_n \sigma
		- \frac 1d g_{mn} \cd^2 \sigma 
\,, \\ \label{jkf3}
V_{mn} &=
	V_{mn}^\perp + \cd_{ [ m } A^\perp_{  n]}
\,, \\ \label{jkf4}
T_{mnr} &= \cd_{  [m } V^\perp_{ nr]  } + \varepsilon_{mnrace} \cd^{ [ a }   W^{\perp\,   ce]  }
\,, 
\qquad
\qquad
(d=6 \text{ only})\\ \label{jkf5}
\psi_m & = 
	 \psi_m^\perp + \Gamma_m \varphi + \cd_m \zeta - \frac1d \Gamma_m \slashed \cd \zeta\,,
\end{align}
\esubeq
where \(A_m\) is a vector, \(h_{mn}\) a symmetric tensor, \(h^0_{mn}\)  a symmetric traceless tensor, \(V_{mn}\) and \(W_{mn}\) 2-forms, \(T_{mnr}\) a 3-form, \(\psi_m\) a  Majorana vector-spinor, \(\sigma\) and \(h\) scalars, \(\varphi\) and \(\zeta\)   Majorana spinors.
The symbol \(\perp\) means that the field is covariantly transverse and (gamma-)traceless. For the 3-form \(T\), which manifestly makes sense only in six dimensions, we  ignore the self-duality condition here, so the exact and co-exact parts are independent. Spinor variables are described in the Dirac gamma-matrix notation. 

We need to find the Jacobian associated to the transformations.
The strategy is the same for all of them and we focus on some relevant examples. We start with the vector transformation and consider the path integral
\begin{equation}\label{fnd}
1= \int \DD{A_1} e^{-\int A_m A^m} 
= J_1 \int \DD{A^\perp_1} \DD{\sigma} e^{- \int A_m^\perp A^{\perp m} +  \sigma [-\cd^2] \sigma }
=  J_1 \Big[\textstyle \det_0 \Delta_0[0] \Big]^{-\frac12}\,,
\end{equation}
where we used that the path integral over nondynamical fields is \(1\) and denoted by \(J_1\) is the Jacobian associated to the transformation \eqref{jkf1}. In order to perform the integral over \(\sigma\) one needs to complexify the field \(\sigma \mapsto {\rm i }\sigma\); this is the typical case and will be  always understood. Therefore, \eqref{fnd} allows us to express \(J_1\) in terms of the (known) determinant of the scalar Laplacian. 

The same  idea applies directly to \(h_{mn}\) and \(V_{mn}\). In the case of the 3-form we have (up to total derivatives)
\begin{equation}\label{fds}
T_{mnr}T^{mnr} 
	= 
			\tfrac13 V^{mn} \, \Delta_{2\text f}[\tfrac 4 {15}] \,V_{mn}  
			+{12} \,  W^{mn} \, \Delta_{2\text f}[\tfrac 4 {15}] \, W_{mn}
	\,,
 \qquad\qquad (d=6)
\end{equation}
and hence the the exact and co-exact parts contribute equally and independently.
For the fermionic field the situation is similar,
\begin{equation}\label{pyk}
1 = \int \DD{\psi_{\frac32}} e^{\int \bar \psi_m   \psi^m}
	= J _\frac32
		\int \DD{\psi^\perp_{\frac32}} 
		   \DD{\varphi}   
		  \DD{\zeta}    e^{\int \bar \psi_m   \psi^m}
	= J_\frac32    \Big[ \textstyle \det_{\frac12} \Delta_{\frac12}[-\frac1{4(d-1)}]  \Big]^{\frac12}
	\,,
\end{equation}
where the exponent is \(\frac12\) because we assume Weyl or Majorana spinors in gamma-matrix notation.

We therefore obtain the following Jacobians
\bsubeq
\begin{align}
\label{jac1}
A_m & \to (A_m^\perp,\sigma)	& : & &
J_1
		& =  \Big[\textstyle \det_{ 0 } \Delta_0[0] \Big]^{\frac12}
  \,,
\\
\label{jac2}	
h_{mn} & \to (h_{mn}^\perp,A_m^\perp,\sigma,h)	&  : & &
	J_2
			& = \Big[
								 \textstyle \det_{1 \perp}\Delta_1[-\tfrac1d]
								 				\	 \det_0 \Delta_0[0]     \ 
								 \textstyle \det_0 \Delta_0[-\tfrac 1{d-1}]    
						\Big]^{\frac12}
      \,,
\\
\label{jac20}
h^0_{mn} & \to (h_{mn}^\perp,A_m^\perp,\sigma)	&  : & &
	J_{2^0}
			& = \Big[
						 \textstyle \det_{1 \perp}\Delta_1[-\tfrac1d] 
						 				\	 \det_0 \Delta_0[0]    \ 
							 \textstyle \det_0 \Delta_0[-\tfrac 1{d-1}]   
						\Big]^{\frac12}
      \,,
\\
\label{jac3}
V_{mn} & \to (V_{mn}^\perp,A_m^\perp)	&  : &  &
	J_{2\text f}
			& =  \Big[\textstyle \det_{1 \perp} \Delta_1[\frac1d] \Big]^{\frac12}
   \,,
\\
\label{jac4}
T_{mnr} & \to (V_{mn}^\perp,W_{mn}^\perp)	&  : &  &
	J_{3\text f}
			& =  \textstyle \det_{2\text f \perp}  \Delta_{\text 2 \text f}[ \tfrac4{15} ]
   \,,
   \qquad\qquad(d=6)
\\
\label{jac5}
\psi_{m} & \to (\psi_{m}^\perp,\varphi,\zeta)	&  : &  &
	J_{\frac32}
			& =   \Big[ \textstyle \det_{\frac12}  \Delta_{\frac12}[-\frac1{4(d-1)}]  \Big]^{-\frac12}
   \,.
\end{align}
\esubeq
The Jacobians \(J_2\) and \(J_{2^0}\) are the same because the contribution  \(h\)  in \eqref{jkf2}, which carries the trace of \(h_{mn}\), does not come with a differential operator.

\subsection{Un-constraining determinants}\label{app:detp}
The changes of variables \eqref{jkf1}-\eqref{jkf5} typically produce determinants restricted to transverse operators, like in \eqref{jac1}-\eqref{jac5}. We want now to relax this restriction. This is standard, but we spell it out for completeness.
The idea is to consider the unconstrained determinant represented as a path integral and then change variables. For this, it is useful to notice (up to total derivatives)
\bsubeq
\begin{align} 
 \label{rpe1}
 A^m \Delta_1[r]	A_{m}
 	& =		A^{\perp m} \Delta_1[r]	A^\perp_{m}
 				 + \sigma \Delta_0[0] \, \Delta_0[r-\tfrac1d ] \,\sigma
\,,
 \\
 \label{rpe2}
h^{mn} \Delta_2[r] h_{mn}	
 	& =  
 			h^{0\,mn} \Delta_2[r] h^0_{mn}	
 				+ 2 A^{\perp m} \Delta_1[ \tfrac{-1}{2 d}]\, \Delta_1[r- \tfrac{1}{ d}]A_m
 				+ \tfrac{d-1}d \sigma \Delta_0[0] \, \Delta_0[\tfrac1{1-d}] \Delta_0[r-\tfrac2d] \sigma
     \,,
 \\
 \label{rpe2f}
V^{mn} \Delta_{2 \text f}[r]	V_{mn}
 	& =  V^{\perp mn} \Delta_{2 \text f}[r]	V^\perp_{mn}
 			 +A^{\perp m} \Delta_1[ \tfrac1d] \  \Delta_1[r - \tfrac{d-3}{d (d-1)}  ]  \ A_m^\perp
     \,,
 \\
 \label{rpe32}
 \psi^m\Delta_{\frac32}[r] \psi_{m}	
 	& =  \psi^{\perp m}\Delta_{\frac32}[r] \psi^\perp_{n}	
 			+ d \bar \varphi \Delta_{\frac12}[r-\tfrac1d] \varphi 
 			- \frac{d-1}{d} \bar \zeta \Delta_{\frac12}[r-\tfrac1d]  \Delta_{\frac12}[\tfrac{-1}{4(d-1)}] \zeta
    \,.
\end{align}
\esubeq
For example, in the case of vector operator \( \Delta_1\) we can consider
\begin{equation}\label{sk}
\Big[
{	\textstyle \det_1 \Delta_1[r]  }
\Big]^{-\frac12}
=
\int \DD{A} e^{-\int A^m \Delta_1[r] A_m}
=
J_1
\Big[
{	\textstyle \det_{1\perp} \Delta_1[r]  } \ 
{	\textstyle \det_0 \Delta_0[0] }\ 
{	\textstyle \det_0 \Delta_0[r-\tfrac1d] }
\Big]^{-\frac12}\,,
\end{equation}
which in turn gives an expression for  \(\det_{1\perp} \Delta_1[r] \) in terms of known  quantities.
In particular, we observe that the Jacobian factor \(J_1\)  given in \eqref{jac1}   cancels against \(r\)-independent contributions from \eqref{rpe1}. This is a typical phenomenon.
One proceeds in the same way for the other cases too and  we  finally have
\bsubeq
\begin{align}  
\label{pr1}
	{	\textstyle \det_{1\perp} \Delta_1[r]  }
		&=   
		 	{	\textstyle \det_1 \Delta_1[r]  } 
			\ \Big[ {	\textstyle \det_0 \Delta_0[r-\tfrac1d]} \Big]^{-1} 
			\,,
\\
\label{pr2}
		{	\textstyle \det_{2\perp} \Delta_2[r]  }
			&=  
		{	\textstyle \det_{2^0 } \Delta_2[r]  }
			\ \Big[ {\textstyle 
				 \det_{1\perp} \Delta_1[r-\frac1d] 
							\   \det_{0} \Delta_0[r-\frac2{d}]
												  } \Big]^{-1}
												  			\,,
\\
\label{pre3}
		{	\textstyle \det_{2\text f\perp} \Delta_{2\text f}[r]  }
					&=  
		{	\textstyle \det_{2\text f } \Delta_{2\text f}[r]  }
			\ \Big[ { 
				\textstyle \det_{1\perp} \Delta_1[r-\frac{d-3}{d(d-1	)}] 
			 } \Big]^{-1}
			 			\,,
\\
\label{pr4}
		{	\textstyle \det_{\frac32 \perp} \Delta_{\frac32}[r]  }
					&=  
							{	\textstyle \det_{\frac32 } \Delta_{\frac32}[r]  }
		\ \Big[ {	 \textstyle
			  \det_\frac12 \Delta_\frac12[ r - \frac1d ]   
				} \Big]^{-2}
							\,.
\end{align}
\esubeq
These formulae allow one to relate transverse determinants to non-transverse ones and determinants of lower-rank representations, so iterating this procedure one ends up with unconstrained quantities. In \eqref{pr2} we expressed the determinant \(\det_{2 \perp}\Delta_2\)   in terms of \(\det_{2^0}\Delta_2\) taken on the space of \emph{traceless} symmetric tensors (as in \eqref{fin}); relaxing this algebraic restriction  to obtain \(\det_{2}\Delta_2\) produces an additional  scalar determinant associated to the trace mode \(h\).

The procedure outlined above does not immediately extend to   the determinant of the Dirac operator acting on the gravitino \(\det_{\frac32\perp}\rm i\slashed \cd\). Indeed, as discussed in \eqref{dop}, the Dirac operator does not preserve transversality and gamma-tracelessness separately, namely it mixes the irreducible representations  \(\varphi,\zeta\) and therefore \(\mathrm i \bar \psi_m\slashed\cd \psi^m \) with \eqref{jkf5} is not diagonal in the fields. However we can consider the square \((\mathrm i\slashed \cd)^2= \Delta_{\frac32}[\frac14]\), which  does have the structure studied above, and thereby use the relation
\begin{align}\label{pyl}
\textstyle
 \det_{\frac32 \perp}  \mathrm i \slashed   \cd 
 & =
 \textstyle
   \Big[  \det_{\frac32 \perp} (\mathrm i  \slashed   \cd )^2  \Big]^{\frac12}
 =  \Big[ \det_{\frac32 } \Delta_\frac32[\tfrac14]  \Big]^\frac12
 \Big[ \det_{\frac12  } \Delta_\frac12[\tfrac14-\frac1d]  \Big]^{-1}\,,
\end{align}
which follows from  \eqref{pr4}.

\section{Conformal invariant four-derivative 3-form}\label{app:T3}

In this appendix we report additional study of non-gauge 3-form fields in six dimensions, mostly regardless of supersymmetry. In analogy with the case of interest in the main text, we consider a 4-derivative theory, so that the 3-form \(T_{mnr}\) has dimension \(1\) and Weyl-weight \(2\).

We first find a basis of   the possible dimension-6 scalars that are quadratic in \(T_{mnr}\) and built out of covariant derivatives and curvature tensors, so that a generic term is expressed in terms of this basis considering symmetries of the tensors and Bianchi identities. 

A careful case-by-case analysis provides the  68 invariants \(\{B_i\}\) given in Table~\ref{tab:inv}.  
We find working with Weyl tensor more convenient than using the Riemann tensor, because the former is an irreducible representation and this facilitates the analysis.
A generic dimension-6 scalar has therefore the structure \(\sum_i b_i B_i\) with coefficients \(b_i\).

From  \(\{B_i\}\) we then find a basis for the possible contribution to the action, namely we also allow for integration by parts. This produces a basis \(\{\Lagr_{T,i}\}\) of  35 terms, listed in Table~\ref{tab:act}. 
A generic action is therefore a combination of the form
\begin{equation}\label{tca}
S = \int \sqrt{g}\ \sum_i a_i \, \Lagr_{T,i}
\,,
\end{equation}
where there are \(35\)   coefficients  \(a_i\) (however one, say \(a_1\), is just an overall scaling).

\begin{table}
\centering 
\begin{tabular}{cc|cc|cc}
\(b_{	1	}\)	&	\(	 T_{mnr}\cd^4 T_{mnr}	\)  &	\(b_{	24	}\)	&	\(	T^{a  m   n } C_{  m   n   r   s }     \cd^{  s }   \cd_{b}T_{a b  r }	\)  &	\(b_{	47	}\)	&	\(	R^{a  m } T_{a   n b}     \cd^2 T_{  m   n b} 	\)  \\
\(b_{	2	}\)	&	\(	T^{mnr}    \cd_{m}   \cd^2   \cd_{a}T^{a nr} 	\)  &	\(b_{	25	}\)	&	\(	T^{a  m   n } C_{  n   r b  s }     \cd^{  s }   \cd^{  r }T_{a  m  b}	\)  &	\(b_{	48	}\)	&	\(	R^{a  m }    \cd_{a}T^{  n b  r }     \cd_{  m }T_{  n b  r } 	\)  \\
\(b_{	3	}\)	&	\(	\cd_{m}T^{mnr}    \cd^2  \cd_{a}T^{a nr} 	\)  &	\(b_{	26	}\)	&	\(	T_{mnp} T^{mnp} C_{abcd}  C^{abcd} 	\)  &	\(b_{	49	}\)	&	\(	R    \cd_{a}T^{a  m   n }     \cd_{b}T_{  m   n  b} 	\)  \\
\(b_{	4	}\)	&	\(	\cd_{a}T^{mnr}    \cd^2   \cd_{m}T^{a nr} 	\)  &	\(b_{	27	}\)	&	\(	T^{mab} T^{n ab} C_{m rsp} C_{nrsp} 	\)  &	\(b_{	50	}\)	&	\(	R    \cd_{  n }T_{a  m b}     \cd^{b}T^{a  m   n } 	\)  \\
\(b_{	5	}\)	&	\(	\cd_{m}T^{anr}    \cd^{m}   \cd^2 T_{anr} 	\)  &	\(b_{	28	}\)	&	\(	T_{a mn} T^{ars} C_{mn pq} C_{rspq} 	\)  &	\(b_{	51	}\)	&	\(	R    \cd_{b}T_{a  m   n }    \cd^{b}T^{a  m   n } 	\)  \\
\(b_{	6	}\)	&	\(	\cd_{m}T^{cnr}    \cd^{m}   \cd_{c}   \cd_{a}T^{a nr}	\)  &	\(b_{	29	}\)	&	\(	T_{a mn} T^{ars} C_{mrpq} C_{ns pq} 	\)  &	\(b_{	52	}\)	&	\(	R^{a  m }    \cd_{  n }T_{a   n b}    \cd_{  r }T_{  m b   r } 	\)  \\
\(b_{	7	}\)	&	\(	\cd^2T^{abc}    \cd^2 T_{abc} 	\)  &	\(b_{	30	}\)	&	\(	T_{a mn} T^{ars} C_{mprq}  C_{n p s q}	\)  &	\(b_{	53	}\)	&	\(	R^{a  m }    \cd_{  m }T_{a   n b}    \cd_{  r }T_{  n b   r } 	\)  \\
\(b_{	8	}\)	&	\(	\cd_{m}   \cd^{n}T^{abc}    \cd_{n}   \cd^{m}T_{abc} 	\)  &	\(b_{	31	}\)	&	\(	T^{abc} T^{rsq} C_{masq} C^{m rbc} 	\)  &	\(b_{	54	}\)	&	\(	R^{a  m }    \cd_{  m }T_{  n b  r }    \cd^{  r }T_{a   n b} 	\)  \\
\(b_{	9	}\)	&	\(	\cd_{a}   \cd^{n}T^{abc}    \cd_{n}   \cd^{m}T_{mbc} 	\)  &	\(b_{	32	}\)	&	\(	R^2 T_{a  m   n } T^{a  m   n } 	\)  &	\(b_{	55	}\)	&	\(	R^{a  m }  \cd_{b}T_{  m   n   r }    \cd^{  r }T_{a   n b} 	\)  \\
\(b_{	10	}\)	&	\(	\cd^{m}   \cd^{n}T^{abc}    \cd_{n}   \cd_{a}T_{mbc} 	\)  &	\(b_{	33	}\)	&	\(	R^{a  m } R T_{a   n b} T_{  m   n b}	\)  &	\(b_{	56	}\)	&	\(	R^{a  m }     \cd_{  r }T_{  m   n b}    \cd^{  r }T_{a   n b} 	\)  \\
\(b_{	11	}\)	&	\(	\cd_{a}   \cd_{r}T^{abc}    \cd_{b}   \cd_{m}T^{mr c} 	\)  &	\(b_{	34	}\)	&	\(	R^{a  m } R^{  n b} T_{a  n    r } T_{  m b  r }	\)  &	\(b_{	57	}\)	&	\(	T^{  m   n b}     \cd_{a}T_{  m   n b}    \cd^{a}R 	\)  \\
\(b_{	12	}\)	&	\(	\cd_{a}   \cd_{r}T^{r bc}    \cd^2 T^{abc} 	\)  &	\(b_{	35	}\)	&	\(	R_{a   n } R^{a  m }  T_{  m  b  r } T_{  n b  r } 	\)  &	\(b_{	58	}\)	&	\(	T_{  n  b  r }     \cd_{  m }T_{ab  r }    \cd^{  n }R^{a  m } 	\)  \\
\(b_{	13	}\)	&	\(	T^{a  m   n }    \cd_{  s }C_{  m   n b  r }    \cd^{  s }T_{a b  r }	\)  &	\(b_{	36	}\)	&	\(	R_{a  m } R^{a  m } T_{  n b  r } T^{  n b  r } 	\)  &	\(b_{	59	}\)	&	\(	T_{a b  r }     \cd_{  m }T_{  n b  r }    \cd^{  n }R^{a  m } 	\)  \\
\(b_{	14	}\)	&	\(	C_{  m   n   r   s }    \cd_{a}T^{a  m   n }    \cd_{b}T^{b  r   s } 	\)  &	\(b_{	37	}\)	&	\(	R^{mn} T^{rac} T^{s ac} C_{mrns} 	\)  &	\(b_{	60	}\)	&	\(	T_{a b  r }     \cd_{  n }T_{  m b  r }    \cd^{  n }R^{a  m } 	\)  \\
\(b_{	15	}\)	&	\(	C_{  m   n   r   s }    \cd_{a}T_{b   r   s }    \cd^{b}T^{a  m   n }	\)  &	\(b_{	38	}\)	&	\(	R_{mn} T^{mab} T^{ncd}  C_{abcd}	\)  &	\(b_{	61	}\)	&	\(	T^{  m   n b}    \cd^{a}R     \cd_{b}T_{a  m   n } 	\)  \\
\(b_{	16	}\)	&	\(	C_{  m   n   r   s }    \cd_{b}T_{a   r   s }    \cd^{b}T^{a  m   n }	\)  &	\(b_{	39	}\)	&	\(	R_{mn} T_{c rs} T^{mac} C^{n ars} 	\)  &	\(b_{	62	}\)	&	\(	T_{a   m   n }    \cd^{a}R     \cd_{b}T_{  m   n  b} 	\)  \\
\(b_{	17	}\)	&	\(	C_{mnrs}    \cd^{m}T^{nac}    \cd^{r}T^{s ac} 	\)  &	\(b_{	40	}\)	&	\(	R T^{mna}  T^{pq a} C_{mnpq} 	\)  &	\(b_{	63	}\)	&	\(	T_{a b  r }    \cd^{  n }R^{a  m }     \cd_{  r }T_{  m   n b} 	\)  \\
\(b_{	18	}\)	&	\(	C_{mrns}     \cd^{m}T^{nac}    \cd^{r}T^{s ac}	\)  &	\(b_{	41	}\)	&	\(	R^{a  m } T^{  n b  r }     \cd_{  m }   \cd_{a}T_{  n b  r } 	\)  &	\(b_{	64	}\)	&	\(	T_{a  n  b}    \cd^{  n }R^{a  m }     \cd_{  r }T_{  m b   r } 	\)  \\
\(b_{	19	}\)	&	\(	C_{  m   n   r   s }     \cd^{b}T^{a  m   n }    \cd^{  s }T_{ab   r } 	\)  &	\(b_{	42	}\)	&	\(	R T^{a  m   n }     \cd_{b}   \cd_{  n }T_{a  m  b} 	\)  &	\(b_{	65	}\)	&	\(	T_{  m   n b} T^{  m   n b}     \cd^2 R 	\)  \\
\(b_{	20	}\)	&	\(	C_{  n   s b  r }     \cd_{a}T^{a  m   n }    \cd^{  s }T_{  m  b  r } 	\)  &	\(b_{	43	}\)	&	\(	R T^{a  m   n }     \cd^2 T_{a  m   n } 	\)  &	\(b_{	66	}\)	&	\(	T_{a   n b} T_{  m   n b}     \cd^{  m }   \cd^{a}R 	\)  \\
\(b_{	21	}\)	&	\(	T^{a  m   n }  C_{  m   n   r   s }    \cd_{a}   \cd_{b}T^{b  r   s } 	\)  &	\(b_{	44	}\)	&	\(	R^{a  m } T^{  n b  r }     \cd_{  r }   \cd_{  m }T_{a  n b}	\)  &	\(b_{	67	}\)	&	\(	T_{a b  r } T_{  m b  r }     \cd^2 R^{a  m }	\)  \\
\(b_{	22	}\)	&	\(	T^{a  m   n } C_{  m   n b  r }     \cd^2 T_{a b  r }	\)  &	\(b_{	45	}\)	&	\(	R^{a  m } T_{a   n b}     \cd_{  m }   \cd_{  r }T_{  n b   r }	\)  &	\(b_{	68	}\)	&	\(	T_{a  n    r } T_{  m b  r }     \cd^{b}   \cd^{  n }R^{a  m }	\) \\
\(b_{	23	}\)	&	\(	T^{a  m   n } C_{  n   s b  r }     \cd^{  s }   \cd_{  m }T_{a b  r }	\)  &	\(b_{	46	}\)	&	\(	R^{a  m } T_{a   n b}     \cd_{b}   \cd_{  r }T_{  m   n    r } 	\)  &							
\end{tabular}
\caption{Algebraically independent dimension-six monomials.}
 \label{tab:inv}
\end{table}

\begin{table}
\centering
\begin{tabular}{cc|cc|cc}
\(	a_1	\)	&	\(	  T^{mnr}  \cd^4T_{mnr}   	\)	&	
\(	a_{13}	\)	&	\(	  T^{abc} T^{rsq} C_{masq} C^{m  rbc}   	\)	&	
\(	a_{25}	\)	&	\(	  R  \cd_{a}T^{a m  n }  \cd_{b}T_{ m  n   b}   	\)	\\
\(	a_2	\)	&	\(	  T^{mnr}  \cd_{m} \cd^2 \cd_{a}T^{a  nr}   	\)	&	
\(	a_{14}	\)	&	\(	  R^{mn} T^{rac} T^{s  ac} C_{mrns}   	\)	&
\(a_{26}	\)	&	\(	  R  \cd_{a}T_{ m  n   b}  \cd_{b}T^{a m  n }   	\)	\\
\(	a_3	\)	&	\(	  T^{mnr} C_{nr  ac}  \cd^2 T_{mac}   	\)	&	
\(a_ {15}	\)	&	\(	  R_{mn} T^{mab} T^{ncd} C_{abcd}   	\)	&	
\(	a_{27}	\)	&	\(	  R  \cd_{b}T_{a m  n }  \cd^{b}T^{a m  n }   	\)	\\
\(	a_4	\)	&	\(	  T^{nac} C_{acrs}  \cd_{n} \cd_{m}T^{mrs}   	\)	&	
\(	a_{16}	\)	&	\(	  R_{mn} T_{c  rs} T^{mac} C^{n  ars}   	\)	&	
\(	a_{28}	\)	&	\(	  R^{a m }  \cd_{ n }T_{a   n b}  \cd_{ r }T_{ m b   r }   	\)	\\
\(	a_5	\)	&	\(	  T^{rsa} C_{amn  c}  \cd^{m} \cd^{n}T_{rsc}   	\)	&	
\(	a_{17}	\)	&	\(	  R T^{mna} T^{pq  a} C_{mnpq}   	\)	&	
\(	a_{29}	\)	&	\(	  R^{a m }  \cd_{ n }T_{ m b   r }  \cd_{ r }T_{a   n b}   	\)	\\
\(	a_6	\)	&	\(	  T^{nrc} C_{nr  sa}  \cd_{a} \cd^{m}T_{msc}   	\)	&	
\(	a_{18}	\)	&	\(	  R^{a m } T^{ n b r }  \cd_{ m } \cd_{a}T_{ n b r }   	\)	&	
\(a_ {30}	\)	&	\(	  R^{a m }  \cd_{ r }T_{ m  n b}  \cd^{ r }T_{a   n b}   	\)	\\
\(a_	7	\)	&	\(	  T^{a m  n }  \cd_{ s }C_{ m  n b r }  \cd^{ s }T_{a  b r }   	\)	&	
\(a_{19}	\)	&	\(	  R T^{a m  n }  \cd_{b} \cd_{ n }T_{a m   b}   	\)	&	
\(	a_{31}	\)	&	\(	  R^2 T_{a m  n } T^{a m  n }   	\)	\\
\(	a_8	\)	&	\(	  T_{mnp} T^{mnp} C_{abcd} C^{abcd}   	\)	&	
\(a_{20}	\)	&	\(	  R T^{a m  n }  \cd_{b} \cd^{b}T_{a m  n }   	\)	&	
\(	a_{32}	\)	&	\(	  R^{a m } R T_{a   n b} T_{ m  n b}   	\)	\\
\(	a_9	\)	&	\(	  T^{mab} T^{n  ab} C_{m  rsp} C_{nrsp}   	\)	&	
\(a_	{21}	\)	&	\(	  R^{a m } T^{ n b r }  \cd_{ r } \cd_{ m }T_{a n b}   	\)	&	
\(	a_{33}	\)	&	\(	  R^{a m } R^{ n b} T_{a n    r } T_{ m b r }   	\)	\\
\(a_	{10}	\)	&	\(	  T_{a  mn} T^{ars} C_{mn  pq} C_{rspq}   	\)	&	
\(	a_{22}	\)	&	\(	  R^{a m } T_{a   n b}  \cd_{ r } \cd_{ m }T_{ n b   r }   	\)	&	
\(	a_{34}	\)	&	\(	  R_{a   n } R^{a m } T_{ m   b r } T_{ n b r }   	\)	\\
\(a_	{11}	\)	&	\(	  T_{a  mn} T^{ars} C_{mrpq} C_{ns  pq}   	\)	&	
\(	a_{23}	\)	&	\(	  R^{a m } T_{a   n b}  \cd_{ r } \cd_{b}T_{ m  n    r }   	\)	&	
\(a_	{35}	\)	&	\(	  R_{a m } R^{a m } T_{ n b r } T^{ n b r } 	\)	\\
\(a_	{12}	\)	&	\(	  T_{a  mn} T^{ars} C_{mprq} C_{n  p  s  q}   	\)	&	
\(a_	{24}	\)	&	\(	  R^{a m } T_{a   n b}  \cd_{ r } \cd^{ r }T_{ m  n b}   	\)	&				&					
\end{tabular}
\caption{Basis of the action, i.e.\ invariants in Table~\ref{tab:inv} up to integration by parts.}
\label{tab:act}
\end{table}

Our starting point is the expression \eqref{tca}. We  explicitly perform a Weyl transformation and we impose invariance of the action at first order. This translates into a set of equations for the \(a_i\)'s.
The Weyl transformation is
\begin{equation}\label{tba}
\delta g_{mn} = 2\sigma   g_{mn}
\,,
\qquad
\delta T_{mnr} =2\sigma  T_{mnr}
\,,
\end{equation}
and impose vanishing \(\delta S\) at first order in \(\sigma(x)\).  There are terms with up to 4 derivatives on \(\sigma\). Terms without derivatives on \(\sigma\) disappear as a consequence of rigid scale invariance. After integration to remove derivatives on \(\sigma\), the rest of the integrand, which can be expressed in terms of the basis \(\{B_i\}\), has to vanish,
\begin{equation}\label{tbb}
0=\delta S  = \int \sum_i a_i \,\Lagr_{T,i}
= \int \sqrt{g} \  \Big(  \sum_i b_i (\{a_j\}) B_i \Big) \sigma(x)
\,,\qquad  
b_i (\{a_j\})=0\,,
\end{equation}
which gives a set of equations. Explicitly,  each of the \(\Lagr_{T,i}\) in \eqref{tbb}  transforms (upon use of identities and symmetries, but not of integration by parts) in a combination of the \(B_j\)'s, producing an expression whose coefficients are combinations of \(a_i\)'s. These combinations of \(a_i\)'s must then vanish.
We find the non-unique solution
\begin{equation}\label{tbc}
\begin{gathered}
  a_{2}    =  -6  a_{1}   	   \,,\qquad
  a_{4}    =  9  a_{1}  +  a_{14}  -     \tfrac{1}{2}  a_{3}  +  a_{7}   	   \,,\qquad
  a_{5}    =  8  a_{1}  +    \tfrac{4}{3}  a_{14}  +   \tfrac{4}{3}  a_{3}  +   \tfrac{4}{3}   a_{7}   	   \,,\\
  a_{6}    =  -36  a_{1}  - 4  a_{14}  - 2  a_{3}  -  4  a_{7}   	   \,,\qquad
  a_{15}    =  -   \tfrac{9}{2}  a_{1}  -     \tfrac{1}{2}  a_{14}  -    \tfrac{7}{4}  a_{3}  +   \tfrac{1}{2}  a_{7}   	   \,,\\
  a_{16}    =  -   \tfrac{5}{2}  a_{1}  -    \tfrac{1}{6}   a_{14}  -    \tfrac{17}{12}  a_{3}  +   \tfrac{11}{6}  a_{7}   	   \,,\qquad
  a_{17}    =    \tfrac{13}{10}  a_{1}  -    \tfrac{1}{30}  a_{14}   +   \tfrac{13}{60}  a_{3}  -    \tfrac{1}{30}  a_{7}   	   \,,\qquad
  a_{18}     =   a_{1}   	   \,,\\
  a_{19}    =    \tfrac{3}{2}  a_{1}  -    \tfrac{3}{10}   a_{14}  +   \tfrac{3}{20}  a_{3}  -    \tfrac{3}{10}  a_{7}   	   \,,\qquad
  a_{20}    =  -   \tfrac{7}{10}  a_{1}  -    \tfrac{1}{10}   a_{14}   	   \,,\qquad
  a_{21}    =   a_{14}  +  a_{3}  +  a_{7}   	   \,,\\
  a_{22}    =  -6  a_{1}  -   a_{14}  -   a_{3}  -   a_{7}   	   \,,\qquad
  a_{23}    =  -9  a_{1}  -   a_{14}  -    \tfrac{3}{2}  a_{3}  +  3  a_{7}   	   \,,\qquad
  a_{24}    =   a_{14}   	   \,,\\
  a_{25}    =    \tfrac{9}{10}   a_{1}  +   \tfrac{1}{10}  a_{14}  +   \tfrac{9}{20}  a_{3}  +  \tfrac{1}{10}  a_{7}   	   \,,\qquad
  a_{26}    =  -   \tfrac{3}{5}  a_{1}  -   \tfrac{2}{5}  a_{14}  -    \tfrac{3}{10}  a_{3}  -    \tfrac{2}{5}   a_{7}   	 \,,\\
  a_{27}    =  -   \tfrac{1}{2}  a_{1}  -    \tfrac{1}{10}   a_{14}   	   \,,\qquad
  a_{28}    =  15  a_{1}  +  a_{14}  +   \tfrac{5}{2}  a_{3}  -   a_{7}   	   \,,\qquad
  a_{29}    =  -  a_{3}  - 2  a_{7}   	    \,,\\ 
    a_{30}    =  3  a_{1}  +  a_{14}   	 \,,\qquad
  a_{31}    =  -  \tfrac{29}{100}  a_{1}  -    \tfrac{3}{50}  a_{14}  -   \tfrac{7}{100}  a_{3}  -    \tfrac{7}{200}  a_{7}   	\,,\\
  a_{32}     =    \tfrac{12}{5}  a_{1}  +   \tfrac{1}{2}  a_{14}  +   \tfrac{1}{2}   a_{3}  +   \tfrac{3}{20}  a_{7}   	   \,,\qquad
  a_{33}    =    \tfrac{3}{4}   a_{1}  -    \tfrac{1}{4}  a_{14}  +   \tfrac{1}{8}  a_{3}  -   \tfrac{3}{4}  a_{7}   	   \,, \\
  a_{34}    =  -   \tfrac{9}{4}  a_{1}  -   \tfrac{5}{4}  a_{14}  -    \tfrac{7}{8}  a_{3}   	   \,,\qquad
  a_{35}    =   \tfrac{1}{4}  a_{1}  +   \tfrac{1}{4}  a_{14}  +   \tfrac{1}{8}   a_{3}  +   \tfrac{1}{8}  a_{7} 	
  \,.
\end{gathered}
\end{equation}
The expressions \eqref{tbc} have  \(10\)    unconstrained parameters (including the overall scaling \(a_1\)):
 \(a_1,a_3,a_7,a_{14}\) appearing on right-hand sides of \eqref{tbc}   and \( a_8,a_9,a_{10},a_{11},a_{12},a_{13}\) corresponding to terms
 \(T^2C^2\)  
 which are conformally invariant by themselves.
 
Notice that \eqref{tbb} produces  \(68\) equations \(b_i=0\)  for \(34\) variables. Many of them are therefore redundant, and the fact that there is indeed a family of solutions is quite reassuring.

Furthermore,   the   flat space limit of the action constructed above  exhibit an interesting property. Indeed, the action becomes
\begin{equation}\label{ioa2}
S
=\int \!   \left[ T_{ mnr }  \partial^4 T_{ mnr} 
- 6  T_{mac } \partial_m  \partial^2  \partial_n T_{nac} 
\right] 
, 
\end{equation}
which can be rewritten as
\begin{equation}\label{ioc2}
S
=\int \!  
3 \,
\partial^{a}\hat T^+_{anr} \partial^2 \partial_c \hat T _-^{cnr} 
\,,
\qquad\quad	
\hat T^{\pm}_{mnr} = \frac 12 \left( T_{mnr} \pm   \frac16 \varepsilon_{mnrabc} T^{abc} \right)\,.
\end{equation}
Notice that \(\hat T^\pm\) are \emph{not} the (anti)-self-dual parts, due to the lack of a factor   \(\rm i\)   in front of \(\varepsilon\), rather they are a mixture of the two components. 
The decomposition  \eqref{ioc2}    mimics  what happens for the 2-form in 4d conformal supergravity (cf.\ \cite{Fradkin:1985am}), where one has \( \Lagr_{T,4d} \sim T_{ mn }  \partial^2 T_{ mn} 
- 4  T_{ma } \partial_m    \partial_n T_{na} \sim \partial_m \hat T^+_{ma} \partial_n \hat T^-_{na}\)  where now
\(\hat T^{\pm}_{mn} = \frac 12 \left( T_{mn} \pm  \frac12 \varepsilon_{mnac} T^{ac} \right)\).
The similarity between the four- and six-dimensional cases is suggestive of a group-theoretic origin.
In the calculation presented in this appendix, the expression \eqref{ioa2} is a consequence of the relation \(a_2 = - 6 a_1\) (appearing in \eqref{tbc}) and is therefore entangled in the complicated system of equations discussed above.

Another interesting background is the six sphere \(S^6\). In this case, restricting the 3-form to be (anti)-self-dual, we obtain a \emph{unique} action (up to overall scaling), 
\begin{align}\label{tuba}
S
=\int \! \sqrt g\ \left[
\cd_a T^{amr } \cd^2 \cd^c T_{cmr}
-\frac 15 R \cd_a T^{amr } \cd^c T_{cmr}
\right] \,. 
\end{align} 
This  action coincides with the explicit evaluation of the \((2,0)\) conformal supergravity case discussed in \eqref{tub}.

Finally, we consider a Ricci-flat background endowed with the parallel curvature condition \(\cd_a C_{mnrs} =0\). We restrict to an (anti)-self-dual form \(T_{mnr}\) expressed in terms of a transverse \(2\)-form \(V_{mn}^\perp\) via the decomposition \eqref{dfn}. Such a 2-form has a six-derivative minimal kinetic term; imposing a factorisation with the structure
\begin{equation}\label{jhd}
S
=\int \! \sqrt g\ 
	V^{  \perp mn}  \Delta_V [\beta_1]
	\   \Delta_V [\beta_2]  \
	   \Delta_V [\beta_3]  \ V^{ \perp}_{  mn}   \,,
	   \quad
	   \Delta_V [\beta]  V^{ \perp}_{  mn} = -\cd^2  V^{ \perp}_{  mn} + \beta C_{mnac}   V^{ \perp   ac} \,,
\end{equation}
provides additional restrictions on the undetermined \(a_i\) coefficients and on the values of the parameters \(\beta_i\).
The calculation is remarkably tedious  and requires working out a number of non-trivial identities relating in particular terms of the form \(V^2C^3\), which come from the parallel-curvature condition. We simply state the result: The only combination of \eqref{tbc} compatible with factorisation of the form \eqref{jhd} on Ricci-flat backgrounds obeying the parallel curvature condition is given by 
\begin{equation}\label{fkl}
\beta_1 = \beta_2 =\beta_3 = -1\,.
\end{equation}
This is the same value discussed in \eqref{efg},\eqref{ife2} as a consequence of preservation of the transversality condition. 
The additional restrictions imposed on the \(a_i\)'s are
\begin{equation}\label{kfd}
a_{3} = 6 a_1 \,,\quad
a_7 = -12 a_1 - a_{14}  \,,\quad
a_9 = -3 a_1 -a_{10} - \tfrac12 a_{11}   \,, \quad
a_{12} =- 12 a_1   \,,\quad
a_{13}=0 \,.
\end{equation}

As a consequence, this analysis shows that the factorisation in general case of Einstein space, if possible, is necessarily of the form \eqref{ife2} discussed in the main text.

\bibliography{biblio} 

\bibliographystyle{utphys}

\end{document}